\documentclass[fleqn,usenatbib]{mnras}

\usepackage{newtxtext,newtxmath}

\usepackage[T1]{fontenc}

\DeclareRobustCommand{\VAN}[3]{#2}
\let\VANthebibliography\thebibliography
\def\thebibliography{\DeclareRobustCommand{\VAN}[3]{##3}\VANthebibliography}

\usepackage{graphicx}	
\usepackage{amsmath}	
\usepackage{multirow}   
\usepackage[utf8]{inputenc} 
\usepackage{tabularx}
\usepackage{comment}
\usepackage{scalerel}
\usepackage{tikz}
\usetikzlibrary{svg.path}

\newcommand{\plotsidesize}[2]
 {\centering \leavevmode \includegraphics[width={#2\textwidth}]{#1}}

\newcommand{\msun}{M_{\sun}}

\newcommand{\GIZMO}{{\small GIZMO}}
\newcommand{\yt}{{\small yt}}
\newcommand{\Hmol}{{\rm H}_{2}}
\newcommand{\fCO}{f_{\rm CO}}

\newcommand{\fdens}{f_{\rm dense}}
\newcommand{\nH}{n_{\rm H}}
\newcommand{\NHn}{N_{\rm H,neutral}}
\newcommand{\tcl}{\tau_{\rm cloud}}

\newcommand{\gizmourl}{\href{http://www.tapir.caltech.edu/~phopkins/Site/GIZMO.html}{\url{http://www.tapir.caltech.edu/~phopkins/Site/GIZMO.html}}}

\newcommand{\datastatement}[1]{\begin{small}\section*{Data Availability Statement}\end{small}{\noindent #1}\vspace{5pt}}

\definecolor{orcidlogocol}{HTML}{A6CE39}
\tikzset{
  orcidlogo/.pic={
    \fill[orcidlogocol] svg{M256,128c0,70.7-57.3,128-128,128C57.3,256,0,198.7,0,128C0,57.3,57.3,0,128,0C198.7,0,256,57.3,256,128z};
    \fill[white] svg{M86.3,186.2H70.9V79.1h15.4v48.4V186.2z}
                 svg{M108.9,79.1h41.6c39.6,0,57,28.3,57,53.6c0,27.5-21.5,53.6-56.8,53.6h-41.8V79.1z M124.3,172.4h24.5c34.9,0,42.9-26.5,42.9-39.7c0-21.5-13.7-39.7-43.7-39.7h-23.7V172.4z}
                 svg{M88.7,56.8c0,5.5-4.5,10.1-10.1,10.1c-5.6,0-10.1-4.6-10.1-10.1c0-5.6,4.5-10.1,10.1-10.1C84.2,46.7,88.7,51.3,88.7,56.8z};
  }
}
\newcommand\orcidicon[1]{\href{https://orcid.org/#1}{\mbox{\scalerel*{
\begin{tikzpicture}[yscale=-1,transform shape]
\pic{orcidlogo};
\end{tikzpicture}
}{|}}}}

\title[Modeling Dust Evolution in FIRE]{The Galactic Dust-Up: Modeling Dust Evolution in FIRE}

\author[C. R. Choban et al.]{
\parbox[t]{\textwidth}{
        Caleb R. Choban\orcidicon{0000-0001-9200-169X}$^{1}$\thanks{email: cchoban@ucsd.edu},
        Du\v{s}an Kere\v{s}\orcidicon{0000-0002-1666-7067}$^{1}$,
        Philip F. Hopkins\orcidicon{0000-0003-3729-1684}$^{2}$,
        Karin M. Sandstrom\orcidicon{0000-0002-4378-8534}$^{1}$,
        Christopher C. Hayward\orcidicon{0000-0003-4073-3236}$^{3}$,
        Claude-Andr{\'e} Faucher-Gigu{\`e}re\orcidicon{0000-0002-4900-6628}$^{4}$
} \vspace*{4pt} \\
$^{1}$ Center for Astrophysics and Space Sciences (CASS), University of California San Diego, 9500 Gilman Drive, La Jolla, CA 92093, USA \\
$^{2}$ TAPIR, Mailcode 350-17, California Institute of Technology, Pasadena, CA 91125, USA \\
$^{3}$ Center for Computational Astrophysics, Flatiron Institute, 162 5th Ave., New York, NY 10010 USA \\
$^{4}$ Department of Physics and Astronomy and CIERA, Northwestern University, 2145 Sheridan Road, Evanston, IL 60208, USA \\
}

\date{Accepted XXX. Received YYY; in original form ZZZ}

\pubyear{2022}

\begin{document}
\label{firstpage}
\pagerange{\pageref{firstpage}--\pageref{lastpage}}
\maketitle

\begin{abstract}
Recent strides have been made developing dust evolution models for galaxy formation simulations but these approaches vary in their assumptions and degree of complexity. Here we introduce and compare two separate dust evolution models (labelled `Elemental' and `Species'), based on recent approaches, incorporated into the \GIZMO\ code and coupled with FIRE-2 stellar feedback and ISM physics. Both models account for turbulent dust diffusion, stellar production of dust, dust growth via gas-dust accretion, and dust destruction from time-resolved supernovae, thermal sputtering in hot gas, and astration. The ``Elemental'' model tracks the evolution of generalized dust species and utilizes a simple, `tunable' dust growth routine, while the ``Species'' model tracks the evolution of specific dust species with set chemical compositions and incorporates a physically motivated, two-phase dust growth routine. We test and compare these models in an idealized Milky Way-mass galaxy and find that while both produce reasonable galaxy-integrated dust-to-metals (D/Z) ratios and predict gas-dust accretion as the main dust growth mechanism, a chemically motivated model is needed to reproduce the observed scaling relation between individual element depletions and D/Z with column density and local gas density. We also find the inclusion of theoretical metallic iron and O-bearing dust species are needed in the case of specific dust species in order to match observations of O and Fe depletions, and the integration of a sub-resolution dense molecular gas/CO scheme is needed to both match observed C depletions and ensure carbonaceous dust is not overproduced in dense environments.

\end{abstract}




\section{Introduction}

Although it only makes up 1\% of the interstellar medium (ISM) by mass in the Milky Way \citep{whittet_2003:DustGalacticEnvironment}, dust is integral to the physics within. It provides a surface for complex astrochemistry such as H$_2$ formation \citep{hollenbach_1971:SurfaceRecombinationHydrogen}, facilitating the formation of molecular clouds and indirectly stars. It is a key coolant in extremely dense environments \citep{goldsmith_1978:MolecularCoolingThermal,burke_1983:GasgrainInteractionInterstellar,goldsmith_2001:MolecularDepletionThermal}, important for star and planet formation, is the primary heat source via the photoelectric effect in neutral phases of the ISM \citep{tielens_1985:PhotodissociationRegionsBasic,hollenbach_1991:LowDensityPhotodissociationRegions,weingartner_2001:PhotoelectricEmissionInterstellar}, and can reduce the abundance of important gas coolants, which allow gas to cool and collapse, by locking away elements from the gas phase. Dust also affects how the ISM reacts to radiation pressure and might help drive galactic winds  \citep{murray_2005:MaximumLuminosityGalaxies,thompson_2015:DynamicsDustyRadiationpressuredriven}, which can be crucial for shaping galaxy evolution. Observationally, dust redistributes the stellar spectral energy distribution (SED) shifting optical-UV light to infrared affecting all observations to varying degree \citep[e.g][]{salim_2020:DustAttenuationLaw}. This fact along with the spatial distribution of dust within galaxies is especially critical for constraining the star formation rate density (SFRD) \citep{madau_2014:CosmicStarFormationHistory}. For example, the exact dust geometry (clumpiness and covering fraction) and relative distribution between dust and stars has dramatic effects on the effective attenuation law and thus the IRX-$\beta_{\rm UV}$ relation \citep{narayanan_2018:TheoryVariationDust, liang_2021:IRXvRelationHighredshift}, which is a useful tool to constrain the attenuation properties at UV wavelengths of high-z galaxies.

In order for these physical processes and observations to be accurately modelled and predicted in simulations a detailed understanding of dust evolution on galactic/cosmological scales is needed. Currently many galaxy formation models treat dust in post-processing or assume a constant dust-to-metals ratio (D/Z) \citep[e.g.][]{hayward_2011:WhatDoesSubmillimeter,narayanan_2015:FormationSubmillimetrebrightGalaxies,camps_2016:FarinfraredDustProperties,trayford_2017:OpticalColoursSpectral,trayford_2020:FadeGreySystematic,liang_2018:SubmillimetreFluxProbe,liang_2019:DustTemperaturesHighredshift,ma_2019:DustAttenuationDust}, and therefore may not accurately predict observed diverse dust scaling relations and their evolution. For example, in the Milky Way, observations of gas-phase element depletions (fraction of elements missing from the gas-phase assumed to be locked in dust) indicate a strong correlation between the total amount of metals in dust and gas density along with a varying dust population composition \citep{jenkins_2009:UnifiedRepresentationGasPhase}. Similar relations have also been found for the Magellanic Clouds but with systematically smaller fractions of metals in dust \citep{jenkins_2017:InterstellarGasphaseElement,roman-duval_2019:METALMetalEvolution}. Outside the MW and its satellites, galaxy surveys of 126 local galaxies by \citet{remy-ruyer_2014:GastodustMassRatios} and $\sim$500 local galaxies by \citet{devis_2019:SystematicMetallicityStudy} found an overall increase of D/Z with metallicity with a large (>1 dex) scatter in the galaxy-integrated D/Z at a given metallicity. Interestingly, these studies disagree on whether this relation extends to high metallicity ($12+\log_{10}{\rm(O/H)}>8.1$) or becomes approximately constant. Recent spatially-resolved studies of individual, local galaxies also show D/Z variation with local gas properties \citep{roman-duval_2017:DustAbundanceVariations,chiang_2018:SpatiallyResolvedDusttometals,chiang_2021:ResolvingDusttoMetalsRatio,vilchez_2019:MetalsDustContent}. Furthermore, the current paradigm of carbonaceous-silicate dust chemical composition based on observed dust extinction curves \citep[e.g.][]{draine_2007:InfraredEmissionInterstellar} does not fully agree with the above observations. In particular, observations of O and Fe gas-phase depletions \citep{whittet_2010:OxygenDepletionInterstellar,dwek_2016:IronKeyElement} cannot be explained by silicate dust alone, suggesting currently unknown dust species. This can be especially important for the expected D/Z since O makes up a large fraction of the metal mass. These all suggest a complex dust system which depends heavily on the local gas properties the dust resides in and possibly on the history of the galaxy as a whole. This necessitates a more detailed modelling of dust in galaxy simulations to accurately account for stellar feedback and the effects of dust on galaxy evolution. Furthermore, detailed dust modelling will help interpret and guide observations such as the predicted amount and composition of dust populations in various gas and galactic environments which determine the expected dust extinction curves and emission spectra.

An accurate model of dust evolution on galactic scales needs to account for the main mechanisms of the dust life cycle. It is generally believed that under typical ISM conditions in Milky Way-like galaxies three processes dominate: i) production from stellar sources that create the initial `seeds' of the dust grain populations, ii) dust growth by gas-phase metal accretion onto preexisting dust grains, and iii) dust destruction by sputtering in supernovae (SNe) shocks and in hot, diffuse gas. An accurate model for these processes must also be coupled with a chemical evolution model since the evolution of dust is directly dependent on the evolution of refractory elemental abundances during a galaxy's life. One of the first detailed analytic models to accomplish this was developed by \citet{dwek_1998:EvolutionElementalAbundances}. They integrated a dust evolution model into a one-zone and one-phase (averaging over properties of the ISM and vertical direction of the disk) chemical evolution model of the Milky Way. Although they were unable to accurately model spatial variations in dust properties or dust formation in molecular clouds, their model reproduced observed galaxy-integrated dust properties (specifically a steady-state D/Z $\approx$ 0.4 similar to that in the MW; \citealt{dwek_1998:EvolutionElementalAbundances,draine_2007:InfraredEmissionInterstellar}).

Owing to the success and simplicity of  \citet{dwek_1998:EvolutionElementalAbundances} model, it has served as the core framework for most recent numerical dust evolution models on galactic scales. One of the first of these was presented in \citet{bekki_2013:CoevolutionDustGas} who modeled dust evolution coupled to gas in smoothed particle hydrodynamics (SPH) models of disk galaxies and added modeling of H$_2$ formation on the surface of dust grains. This model's main limitation was an assumed constant accretion and destruction timescale that was independent of local ISM conditions. In later work \citep{bekki_2015:CosmicEvolutionDust}, the accretion timescale is scaled with the density and temperature of the gas. In \citet{bekki_2015:DustregulatedGalaxyFormation}, a live dust particle model was introduced which decoupled the gas and dust particles. A dust evolution model has also been implemented in the moving-mesh code AREPO  and used in a suite of cosmological zoom-in simulations of Milky Way-like galaxies \citep{mckinnon_2016:DustFormationMilky}. Their model closely follows that of \citet{bekki_2015:CosmicEvolutionDust}, with a coupling of the dust destruction timescale to the local supernova rate being the main modification. Supplemental work added thermal gas-dust sputtering \citep{mckinnon_2017:SimulatingDustContent} and decoupling of gas and dust particles with dust grain size evolution \citep{mckinnon_2018:SimulatingGalacticDust}, with this framework becoming a recent staple for numerous galaxy simulations \citep[e.g.][]{li_2019:DusttogasDusttometalRatio,aoyama_2020:GalaxySimulationEvolution}. In another vein, a more comprehensive analytical dust evolution model was presented by \citet{zhukovska_2008:EvolutionInterstellarDust} which included detailed AGB dust yields, incorporated a molecular cloud evolution model to restrict dust accretion to only molecular clouds, and restricted dust accretion based on the chemical make up of dust species. In later works \citep{zhukovska_2016:ModelingDustEvolution,zhukovska_2018:IronSilicateDust}, the model was incorporated into short-term hydrodynamic simulations of a Milky-Way like disc with molecular clouds and the gas-dust accretion was modified to account for temperature dependent sticking efficiencies and ion-grain interactions for both silicate and metallic iron dust species. A middle ground between these divergent methods has also been presented by \citet{granato_2021:DustEvolutionZoomin} which incorporated chemically distinct dust species, similar to \citet{zhukovska_2008:EvolutionInterstellarDust}, with two-size approximate dust grain size evolution \citep{hirashita_2015:TwosizeApproximationSimple, aoyama_2017:GalaxySimulationDust,gjergo_2018:DustEvolutionGalaxy} and was used in cosmological zoom-in simulations of an isolated disc galaxy with a sub-resolution star
formation and feedback model. While all of these models agree with observations to varying degrees a detailed comparison between them under similar conditions has not been carried out.

In this work, we develop two separate implementations of dust evolution based on the approaches discussed above with a more detailed and varied set of dust physics. One tracks generalized dust species and utilizes a simple, `tunable' dust growth routine and the other tracks specific dust species with set chemical compositions and incorporates a physically motivated, two-phase dust growth routine. We integrate both of these dust evolution routines into the magneto-hydrodynamics meshless-finite mass code \GIZMO\ coupled with the FIRE-2 (Feedback in Realistic Environments)\footnote{See the FIRE project web site: \url{http://fire.northwestern.edu}} model for stellar feedback and ISM physics, which is the first application for both models in simulations which resolve whole galaxies and routinely resolve gas phase structure down to ${\lesssim} 10$ K and molecular cloud core densities of ${\sim} 10^4\;{\rm cm}^{-3}$. We also incorporate a sub-resolution treatment of dense molecular gas chemistry into both routines, and in the case where we track specific dust species, we investigate the inclusion of theoretical dust species to account for the gas-phase depletions of O and Fe.

This paper is organized as follows. We layout the framework for our dust evolution model and describe the separate ``Elemental'' and ``Species'' implementations in Section~\ref{Methods}. In Section~\ref{Results}, we describe the simulations we use for both implementations, testing the sensitivity of our models to free parameters in Section~\ref{Free Parameters} and comparing the results of our models to local Milky Way and extragalactic observations in Section~\ref{Element Depletions} and~\ref{Extragalactic}. Finally, we present our conclusions in Section~\ref{Conclusions}.


\section{Methods} \label{Methods}

To study the evolution of dust in and its effects on galaxies we utilize simulations of an idealized, non-cosmological Milky Way-like galaxy running two separate dust evolution models incorporated into the \GIZMO\ code base \citep{hopkins_2015:NewClassAccurate} and coupled with FIRE-2 stellar feedback and ISM physics. FIRE-2 is an update of the FIRE star-formation and stellar feedback model \citep{hopkins_2014:GalaxiesFIREFeedback}. Detailed description is available in  \citet{hopkins_2018:FIRE2SimulationsPhysics}, but a general overview along with the modifications we made are explained below in Sec.~\ref{GFM}. The initial conditions used in our simulations are presented in Sec.~\ref{ICs}. Our two dust evolution models are based on earlier work by \citet{bekki_2013:CoevolutionDustGas,mckinnon_2016:DustFormationMilky,mckinnon_2017:SimulatingDustContent} or \citet{zhukovska_2008:EvolutionInterstellarDust, zhukovska_2016:ModelingDustEvolution, zhukovska_2018:IronSilicateDust} that we extend, modify, and adjust to the FIRE-2 model as explained in detail in Sec.~\ref{Dust Evo}. 

\subsection{Galaxy Feedback Mechanisms} \label{GFM}

All simulations in this work are run with the GIZMO code base in the meshless finite-mass (MFM) mode with FIRE-2 model of star formation, and stellar feedback. FIRE-2 incorporates multiple sources of stellar feedback, specifically stellar winds (O/B and AGB), ionizing photons, radiation pressure, and supernovae (both Types Ia and II). Gas cooling is followed self-consistently for T = 10 - $10^{10}$ K including free-free, Compton, metal-line, molecular, fine-structure, and dust collisional processes while gas is also heated by cosmic rays, photo-electric, and photoionization heating by both local sources and a uniform but redshift dependent meta-galactic background \citep{faucher-giguere_2009:NewCalculationIonizing}, including the effect of self-shielding\footnote{Note that all cooling and heating processes and radiative transfer modeled in FIRE-2 are not coupled with our dust evolution models. Specifically, dust heating and cooling and radiative transfer assume a constant D/Z ratio and metal-line cooling assumes no metals are locked in dust. In future works, we will fully integrate our dust evolution models with FIRE and investigate the effects on galaxy evolution.}. Star formation is only allowed in cold, molecular, and locally self-gravitating regions with number densities above $\nH = 1000 \, {\rm cm}^{-3}$. 

Each star particle represents a stellar population with a known mass, age, and metallicity assuming a \citet{kroupa_2002:InitialMassFunction} initial mass function (IMF) from $0.1-100\; \rm \msun$. The luminosity, mass loss rates, and SNe II rates of each star particle are calculated based on the STARBURST99 \citep{leitherer_1999:Starburst99SynthesisModels} libraries, and SNe Ia rates following \citet{mannucci_2006:TwoPopulationsProgenitors}. Metal yields from SNe II, Ia, and AGB winds are taken from \citet{nomoto_2006:NucleosynthesisYieldsCorecollapse}, \citet{iwamoto_1999:NucleosynthesisChandrasekharMass}, and \citet{izzard_2004:NewSyntheticModel} respectively. Evolution of eleven species (H, He, C, N, O, Ne, Mg, Si, S, Ca, and Fe) is tracked for each gas cell. Sub-resolution turbulent metal diffusion is modeled as described in \citet{su_2017:FeedbackFirstSurprisingly} and \citet{escala_2018:ModellingChemicalAbundance}. For future reference, FIRE-2 adopts the older \citet{anders_1989:AbundancesElementsMeteoritic} solar metal abundances with $Z\sim 0.02$ so whenever we mention solar abundances we are referring to the Andres \& Gravesse abundances.\footnote{In Appendix~\ref{Appendix_FIRE23} we provide a preliminary comparison of our dust evolution model results incorporated into both FIRE-2 and FIRE-3. While FIRE-3 incorporates numerous improvements, including adoption of the newer \citet{asplund_2009:ChemicalCompositionSun} proto-solar abundances with $Z\sim0.014$, we find our results are primarily sensitive to the adopted nucleosynthesis yields.}

FIRE has been successful in matching a wide range of observations related to galaxies, including the mass-metallicity relation and its evolution over redshift \citep{ma_2016:OriginEvolutionGalaxy} and the Kennicutt–Schmidt star formation law \citep{orr_2018:WhatFIREsStar}. This success is owed to the high resolution, star formation criteria, cooling to low temperatures, and multi-channel stellar feedback of FIRE, all of which result in a reasonable ISM phase structure and giant molecular cloud (GMC) mass function \citep{benincasa_2020:LiveFastYoung}. These also lead to the self-consistent development of galactic winds that eject large amounts of gas \citep{muratov_2015:GustyGaseousFlows, angles-alcazar_2017:CosmicBaryonCycle} and metals \citep{muratov_2017:MetalFlowsCircumgalactic, hafen_2019:OriginsCircumgalacticMedium,pandya_2021:CharacterizingMassMomentum} out of galaxies, preventing excessive star formation and leading to a plausible stellar-halo mass relation.

Furthermore, the FIRE model is ideally suited for examining dust evolution due to its in-depth treatment of the multi-phase ISM and tracking of principle heavy elements that make up carbonaceous and silicate dust in gaseous form as a product of stellar evolution. Also, in contrast to most other cosmological simulations, FIRE time-resolves individual SNe events \citep{hopkins_2018:HowModelSupernovae} and models their interaction with the ISM. This is particularly relevant for dust evolution, since SNe are one of the main creators and destroyers of dust. Being able to time-resolve individual SNe events allows us to track the local variability of dust in the ISM.

For this work, we made a few, specific, changes to the underlying stellar population model in FIRE to enable more accurate treatment of dust production from AGB stars.  Specifically, the age at which a stellar population switches from producing a majority O/B winds to AGB winds is changed from 100 Myr to 37.5 Myr to match the end of SNe II, and the stellar winds mass return is modified to more accurately follow results from \citet{leitherer_1999:Starburst99SynthesisModels} past stellar ages of 3.5 Myr, specifically the IMF-average mass-loss rate of a stellar population of mass $M_*$ and age $t_{\rm Myr}$ in Myr is $\dot{M}_{\rm w}=29.4(t_{\rm Myr}/3.5)^{-1.3} M_*$ Gyr$^{-1}$ for $t_{\rm Myr}>3.5$. Together, these changes increase the cumulative AGB stellar yields by a factor of ${\sim}2.5$. We will later show in Sec.~\ref{Free Parameters} that these changes only affect the early-time dust production, and have no effect on the steady-state dust population.

\subsection{Initial Conditions} \label{ICs}

In this paper, we utilize an idealized, non-cosmological Milky Way-like galaxy. Specifically we initialize a disc galaxy with $M_{\rm disc,gas} = 0.9 \times 10^{10} \, \msun$ and $M_{\rm disc,*}=4.7\times10^{10} \, \msun$ and an exponential gas and stellar density profile $\rho(R,z)\propto e^{-R/R_{\rm d}}e^{-|z|/z_{\rm d}}$ with radial scale lengths of $R_{\rm d,gas} =6.4$ kpc and $R_{\rm d,*} =3.2$ kpc respectively and vertical scale length of $z_{\rm d}= 0.25$ kpc for both. We also include a stellar bulge with $M_{\rm bulge}=1.5\times10^{10} \, \msun$ and a Hernquist profile~\citep{hernquist_1990:AnalyticalModelSpherical}. The galaxy is embedded in a NFW profile \citep{navarro_1996:StructureColdDark} dark matter halo with $M_{\rm vir} = 1.5 \times 10^{12} \, \msun$ and halo concentration of $c=12$. We use a gas cell mass resolution of ${\sim} 2\times10^4 \, \msun$ with adaptive softening lengths, achieving a minimum softening length of $\epsilon^{\rm MIN}_{\rm gas}\approx10$ pc at simulation end. The dark matter particles have a mass resolution of ${\sim} 3\times10^6 \, \msun$ with a universal softening length of $\epsilon_{\rm DM}\approx40$ pc.
All gas cells and star particles start with an initial $Z = Z_{\odot}$, star particles initially have a uniform age distribution over 13.8 Gyr, and gas cells are initially free of dust. The galaxy was simulated for $\sim$1.5 Gyr, long enough for all dust evolution models to reach a steady-state D/Z ratio and dust population composition as we show in Sec.~\ref{Free Parameters}. The galaxy undergoes a roughly constant star formation rate of ${\sim}1 \, \msun/{\rm yr}$ throughout the simulation, producing a median gas metallicity within the galactic disc of $Z\approx1.3 Z_{\odot}$ at simulation end. To test the sensitivity of our results to our initial metallicity and dust population we ran two simulations, using our preferred dust evolution model, with either reduced initial metallicity or an initial dust population. We present the specific initial conditions and their results in Appendix~\ref{Appendix_ICs}, finding little difference in our results beyond small systematic offsets.

\subsection{Dust Evolution} \label{Dust Evo}

In this work we examine two separate implementations of dust evolution whose general methodology can be traced to the semi-analytic model of \citet{dwek_1998:EvolutionElementalAbundances}:

\subsubsection{Dust by Element: ``Elemental''} 

This implementation is motivated by and largely follows numerical hydrodynamic galaxy simulations by \citet{bekki_2013:CoevolutionDustGas} and \citet{mckinnon_2016:DustFormationMilky, mckinnon_2017:SimulatingDustContent}. It follows the evolution of individual elements (C, O, Si, Mg, and Fe) within dust, assuming they comprise carbonaceous and generalized (no set chemical composition) silicate dust species. A major consequence of such an approach, which we discuss in detail later on, is the possibility of locking the entirety of all elements into dust. This implementation also relies on explicit tuning of a few free parameters which encompass a vast range of `sub-grid' dust physics and is effectively single-phase since it does not restrict dust processes to certain gas environments, most notably gas-dust accretion. In this dust evolution model we track the fraction of mass for each element locked up in dust for each gas cell. 

\subsubsection{Dust by Species: ``Species''} \label{Dust_by_Species}

We have also implemented a more physically and chemically motivated dust evolution model based on \citet{zhukovska_2008:EvolutionInterstellarDust, zhukovska_2016:ModelingDustEvolution, zhukovska_2018:IronSilicateDust}. This implementation tracks the evolution of specific dust species (silicates, carbonaceous, and silicon carbide), concentrating on the most abundant species that originate from stars that are also found as presolar dust grains in meteorites. We also consider metallic iron dust since it should theoretically be produced in stellar outflows and SNe and may be needed to explain observed Fe depletions which cannot be explained by Fe in silicate dust alone \citep{dwek_2016:IronKeyElement}. In this implementation we track both the fraction of mass for each element locked up in dust for each gas cell and the mass fraction of each gas cell comprised of each dust species (e.g. $M_{\rm dust,silicate}/M_{\rm gas}$).
This implementation is also effectively two-phase due to the inclusion of a more 
physically motivated dust growth model which is discussed below. Owing to its more realistic accounting of elements locked in dust and complex dust growth model this is our preferred implementation. Several aspects require special attention in this method as discussed below (common approaches between our two methods are discussed as they appear).

{\bf Carbonaceous:} For both implementations the fraction of total carbon locked up in CO molecules ($\fCO$) must be taken into account since it is unavailable for carbonaceous dust growth, limiting the maximum formable amount. Observations indicate a $\fCO$ ratio of 20\% to 40\% \citep[e.g.][]{irvine_1987:ChemicalAbundancesMolecular, vandishoeck_1993:ChemicalEvolutionProtostellar, vandishoeck_1998:ChemicalEvolutionStarForming,lacy_1994:DetectionAbsorptionH2} in Milky Way molecular clouds. It has also been found that atomic C to CO formation progresses rapidly as gas transitions from the diffuse to dense molecular regime \citep{liszt_2007:FormationFractionationExcitation,burgh_2010:AtomicMolecularCarbon}, making CO the dominant host of gas-phase C in dense molecular clouds. Therefore, to accurately determine $\fCO$ beyond assuming a set fraction we must know what gas is in the dense molecular regime and track its evolution. Since typical FIRE-2 simulations only resolve the high-mass end ($>10^5 \; \msun$) of the GMC spectrum \citep{benincasa_2020:LiveFastYoung}, we devised a sub-resolution prescription to track the mass fraction of each gas cell in the dense molecular phase $\fdens$ (where we assume nearly all gas-phase metals are neutral and specifically gas-phase carbon is almost completely molecular in the form of CO), and with it $\fCO$, which is described in Appendix \ref{Appendix_MC}. Note that this prescription takes into account the depletion of gas-phase C into dust (which limits the maximum formable amount of CO) when calculating $\fCO$.

{\bf Silicates:}
Our prescription for silicate dust composition follows that in \citet{zhukovska_2008:EvolutionInterstellarDust} consisting of an olivine 
($[\rm Mg_x Fe_{(1-x)}]_2 Si O_4$) and pyroxene ($\rm Mg_x Fe_{(1-x)} Si O_3$) 
mixture which is assumed to be constant. The fraction of olivine in the mixture is represented as $f_{\rm ol}$. The fraction of the silicate structure (assumed to be the same for olivine and pyroxene for simplicity) incorporating Mg is represented as $x$.\footnote{For clarity, olivine is comprised of both $\rm Mg_2 Si O_4$ and $\rm Fe_2 Si O_4$ and similarly pyroxene is a mixture of $\rm Mg Si O_3$ and $\rm Fe Si O_3$. So $x$ represents the fraction of the olivine and pyroxene structure incorporating Mg, while the fraction incorporating Fe is $(1-x)$.}
Traditionally $x$ and $f_{\rm ol}$ are chosen to reproduce observed depletions of Si and Mg, assuming depleted gas-phase Si and Mg is predominantly locked up in silicate dust. With the observed number abundance ratio of elements Si and Mg bound in dust ($A_{\rm Mg}/A_{\rm Si}$), $f_{\rm ol}$ is related to $x$ through the simple relation $f_{\rm ol}=\frac{A_{\rm Mg}}{x A_{\rm Si}} - 1$. One issue with this method is its sensitivity to the assumed solar abundances (particularly either \citet{lodders_2003:SolarSystemAbundances} or \citet{asplund_2009:ChemicalCompositionSun} solar abundances) which give observed values of $A_{\rm Mg}/A_{\rm Si}=1.02-1.3$\footnote{Assuming \citet{anders_1989:AbundancesElementsMeteoritic} solar abundances gives $A_{\rm Mg}/A_{\rm Si}\sim1.08.$} in the cold neutral medium (CNM) \citep{dwek_2005:InterstellarDustWhat, jenkins_2009:UnifiedRepresentationGasPhase, draine_2011:PhysicsInterstellarIntergalactic}. 
A more recent approach involves direct observation of dust absorption features in the spectra of bright X-ray binaries in combination with direct synchrotron measurements of X-ray absorption fine structure features for numerous silicate compositions. These observations are focused near the Galactic center and probe dust composition and structure in dense environments ($N_{\rm H}\geq0.5\times10^{22} \; {\rm cm}^{-3}$), giving $f_{\text{ol}}$ close to unity and $x\approx0.5$ \citep{zeegers_2019:DustAbsorptionScattering,rogantini_2020:MagnesiumSiliconInterstellar}. These results are complicated by the small sample size and the fact that they probe a much denser phase of the ISM than the CNM. There are also X-ray observations in the CNM which point to entirely iron-free ($x=1$) silicate crystal structure  with separate metallic iron inclusions \citep{costantini_2012:XMMNewtonObservation4U}.\footnote{The true role of iron in silicate chemical composition will be better understood with future instruments allowing for the direct X-ray observation of Mg, Si, and Fe absorption K-edges simultaneously \citep{rogantini_2018:InvestigatingInterstellarDust}} 
Due to these uncertainties we chose to follow the traditional approach, assuming $A_{\text{Mg}}/A_{\text{Si}}=1.06$ and set $x=0.65$, and thus $f_{\text{ol}}=0.63$. We determine this value for $x$ by matching the observed silicate-to-carbon dust mass ratio of 2 in the local diffuse ISM \citep{dwek_2005:InterstellarDustWhat} given our maximum theoretical carbon and silicate dust masses for our assumed solar metal abundances and silicate dust composition. Our choice of $A_{\text{Mg}}/A_{\text{Si}}$ does not greatly affect the maximum amount of formable silicate dust, with only a change of $<10\%$ between $A_{\text{Mg}}/A_{\text{Si}} = 1.02-1.3$, but it does affect the exact depletion patterns of Mg and Si which we comment on in more detail later.

{\bf Oxygen:} Additional discretion must be given with regards to oxygen in silicate dust since observed oxygen depletions in the Milky Way \citep{jenkins_2009:UnifiedRepresentationGasPhase} cannot be accounted for by silicate dust alone \citep{whittet_2010:OxygenDepletionInterstellar}. If we allow oxygen to only deplete into silicate grains, in the ``Species'' implementation, the maximum possible oxygen depletion and resulting D/Z ratio end up markedly lower than what is observed. The question of where this oxygen goes is still open with a plethora of proposed candidates such as thick ice mantles on large dust grains \citep{poteet_2015:CompositionInterstellarGrains}, $\mu$m-sized ice grains \citep{wang_2015:InterstellarOxygenCrisis}, and organic carbonates on the surface of dust grains \citep{jones_2019:EssentialElementsDust}. We thus  opt for a simple and optional inclusion of an Oxygen Reservoir (O-reservoir) dust species set to match observed oxygen depletion which we describe below.

First, we determine the dependence of gas-phase oxygen depletion on the hydrogen number density. To this end, we use observations of oxygen depletion and the derived relation between oxygen depletion and mean sight line neutral hydrogen number density, $\left< n_{\rm H,neutral} \right>$ [cm$^{-3}$], from \citet{jenkins_2009:UnifiedRepresentationGasPhase} to define the fractional amount of O in dust, $D(\rm O)$, as 
\begin{equation} \label{O_depletion}
    D(\rm O)
    =
    1 - 0.654 \left(\frac{1\;{\rm cm}^{-3}}{\left< n_{\rm H,neutral} \right>}\right)^{0.1} .
\end{equation}

This relation is observed up to $\left< n_{\rm H,neutral} \right> \approx 10$ cm$^{-3}$, but we extrapolate it to higher densities. This also does not consider O in CO or in O-bearing ``ices'' as neither of these depletion sources exist over the observed range. With this in mind, we set the maximum of this relation to $1-f_{\text{O in CO}}$ where $f_{\text{O in CO}}$ is the fractional amount of O in CO derived from our prescription for tracking C in CO discussed earlier. Since converting mean sight line density $\left< n_{\rm H,neutral} \right>$ to physical 3D density $n_{\rm H,neutral}$ is a complicated, multi-faceted problem we assume $\left< n_{\rm H,neutral} \right>\approx n_{\rm H,neutral}$ for simplicity, which can be taken as an upper bound on the expected depletion at a given number density in our simulations since $\left< n_{\rm H,neutral} \right>$ will always be significantly lower than the true physical density. 
 
In addition, assuming this O-reservoir dust species is tied to the local amount of dust, and to enable variation in $D(\rm O)$, we scale $D(\rm O)$ by the fraction of the maximum formable amount of silicate dust currently present in the gas cell, $f_{\rm sil}$, given local element abundances.
Thus the fraction of oxygen in the gas we put into the O-reservoir is
\begin{equation} \label{f_ORes}
    f_{\text{O-res}} = f_{\rm sil} \, D({\rm O}) - f_{\text{O in sil}}
\end{equation}
where $f_{\text{O in sil}}$ is the maximum fraction of oxygen that can be trapped in silicate dust. Note we only manually set $f_{\text{O-res}}$ to match Eq. \ref{f_ORes} when silicate dust grows through gas-dust accretion. Otherwise the O-reservoir is treated as its own distinct dust population experiencing the same destruction processes as the other dust species. The ``Elemental'' routine avoids this unidentified oxygen depletor issue by allowing oxygen to accrete freely onto dust grains assuming it is in the form of water ice as stated in \citet{dwek_1998:EvolutionElementalAbundances}.

{\bf Iron:} The exact form of solid-phase iron dust is unknown and no easily identifiable spectroscopic features exist making direct observation of such species difficult. Two prominent theories for solid-phase iron are free-flying iron nanoparticles \citep{gioannini_2017:NewGalacticChemical,hensley_2017:ThermodynamicsChargingInterstellar} and iron and FeS inclusions in silicate grains \citep{min_2007:ShapeCompositionInterstellar,jones_2013:EvolutionAmorphousHydrocarbons}, with in situ studies on interstellar grains demonstrating silicate particles containing iron and the existence of  individual iron particles \citep{westphal_2014:EvidenceInterstellarOrigin,altobelli_2016:FluxCompositionInterstellar}. Also, as previously mentioned, the exact role of atomic iron in silicate dust chemical composition, either directly integrated in the silicate crystal structure, incorporated purely as metallic iron inclusions, or some mixture of the two, is unclear. With this in mind we examine two separate prescriptions for iron dust. 1) {\bf Normal-iron} assumes entirely free-flying metallic iron dust with the same grain size distribution as silicates such as that implemented in \citet{zhukovska_2008:EvolutionInterstellarDust}, and 2) {\bf Nano-iron} assumes free-flying metallic iron nanoparticles which can be locked into silicate dust as inclusions. Specifically, if any silicate dust is present we lock a set fraction $f_{\rm incl}=0.7$ of this nanoparticle dust into silicates as inclusions which are protected from SNe destruction, unless all dust is destroyed, and unavailable for gas-dust accretion as implemented in \citet{zhukovska_2018:IronSilicateDust}. For simplicity, we assume that the metallic iron dust inclusions contribute to the atomic iron needed in the aforementioned silicate dust composition. This means silicate dust growth via gas-dust accretion will not be hampered by the depletion of gas-phase iron into metallic iron dust since said dust is effectively accreting onto silicate dust and then locked into the dust structure as inclusions. One caveat of this prescription is the total amount of iron in the silicate dust structure can exceed the amount given by our choice of $f_{\rm ol}$ and $x$ discussed earlier.\footnote{Note that the difference in optical properties of silicate dust with iron inclusions versus iron being directly included in the silicate structure are small \citep[][see Appendix A]{jones_2013:EvolutionAmorphousHydrocarbons}} Any differences between these prescriptions will be noted in the proceeding sections.

\begin{figure*}
    \plotsidesize{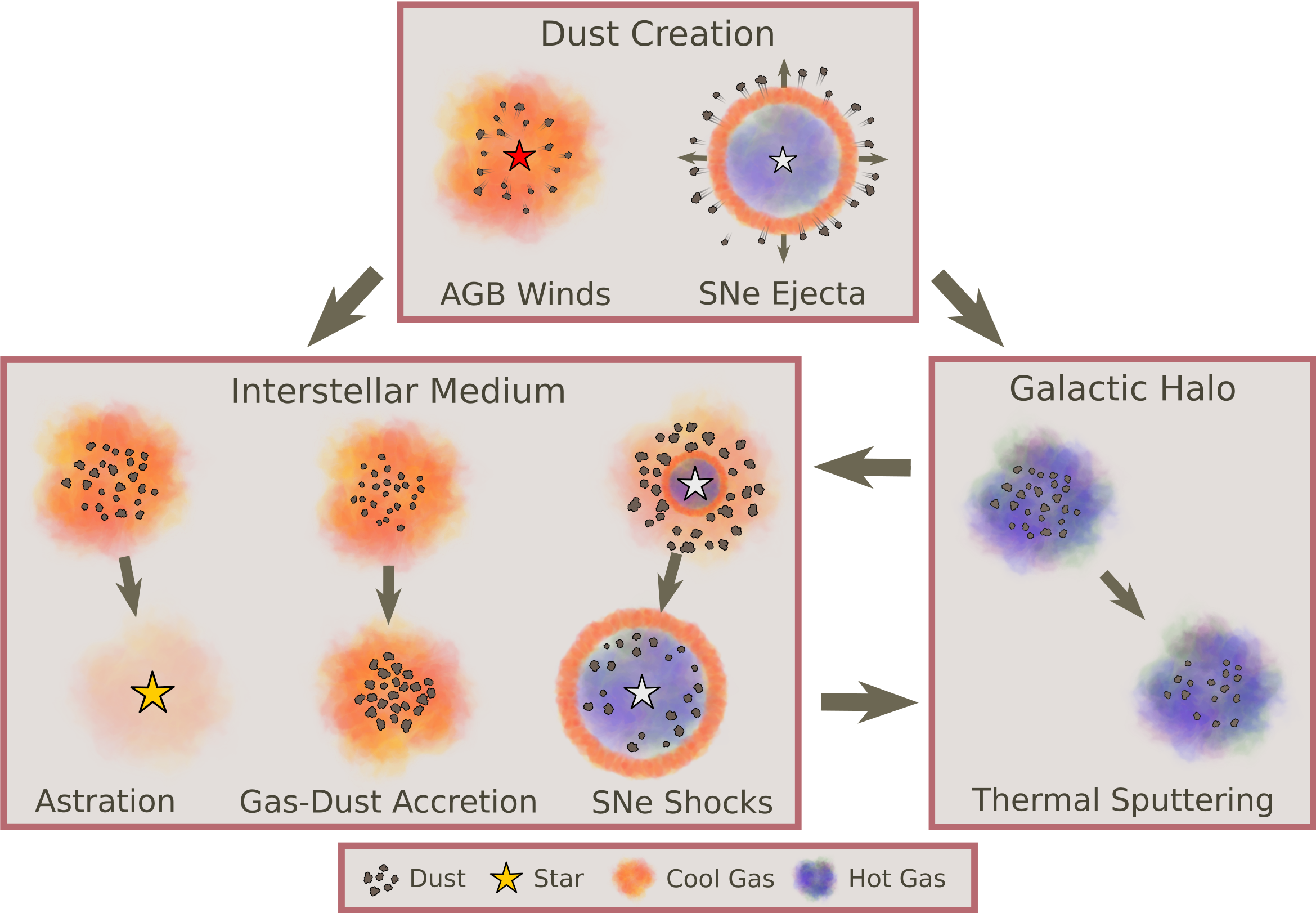}{0.9}
    \vspace{-0.25cm}
    \caption{Pictorial representation of mechanisms composing the dust life cycle included in our dust evolution models. {\bf Dust Creation:} The initial `seeds' of the dust population are created in the stellar ejecta of SNe and AGB winds where a portion of ejected metals condense into dust. Once these `seeds' have been created they spend their life in the ISM and/or the galactic halo where they are exposed to various processes. {\bf Gas-Dust Accretion:} In cool, dense phases of the ISM, gas-phase metals accrete onto the surface of preexisting dust grains growing the grains over time. This is believed to be the main source of dust mass in the MW. {\bf SNe Shocks:} As supernovae remnants propagate through the ISM they destroy and shatter dust grains residing in the ISM via grain-grain collisions, thermal sputtering, and non-thermal sputtering. This is believed to be the main destroyer of dust in the MW. {\bf Astration:} As gas cools and collapses forming stars, dust residing in said gas is also destroyed and contributes to the stellar metallicity. {\bf Thermal Sputtering:} Dust grains residing in hot gas, such as in the galactic halo, are eroded and destroyed by energetic atoms.}
    \label{fig:dust_lifecycle}
\end{figure*}

Both implementations include the dominant sources of dust production and the dominant dust destruction mechanisms. Specifically, we track and differentiate between dust created from SNe Ia and II, AGB stars, and gas-phase accretion in the ISM and account for dust destroyed by SNe shocks, thermal sputtering, and astration. We also incorporate sub-resolution turbulent dust diffusion in the same manner as the turbulent metal diffusion already in FIRE-2. Other mechanisms, such as dust shattering and coagulation, will be left to future work. An illustration of these mechanisms can be seen in Figure~\ref{fig:dust_lifecycle}. We now describe these processes in more detail.

\subsection{Dust Creation}

\subsubsection{Dust Creation by SNe}

 The exact dust yields from SNe are not well known and so both of our implementations follow the same simplified prescriptions from \citet{dwek_1998:EvolutionElementalAbundances} which assumes a set fraction of the metal yields from SNe condense into dust. 

{\bf Elemental:} For this implementation we follow the typical approach used in existing galaxy formation models. Specifically, for both SNe Ia and SNe II the dust mass for a given element $i$ is given by
\begin{equation}
\Delta M_{i,{\rm elem}}  =   
\begin{cases}
      \delta_{\rm C}^{\rm SN} \Delta M_{\rm C} &  {\rm if} \; i = {\rm C} \\[5pt]
      \mu_{\rm O} \sum\limits_{k={\rm Mg,Si,Fe}} \delta^{\rm SN}_{k} \Delta M_{k} / \mu_{k} & {\rm if} \; i = {\rm O} \\[10pt]
    \delta^{\rm SN}_{i} \Delta M_{i} & {\rm if} \; i = {\rm Mg,Si,Fe}
\end{cases}
\end{equation}
where $\delta^{\rm SN}_{i}$ is the dust condensation efficiency for element $i$ whose value can be found in Table~\ref{tab:CondensationEfficiencies}. These choices for SNe efficiencies are originally stated by \citet{dwek_1998:EvolutionElementalAbundances} as arbitrary, but they do reproduce similar results to more detailed theoretical dust yields modeled in \citet{todini_2001:DustFormationPrimordial} and produce similar dust masses as some observations \citep{chawner_2019:CatalogueGalacticSupernova,delooze_2019:DustContentCrab}, but are in contention with others \citep{sugerman_2006:MassiveStarSupernovaeMajor,rho_2008:FreshlyFormedDust,lau_2015:OldSupernovaDust}. For silicates, a majority of the refractory elements, Mg, Si, and Fe, and an equal amount of O by number are assumed to condense into dust. In the rare case this prescription requires more O for silicate dust than the total available O, we scale down silicate dust production to not exceed this O limit.

{\bf Species:} This implementation is similar, but makes a distinction between SNe Ia and SNe II, assuming SNe II produce all dust species while SNe Ia may only theoretically produce some iron dust. This is due to recent observations and modelling that suggests SNe Ia produce little, if any, dust \citep{nozawa_2011:FormationDustEjecta, gomez_2012:DustHistoricalGalactic}. In either case, the amount of dust species returned in one SNe event is tied to the total mass return of the key element\footnote{Here key element refers to the element for which $N/i$ has the lowest value, where $N$ is the number of atoms of the element in the initial SNe ejecta and $i$ is the number of atoms of the element in one formula unit of the dust species under consideration.} required to form the given dust species. The dust condensation efficiencies for silicates, carbon, and SiC are determined by comparing to observed abundance ratios of presolar dust grains from supernova and AGB found in meteorites. This process is explained in detail in \citet{zhukovska_2008:EvolutionInterstellarDust}, but it should be noted that the observations for some of the dust species are somewhat limited and produce relatively low condensation efficiencies, which contradict some observations (e.g. SNe 1987a is observed to have near all ejecta condensed into dust \citealt{matsuura_2011:HerschelDetectsMassive,matsuura_2015:StubbornlyLargeMass}). The condensation efficiencies for iron dust are arbitrarily set to a low nonzero values, but they very well could be zero. All species condensation efficiencies can be found in Table~\ref{tab:CondensationEfficiencies}.

Thus, for a single SNe event the dust mass returned for a given species $j$ is given by
\begin{equation}
    \Delta M_{j,{\rm spec}} 
    = 
    \delta^{\rm SN}_{j} \Delta M_{{\rm key},j} \frac{A_{j}}{A_{{\rm key},j}},
\end{equation}
where $\Delta M_{{\rm key},j}$ and $A_{{\rm key},j}$ are the returned mass and atomic mass of the key element for species $j$ and $A_{j}$ is the atomic mass of one formula unit of species $j$. The dust masses contributed by each element are then updated based on their mass fraction in species $j$.

\begin{figure*}
    \plotsidesize{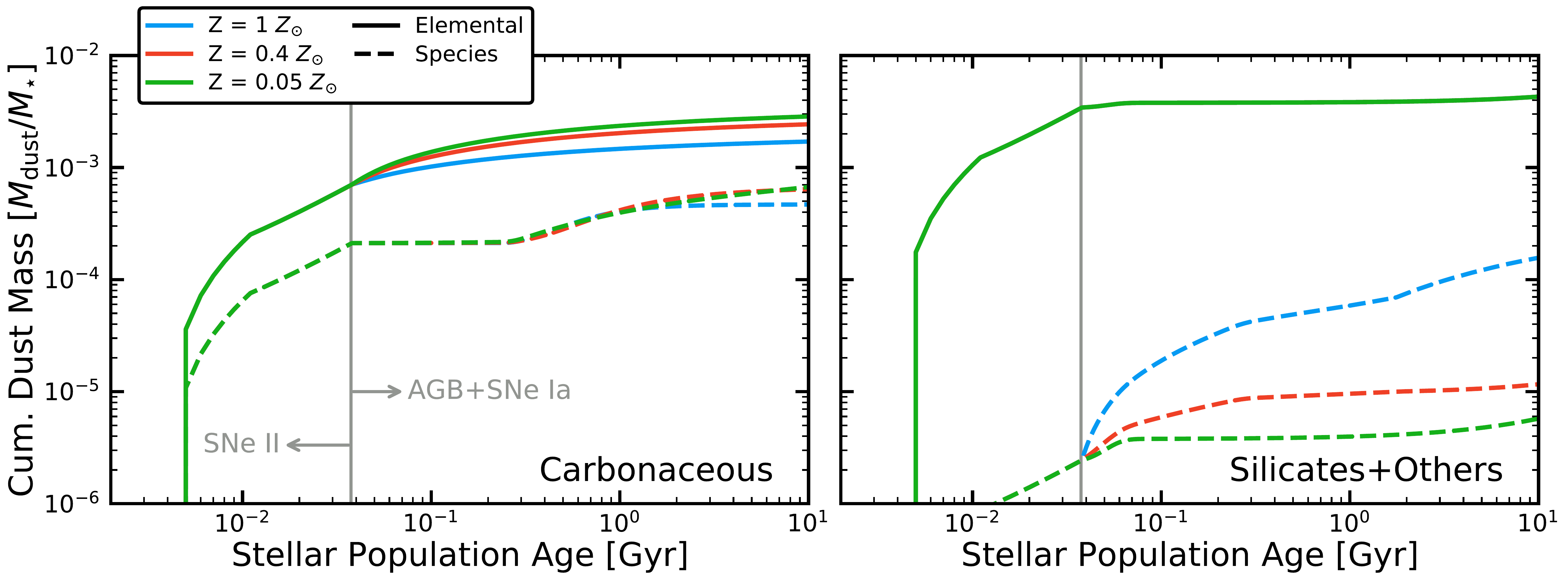}{0.99}
    \vspace{-0.25cm}
    \caption{Comparison of the cumulative dust production (from all stellar sources) per stellar mass of carbonaceous dust ({\it left}) and all other dust (dominated by silicates) ({\it right}) over the lifetime of a stellar population for the ``Elemental'' ({\it solid}) and ``Species'' ({\it dashed}) implementations with various initial stellar metallicities. The grey line shows when the transition between SNe II and SNe Ia/AGB dust production occurs at $t\approx 0.0375$ Gyr. Note the drop in AGB carbonaceous dust with higher metallicity is due to the stellar surface C/O ratio that determines the type of dust that formed. All stars have initial C/O $<1$ and increase their stellar surface C abundance via mixing. This means higher metallicity stars have higher initial surface C and O content, requiring more mixing to achieve C/O $>1$, which may not be possible before the end of the AGB phase. Ultimately, this leads to less carbonaceous dust being formed when compared to low metallicity stars. A more detailed breakdown of these results can be found in \citet{ferrarotti_2006:CompositionQuantitiesDust} and Sec. 2.3 of \citet{zhukovska_2008:EvolutionInterstellarDust}. We also note that the ``Elemental'' AGB dust routine only produces carbonaceous dust since it is coupled with FIRE-2 time- and IMF-averaged AGB metal yields which never produce surface C/O $<1$, effectively making the ``Elemental'' silicate dust production metallicity independent since it is only produced by SNe.}
    \label{fig:stellar_dust_comparison}
\end{figure*}

The overall SNe dust species production for both routines is shown in Figure~\ref{fig:stellar_dust_comparison}. The ``Elemental'' SNe routine produces more carbonaceous and silicate dust overall and is dominated by silicate dust in contrast to the ``Species'' SNe routine which is dominated by carbonaceous dust.

\subsubsection{Dust Creation by AGB Stars}

Near the end of an AGB star's life a certain amount ($\Delta M_{{\rm dust}}$) of the stellar wind injected into the surrounding gas condenses into dust. The type of dust that forms at this end stage depends on the evolution of the stellar surface carbon-to-oxygen number ratio (C/O) during the AGB phase, which is governed by the initial mass and metallicity of the star. In carbon rich outflows most oxygen is tied up in CO molecules, so mainly carbonaceous dust is produced. Conversely, in oxygen rich outflows most carbon is tied up in CO molecules, so mainly silicate dust is produced \citep{draine_1990:EvolutionInterstellarDust,ferrarotti_2006:CompositionQuantitiesDust}.

{\bf Elemental:} For this implementation we take a simple ``leftover'' approach for carbonaceous and silicate dust production, which depends on the carbon-to-oxygen number abundance ratios $\rm C/O>1$ or $\rm C/O<1$ in the stellar outflow. It is assumed that either all the O or C in AGB outflows is locked up into newly formed CO molecules depending on which is less abundant by number, while the remainder of the more abundant element forms into dust. 
Specifically, if $\rm C/O>1$ the excess C condenses into carbon dust while for $\rm C/O<1$, a majority of the refractory elements, Mg, Si, and Fe, and an equal amount of O by number condense into silicate dust, similar to the SNe prescription. 
Thus for a given stellar population and $\Delta M_i$, the $i$th element stellar outflow metal mass which is tabulated from standard stellar population models (STARBURST99; \citet{leitherer_1999:Starburst99SynthesisModels}) assuming a \citet{kroupa_2001:VariationInitialMass} IMF, we calculate the dust mass produced by AGB stars accordingly. For AGB stars with C/O $>$ 1 in their stellar outflows, the amount of dust of element \textit{i} produced is given by 
\begin{equation} \label{eq:AGB_CO>1}
\Delta M_{i,{\rm elem}}  =   
\begin{cases}
      \delta^{{\rm AGB,C/O}>1}_{\rm C} (\Delta M_{\rm C} - 0.75\Delta M_{\text{O}}) &  {\rm if }\; i = {\rm C} \\[5pt]
      0 & {\rm otherwise.} \\
\end{cases} 
\end{equation}
For AGB stellar outflows with C/O $<$ 1 it is given by

\begin{equation} \label{eq:AGB_CO<1}
\Delta M_{i,{\rm elem}}  =   
\small
\begin{cases}
      \delta^{\text{AGB,C/O}<1}_{i} \Delta M_{i} &  {\rm if }\; i = {\rm Mg,Si,Fe} \\[5pt]
      \mu_{\rm O} \sum\limits_{k={\rm Mg,Si,Fe}} \delta^{{\rm AGB,C/O}<1}_{k} \Delta M_{k} / \mu_{k} & {\rm if }\; i = {\rm O} \\[5pt]
      0 & {\rm otherwise,} \\
\end{cases} 
\end{equation}
\normalsize
where $\mu_{k}$ is the atomic mass in AMU of element $k$ and $\delta^{{\rm AGB,C/O}>1}_{\rm C}$ and $\delta^{{\rm AGB,C/O}<1}_{i}$ are the dust condensation efficiencies, whose values can be found in Table~\ref{tab:CondensationEfficiencies}.
In the rare case this prescription requires more O for silicate dust than the total available O not locked in CO, we scale down silicate dust production to not exceed this O limit. We emphasize that for C/O and $\Delta M_i$ in Eq.~\ref{eq:AGB_CO>1} \&~\ref{eq:AGB_CO<1} we use the time- and IMF-averaged stellar metal yields in FIRE-2 and not the metal yields for individual stars which can under- or over-predict the total amount of dust formed. Specifically only one dust species is allowed to form for each stellar wind event and the IMF-averaged C/O ratio can be markedly different from the individual stellar C/O. Furthermore, the FIRE-2 metal yields never satisfy C/O $<1$, so we should expect AGB stars to only produce carbonaceous dust for this implementation.

{\bf Species:} This implementation uses the \citet{zhukovska_2008:EvolutionInterstellarDust} AGB dust production results which they extended from \citet{ferrarotti_2006:CompositionQuantitiesDust} to include a finer grid of metallicities and initial stellar masses. We interpolate this grid to calculate the total dust mass by species produced by AGB stars over their lifetime for a range of metallicities and stellar masses. Using this and averaging over a \citet{kroupa_2001:VariationInitialMass} IMF, we calculate the total AGB production of dust species $j$ for a star particle of a given age, $t_{\rm age}$, and metallicity, $Z$, over a time step $\delta t$ as
\begin{equation} \label{eq:spec_AGB_prod}
    M_{j,{\rm spec}}(t_{\rm age}+\delta t,Z) 
    = 
    \frac{M_{*}}{\msun} 
    \int^{m(t_{\rm age}+\delta t)}_{m(t_{\rm age})}{\Phi(m)\, M_{j,{\rm spec}}^{{\rm AGB}}(m,Z) \, dm}.
\end{equation}
Here $\Phi(m)$ is the \citet{kroupa_2001:VariationInitialMass} IMF normalized such that $\int m \, \Phi(m) \, dm = 1 \, \msun$, $m(t)$ is the inverse of the stellar lifetime function which we take to be the main sequence lifetime, giving $m(t) [\msun] \approx 2.51 \, t_{{\rm Gyr}}^{-0.4}$ which approximately defines the mean mass of stars going through the peak of the AGB dust-production phase, for a well sampled stellar population of age $t_{\rm Gyr}$ in Gyr, $M_{j,{\rm spec}}^{{\rm AGB}}(m,Z)$ is the given species dust mass returned by a single star of a given mass and metallicity interpolated from the data table given in \citet{zhukovska_2008:EvolutionInterstellarDust}\footnote{As noted by \citet{zhukovska_2008:EvolutionInterstellarDust}, most of the dust mass created over the lifetime of an AGB star is formed and expelled at the very end of its life when mass loss rates are the highest but the timescales are vanishingly small compared to galaxy dynamical times. This means AGB dust production can be assumed to occur instantaneously at the end of an AGB star's main sequence lifetime.}, and $M_{*}$ is the total mass of the star particle. Since the \citet{ferrarotti_2006:CompositionQuantitiesDust} AGB dust production model uses an assumed AGB metal yields and mass-loss rate prescription different from that in FIRE-2, this can cause more dust to be produced than metals available. If this occurs, we scale down the amount of dust produced for any `over-budget' dust species.

The overall AGB dust species production for both routines can be seen in Figure~\ref{fig:stellar_dust_comparison} for $t>0.0375$ Gyr, which is when the transition between SNe II and SNe Ia/AGB dust production occurs. Owing to high condensation efficiencies assumed in the ``Elemental'' implementation, it produces far more carbonaceous dust compared to the ``Species'' implementation. This large difference is not immediately reconcilable since even detailed AGB dust production models calculated from either analytic or full integrated stellar evolution models vary considerably in the amount and type of dust produced across initial stellar masses and metallicities while still producing plausible results. Specifically, both models produce reasonable dust masses and present-time AGB dust production rates compared to those observed for the Magellanic Clouds \citep[e.g.][]{zhukovska_2013:DustInputAGB, schneider_2014:DustProductionRate}.

\begin{table*}
	\centering
		\begin{tabular}{| l | l | l |}
			\hline
			Variable & Elemental & Species \\ [0.5ex] 
			\hline\hline
            \multirow{4}*{$\delta^{\rm SN II}_{i}$} 
			    & \multirow{2}*{0.0 for $i=$ H,He,N,Ne,S,Ca} & 0.00035 for $i=$ silicate \\
			    & \multirow{2}*{0.5 for $i=$ C} & 0.15 for $i=$ carbon \\
			    & \multirow{2}*{0.8 for $i=$ O,Mg,Si,Fe} & 0.001 for $i=$ iron \\
			    & & 0.0003 for $i=$ SiC \\
			\hline
			\multirow{2}*{$\delta^{\rm SN Ia}_{i}$} 
			    & \multirow{2}*{same as $\delta^{\rm SN II}_{i}$} 
			        & 0.005 for $i=$ iron \\
			        & &  0.0 otherwise \\
			\hline
			\multirow{2}*{$\delta^{{\rm AGB,C/O}>1}_{i}$} 
			    & 0.0 for $i=$ H,He,N,O,Ne,Mg,Si,S,Ca,Fe & \multirow{2}*{refer to Eq.~\ref{eq:spec_AGB_prod}}\\ 
                & 1.0 for $i=$ C & \\
			\hline
			\multirow{2}*{$\delta^{{\rm AGB,C/O}<1}_{i}$} 
			    & 0.0 for $i=$ H,He,N,C,Ne,S,Ca & \multirow{2}*{refer to Eq.~\ref{eq:spec_AGB_prod}}\\ 
                & 0.8 for $i=$ O,Mg,Si,Fe & \\
			\hline
			
		\end{tabular}
	\caption{Condensation efficiencies for ``Elemental'' and ``Species'' implementations taken from \citet{dwek_1998:EvolutionElementalAbundances} and \citet{zhukovska_2008:EvolutionInterstellarDust}. Note that ``Elemental'' has efficiencies for each elemental species while ``Species'' has efficiencies for common dust types.}
	\label{tab:CondensationEfficiencies}
\end{table*}

\subsection{Dust Growth in the ISM} \label{Accretion}

In dense interstellar clouds, dust grains can grow by the accretion of gas-phase metals onto preexisting dust cores \citep{draine_1990:EvolutionInterstellarDust}. Following work done by \citet{dwek_1998:EvolutionElementalAbundances} and \citet{hirashita_1999:DusttogasRatioPhase} we track the instantaneous fractional dust growth of element $i$ via accretion for each gas cell as
\begin{equation}
\label{eqn:accretion}
    \left( \frac{\dot{M}_{i,{\rm dust}}}{M_{i,{\rm dust}}} \right)_{\rm growth} 
    = 
    \left( 1 - \frac{M_{i,{\rm dust}}}{M_{i,{\rm metal}}} \right) 
    \left( \frac{1}{\tau_{\rm g}}\right),
\end{equation}
where $M_{i,{\rm dust}}$ and $M_{i,{\rm metal}}$ are the corresponding element $i$ total dust mass and total metal mass (gas-phase and dust) in the cell respectively and $\tau_{\rm g}$ is the characteristic growth time-scale. The $\left( 1 - \frac{M_{i,{\rm dust}}}{M_{i,{\rm metal}}} \right)$ term represents the free gas-phase metal mass fraction.

{\bf Elemental:} For this implementation each element $i$ accretes independently of one other and adopts a set growth time-scale for all elements, following derivations from \citet{hirashita_2000:DustGrowthTimescale} to calculate the time-scale $\tau_{\rm g}$. This assumes a set sticking efficiency, computing the growth time-scale as
\begin{equation}
\tau_{\rm g} = \tau^{\rm ref}_{\rm g} \left( \frac{\rho^{\rm ref}}{\rho} \right) \left( \frac{T^{\rm ref}}{T} \right)^{1/2},
\end{equation}
where $\rho$ and $T$ are the density and temperature of the gas cell, $\rho^{\rm ref}$ and $T^{\rm ref}$ are reference values for density and temperature, and $\tau^{\rm ref}_{\rm g}$ is an overall normalization factor influenced by the atom-grain collision sticking efficiency, grain cross-section, grain density, clumping factors in the ISM, and many other `sub-grid' physical processes. Here these are simply taken to be constants, with values similar to those in \citet{mckinnon_2016:DustFormationMilky}, specifically $\rho^{\rm ref}=$ 1 H atom ${\rm cm}^{-3}$, $T^{\rm ref} = 20$ K and $\tau_{\rm g}^{\rm ref} = 2$ Gyr. We note that \citet{mckinnon_2016:DustFormationMilky} uses $\tau_{\rm g}^{\rm ref} = 0.2$ Gyr which they `tuned' to Milky Way-sized galaxies, but we will later show that this timescale is far too short for the MW-mass galaxy we simulate.

{\bf Species:} For this implementation the accretion rate for each dust species $j$ is limited by the key element for that species. This means Eq.~\ref{eqn:accretion} applies to the dust growth of said key element\footnote{Once  $\dot{M}_{i,{\rm dust}}$ is determined for the key element we assume appropriate masses for all other elements which comprise said dust species are also accreted. These masses are determined by the dust species' chemical composition mentioned previously.} (which is calculated at each time step) and each dust species $j$ has its own characteristic growth time-scale. From \citet{zhukovska_2008:EvolutionInterstellarDust} we use the characteristic accretion growth time-scale
\begin{equation}  \label{eq:spec_growth_timescale}
    \tau_{{\rm g},j} 
    =
    \left(\rho_{\rm c} V_{\rm grain}\right) 
    \left( \frac{1}{\varv_{j,{\rm th,m}} \, n_{\rm m} \, \sigma_{\rm grain} }\right)
    \left(\frac{1}{\xi_{\rm m}\,m_{\rm added}}\right),
\end{equation}
where the terms in this equation from left to right are the mass of the dust grain, the interaction rate between the growth species\footnote{The growth species is the atomic or molecular species in the gas-phase which carries most of the key element for the dust species under consideration.} and dust grain, and the overall mass added to the dust grain after each interaction. The variables are as follows: $\rho_{\rm c}$ is the mass density of the solid dust grain (taken from \citealt{zhukovska_2008:EvolutionInterstellarDust}), $V_{\rm grain}$ is the volume of the dust grain, $v_{j,{\rm th,m}}=\sqrt{\frac{8\, k\, T}{\pi A_{j,{\rm m}}m_{\rm H}}}$ is the thermal velocity of the growth species $m$, $n_{\rm m}$ is the maximum number density of the growth species (i.e. assuming no elements are locked up in dust), $\sigma_{\rm grain}$ is the surface area of the dust grain, $\xi_{\rm m}$ is the sticking efficiency for each collision, $m_{\rm added}=A_{j,{\rm c}} \, m_{\rm H} / \alpha_{j,{\rm c}}$ is the mass added to the dust grain with each collision, $A_{j,{\rm c}}$ and $A_{j,{\rm m}}$ are the atomic weight of one formula unit of the dust material under consideration and of the growth species respectively, and $\alpha_{j,{\rm c}}$ is the number of atoms of the key element contained in one formula unit of the condensed phase. Note, even with this complexity Eq.~\ref{eq:spec_growth_timescale} makes many strong assumptions. It neglects clumping/cross-correlation factors and gas-dust kernel collision enhancement terms, and it assumes negligible dust drift velocity throughout the gas, a uniform internal grain density, and hard-sphere type encounters.

For simplicity we assume that the growth species is just the free atoms of the key element and that dust grains are spherical. With these assumptions, and averaging over the dust grain size distribution, the growth timescale can be written as
\begin{equation}  \label{growth_timescale}
    \tau_{{\rm g},j} 
    = 
    \frac{\rho_{\rm c} \left<a\right>_{3} \alpha_{j,{\rm c}}}
    {3 \xi_{\rm m} \varv_{j,{\rm th,m}} A_{j,{\rm c}} m_{\rm H} \, n_{\rm m}},
\end{equation}
where $\left<a\right>_{3}$ is the average grain size given by
\begin{equation} \label{eq:avg_grain_size}
    \left<a\right>_{3} 
    =  
    \frac{\left<a^{3}\right>}{\left<a^{2}\right>} 
    = 
     \frac{\int^{a_{\rm max}}_{a_{\rm min}} \frac{dn_{\rm gr}(a)}{da} \; a^3 \; da}{\int^{a_{\rm max}}_{a_{\rm min}} \frac{dn_{\rm gr}(a)}{da} D(a) \; a^2 \; da},
\end{equation}
where $D(a)$ is the grain size dependent electrostatic enhancement factor which accounts for the change in cross section of an interaction between ionized gas-phase metals and charged dust grains (Coulomb enhancement) \citep{weingartner_1999:InterstellarDepletionVery} and $n_{\rm gr}(a)$ is the grain size distribution with minimum and maximum grain sizes $a_{\rm min}$ and $a_{\rm max}$ respectively.

For simplicity we adopt a MRN size distribution $ \frac{dn_{\rm gr}(a)}{da} \propto a^{-3.5}$ \citep{mathis_1977:SizeDistributionInterstellar} for all dust species with $a_{\rm min}=4$ nm, and $a_{\rm max}=250$ nm for all dust species besides Nano-iron, which has $a_{\rm min}=1$ nm and $a_{\rm max}=10$ nm. In diffuse ISM gas (CNM and diffuse molecular), for silicates and carbonaceous dust we adopt their respective CNM enhancement factors $D(a)$ from \citet{weingartner_1999:InterstellarDepletionVery}, for Nano-iron dust we adopt the enhancement factor for iron nanoparticles from \citet{hensley_2017:ThermodynamicsChargingInterstellar}, and for Normal-iron dust we assume the same $D(a)$ as silicate dust. In dense molecular gas, where gas-phase metals are neutral, $D(a)=1$ for all dust species.

For the sticking efficiency, we follow \citet{zhukovska_2016:ModelingDustEvolution} and take a simple step function with $\xi_{\rm m} = 1$ for $T_{\rm gas}<300K$ and $\xi_{\rm m} = 0$ for $T_{\rm gas}>300K$, where $T_{\rm gas}$ is the overall temperature of the gas cell.\footnote{\citet{zhukovska_2016:ModelingDustEvolution} notes that expected depletion trends are not sensitive to the exact shape of the sticking efficiency relation with gas temperature as long as it decreases, but our cutoff choice of $T=300$K is somewhat arbitrary and prone to uncertainty due to the lack of any experimental data or theoretical calculations for most refractory elements.}

Using the values above, Eq.~\ref{growth_timescale} numerically evaluates to
\begin{equation}
    \tau_{{\rm g},j} 
    = 
    \tau_{{\rm ref},j} 
    \frac{\alpha_{j,{\rm c}} \; A_{j,{\rm m}}^{1/2}}{\xi_{\rm m}A_{j,{\rm c}}} 
    \left( \frac{\rho_{\rm c}}{3 \; {\rm g \; cm}^{-3}}\right) 
    \left( \frac{10^{-2} \; {\rm cm}^{-3}}{n_{\rm m}}\right)
    \left(\frac{300 \; {\rm K}}{T} \right)^{1/2},
\end{equation}
where $\tau_{{\rm ref},j}$ is the normalization calculated as given in Table~\ref{tab:AccretionSummary} alongside $\rho_{\rm c}$.

Since we track the mass fraction of each gas cell which is in the dense molecular phase ($\fdens$), we replace $\tau_{{\rm ref},j}$ with an effective reference timescale $\tau_{{\rm ref},j}^{\rm eff}$ which is defined as
\begin{equation}
    \left(\tau_{{\rm ref},j}^{\rm eff}\right)^{-1} = \frac{\fdens}{\tau^{\rm dense}_{{\rm ref},j}} + \frac{1-\fdens}{\tau^{\rm diffuse}_{{\rm ref},j}}
\end{equation}
where $\tau^{\rm diffuse}_{{\rm ref},j}$ is the reference timescale for dust species $j$ in gas in the CNM and diffuse molecular phase where the electrostatic enhancement factor has to be taken into account and free gas-phase C atoms exist, and $\tau^{\rm dense}_{{\rm ref},j}$ is the reference timescale for dust species $j$ in gas in the dense molecular phase where there is no electrostatic enhancement factor and all gas-phase C is locked into CO.

It should be noted, that for gas in cold ISM phases, our hydro-solver time steps are much shorter than the growth timescales of any dust species which means we can accurately time-resolve gas-dust accretion. Specifically, in molecular gas (with e.g. $\nH=10^3$ cm$^{-3}$, $T=30$ K) with solar metal abundances the ``Species'' growth timescales for each of the dust species without Coulomb enhancing are for silicates $\tau_{\rm g} \approx 0.66$ Myr, for Normal-iron $\tau_{\rm g} \approx 8.4$ Myr, and for Nano-iron $\tau_{\rm g} \approx 0.85$ Myr and for ``Elemental'' is $\tau_{\rm g} \approx 1.6$ Myr for all elements. For CNM gas (with e.g. $\nH=30$ cm$^{-3}$, $T=100$ K) with solar metal abundances, the ``Species'' growth timescales for each of the dust species with Coulomb enhancing are for silicates $\tau_{\rm g} \approx 2.2$ Myr, for Normal-iron $\tau_{\rm g} \approx 28$ Myr, for Nano-iron $\tau_{\rm g} \approx 0.19$ Myr, and for carbonaceous $\tau_{\rm g} \approx 12$ Myr and for ``Elemental'' is $\tau_{\rm g} \approx 30$ Myr for all elements. The simulation time steps, in contrast, range from ${\sim}10^2-{\sim}10^4$ yr under these conditions.

\begin{table*}
    \renewcommand{\arraystretch}{1.2}
	\begin{tabular}{|l l l l l l l|}
		 \hline
			Physical Quantity & Silicate & Carbon & Normal-iron & Nano-iron & SiC & O-reservoir\\ [0.5ex] 
			\hline\hline
			$\rho_{\rm c} \text{ (g cm}^{-3})$ & 3.13 & 2.25 & 7.86 & 7.86 & 3.21 & \textemdash \\
			$a_{\rm min}$ (nm) & 4 & 4 & 4 & 1 & 4 & \textemdash \\
			$a_{\rm max}$ (nm) & 250 & 250 & 250 & 10 & 250 & \textemdash \\
			$\left<a\right>^{\rm diffuse}_{3}$ (nm) & 5.8 & 35.4 & 5.8 & 0.038 & 31.6 & \textemdash\\
			$\left<a\right>^{\rm dense}_{3}$ (nm) & 31.6 & 31.6 & 31.6 & 3.2 & 31.6 & \textemdash\\
			$\tau^{\rm diffuse}_{\rm ref}$ (Myr) & 4.4 & 26.7 & 4.4 & 0.029 & $\infty$ & \textemdash \\
			$\tau^{\rm dense}_{\rm ref}$ (Myr) & 23.9 & $\infty$ & 23.9 & 2.42 & $\infty$ & \textemdash \\
			$A_{\rm c}$ & 143.8 & 12.0 & 55.9 & 55.9 & 30.1 & 16.0\\
		\hline
	\end{tabular}
	\centering
	\caption{Summary of input constants assumed in our ``Species'' gas-dust accretion model. We assume SiC does not grow in the ISM and O-reservoir dust follows the prescription outlined in Sec.~\ref{Dust_by_Species}. The `diffuse' label denotes the atomic or diffuse molecular gas regime where gas-phase metals are ionized and so accretion includes Coulomb enhancement and C is primarily atomic. The `dense' label denotes the dense molecular gas regime where we assume essentially all gas-phase metals are neutral and so accretion does not include Coulomb enhancement; meanwhile we assume all gas-phase C is locked into CO, so gas-dust accretion of C cannot occur (the $\infty$ here).}
	\label{tab:AccretionSummary}
\end{table*}

\subsection{Dust Destruction}

Dust that has been injected and grown in the ISM is subjected to numerous destructive processes which destroy and shatter dust grains, shifting the grain size distribution and reducing the total dust mass. Since we do not evolve the grain size distribution, we explicitly follow destruction only. One process we intrinsically track is astration, the destruction of dust in gas which condenses into stars. Specifically, as star particles form from gas cells or fractions thereof, the corresponding (cell-averaged) amount of dust is removed (added to the stellar metallicity). The other major dust destruction processes we track are described below.

\subsubsection{Thermal Sputtering}
Dust grains residing in hot gas in the galactic halo can undergo thermal sputtering, which causes erosion of dust grains by energetic atoms and can limit the depletion of gas-phase metals onto grains. Protons and helium ions are the main sputtering agents, and predictions of thermal sputtering rates indicate that sputtering overwhelms dust growth via accretion for $T \gtrsim 10^{5}$K \citep{draine_1979:PhysicsDustGrains}.

For both the ``Elemental'' and ``Species'' implementations we follow the prescription for thermal sputtering from \citet{tsai_1995:InterstellarGrainsElliptical}. The sputtering rate for a grain of radius $a$ in gas of density $\rho$ and temperature $T$ is then approximated by
\begin{equation}
    \frac{da}{dt} 
    = 
    -(3.2 \times 10^{-18} {\rm cm}^{4} {\rm s}^{-1}) 
    \left( \frac{\rho}{m_{\rm p}}\right) 
    \left[ \left( \frac{T_{0}}{T} \right)^{\omega} + 1\right]^{-1},
\end{equation}
where $m_{\rm p}$ is the proton mass, $\omega=2.5$ controls the low-temperature scaling of the sputtering rate and $T_{0}=2 \times 10^{6}$ K is the temperature above which the sputtering rate is approximately constant. It is important to note, this makes similar assumptions to Eq.~\ref{eq:spec_growth_timescale} ignoring clumping/cross-correlation factors, unresolved phase structure, non-trivial grain compositions, or geometric structures, strong charge effects, dust drift, and many other terms which can significantly alter $da/dt$. The associated sputtering time-scale for the grain is
\begin{equation} \label{sputtering_timescale}
    \tau_{\rm sp} 
    = 
    a \left| \frac{da}{dt} \right|^{-1}
    \approx 
    (0.17 \, {\rm Gyr}) 
    \left(\frac{a_{-1}}{\rho_{-27}}\right)
    \left[\left(\frac{T_{o}}{T}\right)^{\omega}+1\right],
\end{equation}
where $a_{-1}$ is the grain size in units of 0.1 $\mu$m, and $\rho_{-27}$ is the gas density in units of $10^{-27}$g cm$^{-3}$. 

Assuming a constant solid grain mass density $\rho_{\rm g}$ and grain mass $m_{\rm g} \propto a^3 \rho_{\rm g}$, Equation~\ref{sputtering_timescale} implies that grain mass changes according to the timescale $|m/\dot{m}|=\tau_{\rm sp}/3$. Averaging over the grain size distribution gives an average grain size similar to Eq.~\ref{eq:avg_grain_size}, but with no Coulomb enhancement ($D(a)=1$). We again assume an MRN size distribution, $ \frac{dn_{\rm gr}(a)}{da} \propto a^{-3.5}$, with $a_{\rm min}$ and $a_{\rm max}$ given in Sec.~\ref{Accretion} which gives an average grain size of $a=\sqrt{a_{\rm min}\,a_{\rm max}}=0.032$ $\mu$m for carbonaceous, silicate, and Normal-iron dust and $a=\sqrt{a_{\rm min}\,a_{\rm max}}=0.0032$ $\mu$m for Nano-iron dust. For ``Elemental'' we assume $a=0.032$ $\mu$m for all elements in dust.

Thus for each time step we calculate the fractional change in element or species $i$ dust mass for every gas cell due to thermal sputtering as
\begin{equation}
    \left( \frac{\dot{M}_{i,{\rm dust}}}{M_{i,{\rm dust}}} \right)_{\rm sp} 
    = 
    -\frac{1}{\tau_{\rm sp}/3}.
\end{equation}

\subsubsection{SNe Shocks}
Supernovae remnants (SNR) destroy and shatter dust grains as they propagate through the ISM through grain-grain collisions, thermal sputtering, and non-thermal sputtering \citep[e.g.][]{jones_1996:GrainShatteringShocks}. As the SNR expands into the ISM and shock-heats the gas it destroys a fraction of the dust grains in the gas. This dust destruction efficiency, $\epsilon$, depends on the speed of the shock.

For both the ``Elemental'' and ``Species'' implementations we follow the results from \citet{cioffi_1988:DynamicsRadiativeSupernova} which consider a radiative SNR. Assuming a homogeneous, solar metallicity medium we calculate the amount of gas shocked to velocity of at least $v_s$ for each SNe event as
\begin{equation} \label{eq:mass_cleared}
    M_{\rm s}(v_{\rm s})
    =
    2460 \frac{E_{\rm 51}^{0.95}}
    {n_{0}^{0.1} v_{\rm s7}^{9/7}} 
    \msun,
\end{equation}
where $n_{0}$ is the number density of the surrounding medium, $E_{\rm 51}$ is the energy released in a typical supernova in units of $10^{51}$ erg, and $v_{\rm s7}$ is the shock velocity in units of 100 km ${\rm s}^{-1}$. We take the shock velocity to be $v_{s7}=1$, i.e. the destruction is only efficient when the shock velocity is larger than $\sim$100 km/s which also roughly corresponds to when the SNR begins to rapidly cool, and assume an average dust destruction efficiency for the shocked gas to be $\bar{\epsilon}\approx0.4$, meaning typically all dust is destroyed in $980 \; \msun$ of gas by a single SNR. Note this gas mass cleared of dust (and in particular $\bar{\epsilon}$) encompasses numerous parameters such as the detailed SNR structure, grain physics, grain size distribution, etc. which can have large uncertainties, but our prescription is roughly consistent with detailed hydrodynamical simulations of dust destruction via thermal and non-thermal sputtering in SN shocks assuming an MRN grain size distribution \citep{hu_2019:ThermalNonthermalDust} for all but the most diffuse gas ($\nH<10^{-2}$ cm$^{-3}$). We then couple the amount of gas cleared of dust to the surrounding gas cells by using the weights calculated in \citet{hopkins_2018:HowModelSupernovae} for mechanical SNe feedback. We also assume the dust is thoroughly mixed so all dust elements/species have equal fractions destroyed. To account for the possible double counting of destruction via thermal sputtering due to our separate thermal sputtering and SNe dust destruction routines, we prevent thermal sputtering from occurring in gas which has been affected by an SNe event in the past 0.3 Myr, which is the typical time it takes for all dust destruction process to cease after a single SNe event \citep[][see Appendix A]{hu_2019:ThermalNonthermalDust}.

It should be noted that while we keep this prescription for all resolutions, the highest resolution FIRE simulations $m_{\rm gas}\ll2460 \; \msun$ (which is the case for many simulations of dwarf galaxies, e.g.~\citet{wheeler_2019:BeItTherefore}) can resolve the SNe cooling radius. While actually resolving the processes that destroy dust in SNR \citep{hu_2019:ThermalNonthermalDust} is beyond the current resolution of FIRE, these high resolution simulations will need a more detailed, time-resolved prescription which tracks the dust destruction efficiency based on the individual shock velocities for surrounding gas cells, but we save this for future work.

\section{Results} \label{Results}

To test both implementations, we simulate an idealized, non-cosmological Milky Way-like galaxy as described in Sec.~\ref{ICs}. The galaxy was simulated for $\sim$1.5 Gyr, long enough for all dust evolution models to reach close to a steady-state D/Z ratio and dust population composition. For all results listed we only consider gas cells within the galactic disc with r < 20 kpc from the galactic center and |z| < 2 kpc from the disc plane.

We first investigate the sensitivity of the steady-state D/Z of each model to free parameters in Sec.~\ref{Free Parameters}. We then analyze the resulting relation between gas-phase element depletions and D/Z with column and physical gas density along with the effects of our O-reservoir and Nano-iron dust prescriptions and compare with MW observations in Sec.~\ref{Element Depletions}. Lastly, we compare to extragalactic observations of spatially-resolved D/Z in Sec.~\ref{Extragalactic}.

\subsection{Testing Free Parameters} \label{Free Parameters}

Due to the numerous uncertainties and assumptions made in both implementations we first evaluate the sensitivity of each fiducial implementation to free parameters and variations in each stage in the dust life cycle. Specifically, we individually vary, by an order of magnitude, the accretion rate ($\tau_{\rm ref, g}^{-1}$), SNe destruction efficiency ($\bar{\epsilon}$), and stellar dust production efficiencies ($\delta^{\rm AGB}_{i}$ and $\delta^{\rm SN}_{i}$), including switching efficiencies between the two implementations, and individually turning off each creation and destruction mechanism. We then compare each of these variations by analyzing their resulting steady-state galaxy-integrated D/Z, a commonly used `0$^{\rm th}$ order' metric for comparing dust evolution models. Of these variations, we found turning off thermal sputtering has negligible effects and changes to the stellar dust creation efficiencies only affect the initial build up of dust but has little effect on the steady-state D/Z with results for these shown in Appendix~\ref{Appendix_Stardust}. However, variations to accretion and SNe destruction processes have noticeable effects on the resulting D/Z for both implementations. This reinforces the paradigm that stellar dust production provides the `seeds' for dust growth, while the steady-state D/Z is determined by the balance between gas-dust accretion and SNe destruction \citep{draine_1990:EvolutionInterstellarDust} and suggests resolving the ISM phase structure, to accurately track gas-dust accretion, is crucial for dust studies.\footnote{Strictly speaking, the galaxy steady-state D/Z changes with times since gas-dust accretion scales with $Z$ which increases with time. However, for MW-like galaxies with $Z\sim Z_{\sun}$ and steady star formation, the galactic metal enrichment timescale is significantly longer than the time it takes for the galactic dust population to reach a steady-state.} Below we evaluate these variations for each implementation and show how they both can be predicted by analytic models.

\subsubsection{Elemental} \label{Elemental Free Parameters}

\begin{figure}
    \centering
    \includegraphics[width=0.99\columnwidth]{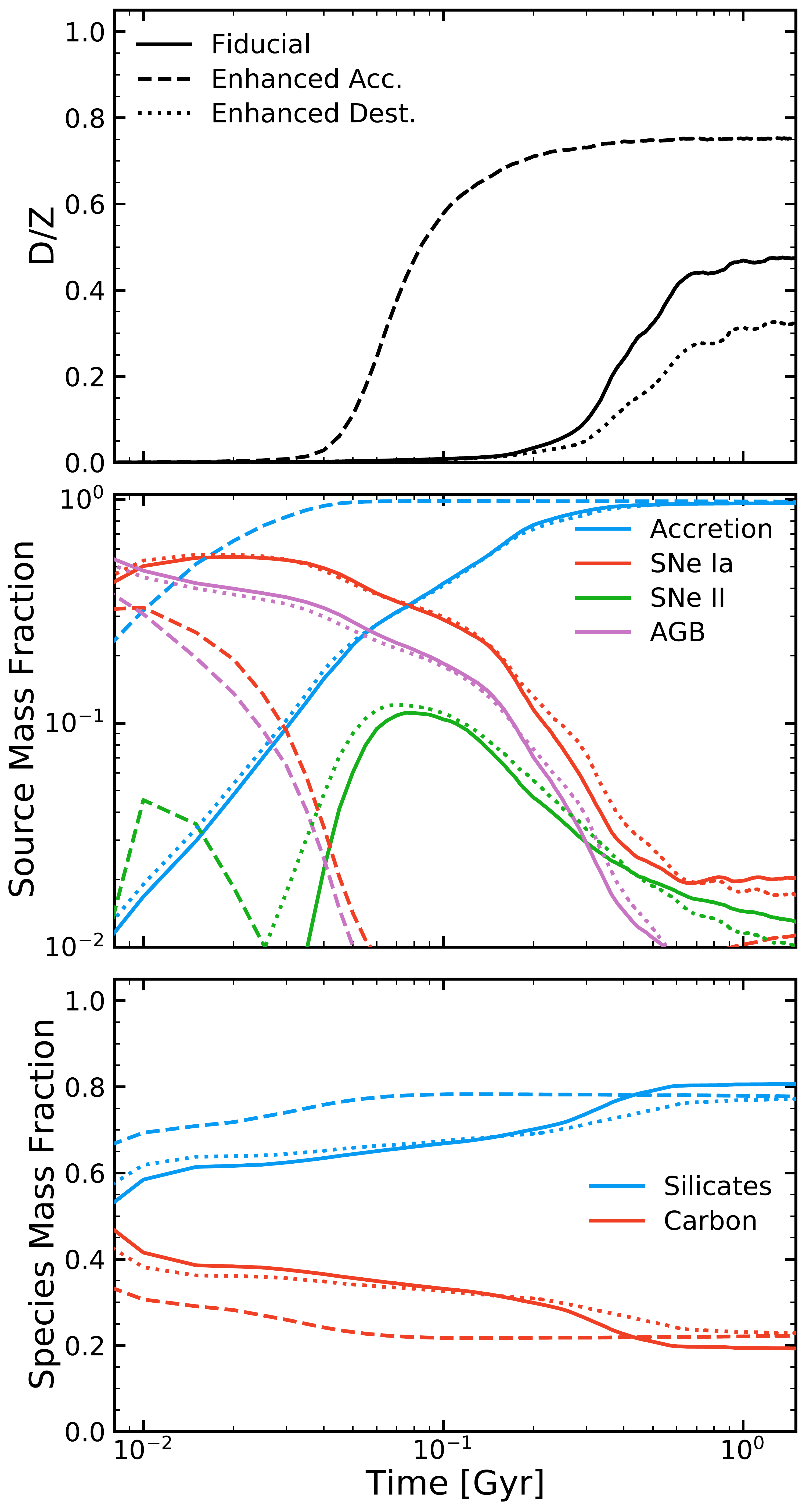}
    \vspace{-0.25cm}
    \caption{``Elemental'' time evolution of galaxy-integrated D/Z ratio ({\it top}), fraction of total dust mass from each source ({\it middle}), and fraction of total dust mass composed of each dust species ({\it bottom}) for an idealized Milky Way-like galaxy. We ran the ``Elemental'' implementation with the fiducial model ({\it solid}), order of magnitude increased gas-phase accretion rates ({\it dashed}), and doubled SNe dust destruction efficiency ({\it dotted}). In all cases the dust population reaches a steady-state by simulation end, with accretion becoming the dominant source of dust mass. Note the fiducial model's accretion rate is `tuned' to produce reasonable D/Z as noted in Sec.~\ref{Accretion}}
    \label{fig:elem_time_evo}
\end{figure}

For the ``Elemental'' implementation noticeable changes were found when increasing accretion rates by a factor of 10 or enhancing the SNe dust destruction efficiency by a factor of 2 or more. The time evolution of the galaxy-integrated D/Z, dust creation source contribution, and dust species composition for these tests are shown in Fig.~\ref{fig:elem_time_evo}. In all cases accretion quickly takes over as the dominant source of dust mass, producing a steady-state D/Z and dust population. Since accretion occurs in all gas environments for this implementation, we can analytically predict the steady-state D/Z by determining the equilibrium between accretion and SNe destruction \citep[e.g][]{mattsson_2012:DustAbundanceGradientsa,aniano_2020:ModelingDustStarlight}. The fractional change in dust mass for a given element $i$ in a gas cell, ignoring stellar dust creation, is then given by
\begin{equation} \label{eq:elem_analytic_model}
   \frac{\dot{M}_{i,{\rm dust}}}{M_{i,{\rm dust}}} = \frac{1-f_{\rm i}}{\tau_{\rm acc}} - \frac{1}{\tau_{\rm d}}
\end{equation}
where $f_i=M_{i,{\rm dust}}/M_{i,{\rm metal}}$ is the degree of condensation of a given element $i$ bound in dust, $\tau_{\rm acc}$ is the median accretion growth timescale, and $\tau_{\rm d}=\frac{M_{\rm ISM}\tau_{\rm SN}}{\bar{\epsilon} M_{\rm s}(1)}\sim0.77$ Gyr is a characteristic dust destruction timescale due to SNe taken from Eq. 18 in \cite{mckee_1989:DustDestructionInterstellar} where $M_{\rm ISM}=6.5\times10^9\msun$ is the gas mass of the ISM and $\tau_{\rm SN}^{-1}\sim\frac{1}{120\;{\rm yr}}$ is the galactic SNe rate in our simulation at its final time, and $\bar{\epsilon} M_{\rm s}(1)$ is the ISM mass wherein all dust is destroyed per SNe, on average given as  $\sim980 \; \msun$ in Eq.~\ref{eq:mass_cleared}. Putting these together, one obtains the very simple expectation $ f_i \sim \max[(1-\frac{\tau_{\rm acc}}{\tau_{\rm d}}),0]$. Given $\tau_{\rm acc}=330$ Myr from our simulations, the equilibrium degree of condensation is $f_{i} = 0.57$ for each refractory element $i$, which yields D/Z $=0.45$ assuming solar metal abundances. Increasing the accretion rates by a factor of 10 or increasing the SNe dust destruction efficiency by a factor of 2 predict a steady-state D/Z $=0.75$ and D/Z $=0.11$ respectively, but only the former matches well with our resulting D/Z shown in Fig.~\ref{fig:elem_time_evo}. This discrepancy is most likely due to this model's assumption that all SNe destroy dust equally. Since we time-resolve SNe events, SNe that occur in the bubbles of previous SNe will destroy less dust causing the `true' dust destruction timescale to be longer than $\tau_{\rm d}$ \citep[][]{hu_2019:ThermalNonthermalDust}.

\begin{figure}
  \centering
    \includegraphics[width=0.99\columnwidth]{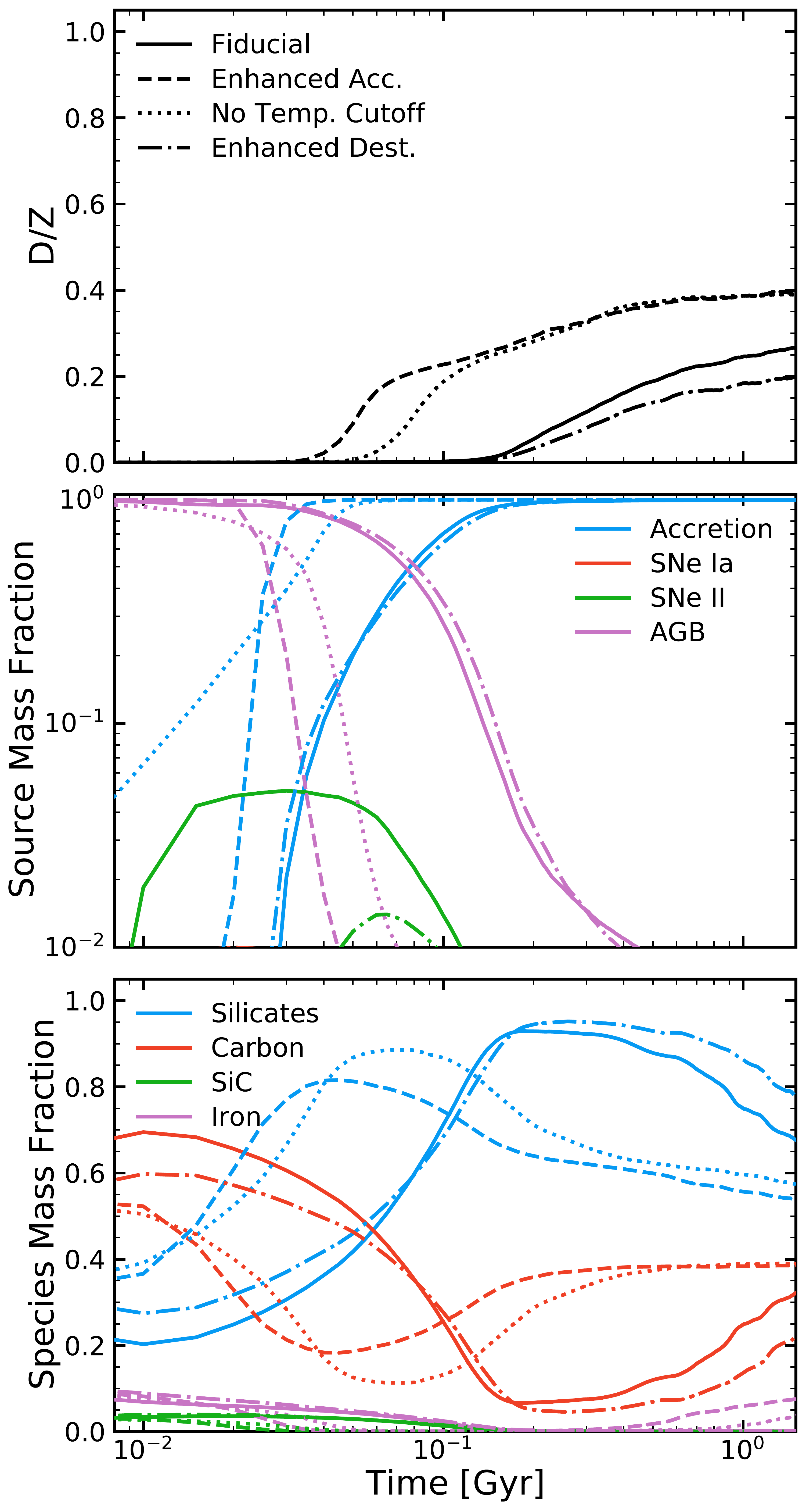}
    \vspace{-0.25cm}
    \caption{``Species'' time evolution of galaxy-integrated D/Z ratio ({\it top}), fraction of total dust mass from each source ({\it middle}), and fraction of total dust mass composed of each dust species ({\it bottom}) for an idealized Milky Way-like galaxy. We ran the ``Species'' implementation with the fiducial model ({\it solid}), order of magnitude increased gas-phase accretion rates ({\it dashed}), no temperature restriction for gas-dust accretion ({\it dotted}), and doubled SNe dust destruction efficiency ({\it dash-dotted}). In all cases they reach a near steady-state D/Z by simulation end, with accretion becoming the dominant source of dust, but the composition of the dust population differs. This difference is mainly due to the amount of carbonaceous dust that can form before CO takes up any remaining gas-phase C in dense environments.}
    \label{fig:spec_time_evo}
\end{figure}

\subsubsection{Species} \label{Species Free Parameters}

For the ``Species'' implementation, we find noticeable changes if we remove the accretion temperature cutoff, or enhance the SNe dust destruction efficiency by a factor of 2 or more, or increase the accretion rate by a factor of 10. The time evolution of the galaxy-integrated D/Z, dust creation source contribution, and dust species composition for these tests are shown in Fig.~\ref{fig:spec_time_evo}. Similar to the ``Elemental'' implementation, accretion prevails as the dominant source of dust mass growth, producing a steady-state D/Z and dust population. Since accretion only occurs in relatively dense environments with T < 300 K for this implementation, we can analytically predict the steady-state D/Z ratio based on the average D/Z between gas inside and outside of cold neutral regions. To determine these values we need to determine the degree of condensation of the key element for dust species $j$ inside and outside of cold gas/clouds ($f_{{\rm in,} j}$ and $f_{{\rm out,} j}$). To determine $f_{{\rm in,} j}$ we use Eq. 45 in \cite{zhukovska_2008:EvolutionInterstellarDust} which is a simple fit to the average degree of condensation of the key element, assuming it condenses in cold clouds which form and disperse over some timescale $\tcl$, maintaining a statistically steady-state abundance,
\begin{equation}\label{f_MC}
    f_{{\rm in,} j} = \left( \frac{1}{f_{{\rm o,}j}^2 (1+\tcl/\tau_{{\rm g,} j})^2} + 1 \right)^{-1/2}
\end{equation}
where $\tcl\sim10$ Myr is the typical mean cloud lifetime in similar simulations as estimated in \citet{benincasa_2020:LiveFastYoung}, $f_{{\rm o,}j}=f_{{\rm out,} j}$ is the average initial degree of condensation when the gas enters the cold cloud, and $\tau_{{\rm g,} j}$ is the median effective accretion growth timescale for dust species $j$ given the gas properties at simulation end. Specifically in our fiducial model implementation, we have approximately $\tau^{\rm eff}_{\rm g,sil}\sim10$ Myr, $\tau^{\rm eff}_{\rm g,carbon}\sim40$ Myr, and $\tau^{\rm eff}_{\rm g,iron}\sim100$ Myr.

To determine the average degree of condensation outside of cold clouds we use
\begin{equation}\label{f_ISM}
    f_{{\rm out,} j} = \left( 1-\frac{\tau_{\rm cycle}}{\tau_{\rm d}} \right) f_{{\rm in}, j}
\end{equation}
where $\tau_{\rm cycle}=\tcl \frac{1-X_{\rm cloud}}{X_{\rm cloud}}$ is the average time it takes to cycle all ISM material from the cold cloud phase through the diffuse/warm ISM phases and back into cold clouds required to give a steady-state fraction $X_{\rm cloud}$ of the mass of the ISM with T<300K (where we have $X_{\rm cloud}\sim0.30$ at simulation end), and $\tau_{\rm d}$ is again the characteristic SNe dust destruction timescale defined above.

With Eq.~\ref{f_MC} and~\ref{f_ISM} we find that the average mass weighted degree of condensation $f_{{\rm avg,}j}=(1-X_{\rm cloud})f_{{\rm out,}j}+X_{\rm cloud}f_{{\rm in,}j}$ which gives an average D/Z $ = 0.29$ assuming solar metal abundances and silicate composition given in Sec.~\ref{Dust Evo}. Increasing the growth rate by a factor of 10 or increasing the SNe dust destruction efficiency by a factor of 2 yield a steady-state D/Z $ = 0.42$ and D/Z $=0.27$ respectively, all of which match well with our simulated D/Z. For removing the accretion temperature cutoff, we can simply approximate D/Z by Eq.~\ref{eq:elem_analytic_model} (but using the rates only for the key element of each dust species) giving D/Z $= 0.44$.

While this model produces lower steady-state D/Z compared to the ``Elemental'' model, it is more robust to changes in $\tau_{\rm g}$ and $\tau_{\rm d}$ because of the model's `two-phase' scheme, with efficient growth within cold clouds and efficient destruction outside them. So even if dust accretion growth is infinitely efficient, D/Z will not increase more than $X_{\rm cloud}$ as long as destruction is efficient. On the other hand if dust destruction is infinitely efficient, D/Z cannot decrease below $X_{\rm cloud}$ if growth there is still efficient. 

As shown, both models can produce galaxy-integrated D/Z values near the canonical MW D/Z $\sim0.4$, depending primarily on the gas-dust accretion and SNe dust destruction timescales. However, both dust evolution models depend directly on the local gas environment, which can produce large D/Z variations within a galaxy compared to the galaxy-integrated value. A better gauge we can use to further analyze and test these implementations is the resulting relationship between D/Z and local gas properties compared with observations.

\subsection{Element Depletions and D/Z} \label{Element Depletions}

\subsubsection{Sight Line Element Depletions}

Element depletions are a commonly used method for estimating interstellar dust abundances. The gas-phase abundance of refractory elements are compared to an assumed reference abundance, with any elements missing from the gas-phase assumed to be locked in dust. The gas-phase depletion of element X assuming solar reference abundances, $\left(N_{\rm X}/N_{\rm H} \right)_{\odot}$, is usually represented logarithmically as
\begin{equation} \label{eq:NH_depletion}
    \left[ \frac{\rm X}{\rm H} \right]_{\rm gas} = \log \left( \frac{N_{\rm X}}{N_{\rm H}} \right)_{\rm gas} - \log \left( \frac{N_{\rm X}}{N_{\rm H}} \right)_{\odot}.
\end{equation}
where $N_{\rm X}$ and $N_{\rm H}$ are the gas-phase column density of element X and column density of neutral hydrogen ($\NHn=N_{\rm \textsc{H\,i}}+2N_{\Hmol})$ respectively. Similarly, the linear depletion of element X is
\begin{equation}
    \delta_{\rm X} = 10^{\left[ {\rm X}/{\rm H} \right]_{\rm gas}}.
\end{equation}

\begin{figure*}
    \plotsidesize{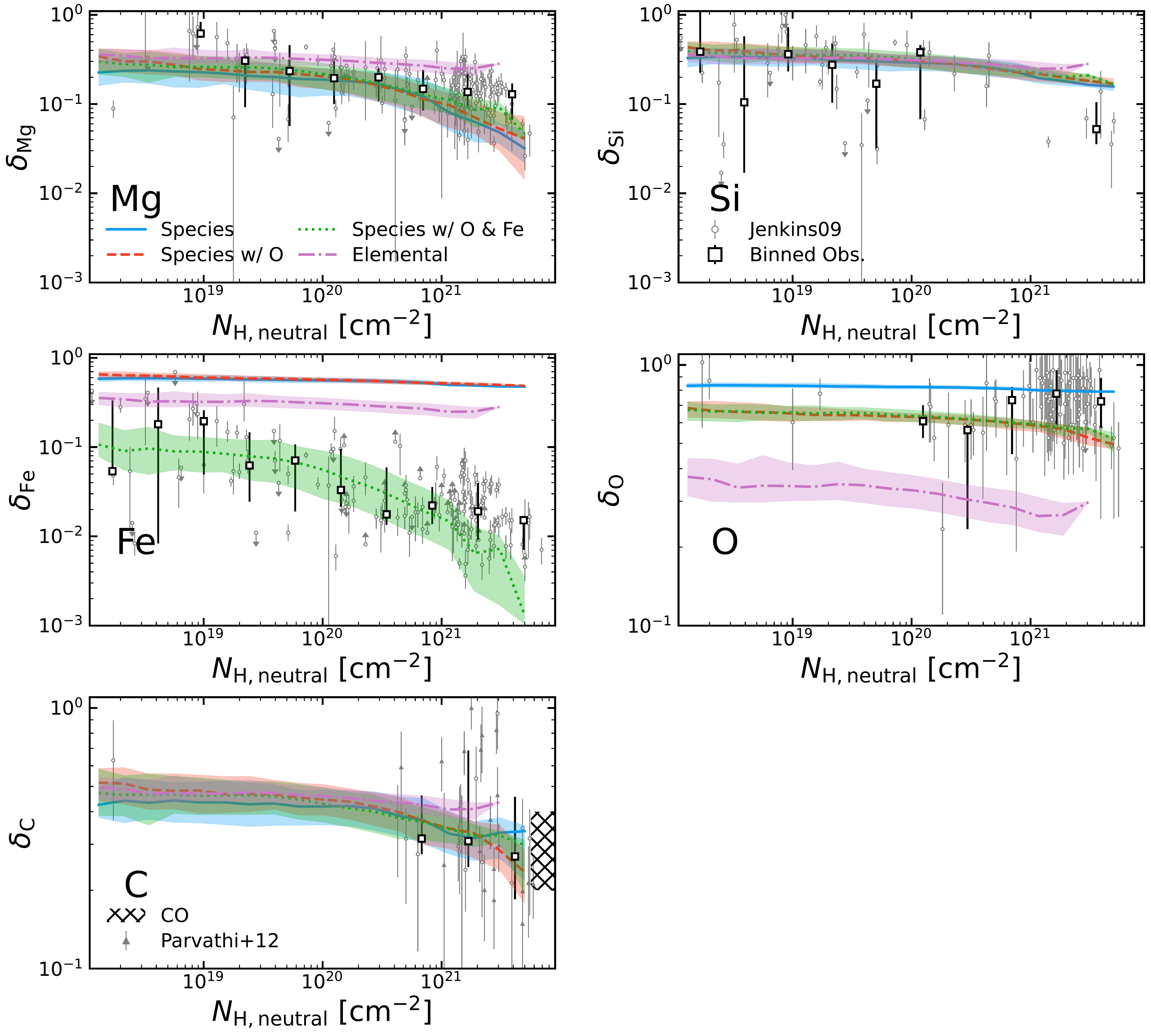}{0.99}
    \vspace{-0.25cm}
    \caption{Predicted sight line C, O, Mg, Si, and Fe depletion versus $N_{\rm H,neutral}$ from 10,000 sight lines at the solar galactic radius in an idealized Milky Way-like galaxy for our fiducial ``Species'' model, ``Species'' with O-reservoir dust species, ``Species'' with O-reservoir and Nano-iron dust species, and the fiducial ``Elemental'' model. For each, 16-/84-percentile ranges are represented by shaded regions. We compare with observed elemental depletion along sight lines in the Milky Way from \citet{jenkins_2009:UnifiedRepresentationGasPhase} ({\it circles}). For C we decreased the Jenkins data by a factor of 2 based on observations from \citet{sofia_2011:DeterminingInterstellarCarbon} and \citet{parvathi_2012:ProbingRoleCarbon}. We also include 21 sight line observations from \citet{parvathi_2012:ProbingRoleCarbon} ({\it triangles}) along with a range of expected minimum depletions in dense environments ({\it hatched}) based on observations of 20\%\ to 40\%\ of C in CO. We also show the binned median and 16-/84-percentile ranges for the Jenkins and Parvathi data ({\it squares}). The ``Elemental'' model produces a relatively flat depletion relation that is near identical for all elements and under-depletes Mg, Si, and Fe while it over-depletes O at high $N_{\rm H,neutral}$. The ``Species'' model produces a more complex relation, transitioning from a shallow to steep slope for increasing $N_{\rm H,neutral}$, which matches well with observations. Si is the only element that does not show this transition, but this is due to the metal yield prescription in FIRE-2 as shown in Appendix~\ref{Appendix_FIRE23}. For "Species" model, the inclusion of O-reservoir and Nano-iron dust is needed to match the observed depletions of  O and Fe.}
    \label{fig:depl_vs_NH} 
\end{figure*}

Observationally, measuring individual element depletions necessitates obtaining high-resolution UV spectroscopy with a high signal-to-noise ratio (S/N), and so detailed observations are mainly limited to the Milky Way. Also since depletions require observing spectral absorption features they mainly sample dust in low density environments. The most comprehensive review of Milky Way element depletions was compiled by \citet{jenkins_2009:UnifiedRepresentationGasPhase}, including over 243 lines of sight probing a wide range of physical conditions. To compare directly to these observations, we created a set of 10,000 sight lines for each simulation, deriving $N_{\rm X}$ and $\NHn$ from the total element abundances, amount of each element in dust, and neutral H number densities for gas cells intersected along each line of sight assuming these properties are uniform within each cell. Similar to the sight lines compiled in Jenkins, each simulated sight line ends at the solar galactic radius ($r \sim 8$ kpc) with a sight line distance chosen from a uniform distribution of 0.2 to 2 kpc. For simplicity each sight line was orientated parallel with the galactic disk in a random direction\footnote{We leave further investigation of the sensitivity of these sight line results to free parameters (e.g. sight line inclination with disk, choice of sight line start points, gas mass resolution, etc.) to fully cosmological simulations.}. We binned these sight lines in logarithmic neutral H column density bins and calculated the median and 16-/84-percentile for C\footnote{We include C in CO in our measured gas-phase C as is done with observations.}, O, Mg, Si, and Fe depletions. Note we use the local element abundances tracked in our simulation along each sight line when calculating depletions instead of assuming solar element abundances as is done with observations. The resulting relation between sight line element depletion and $\NHn$ for each element can be seen in Fig.~\ref{fig:depl_vs_NH} resulting from the fiducial ``Elemental'' model, the fiducial ``Species'' model along with optional O-reservoir and Nano-iron dust species included. We also include the observed sight line depletions from \citet{jenkins_2009:UnifiedRepresentationGasPhase}.

Special attention is given when comparing to C depletions from \citet{jenkins_2009:UnifiedRepresentationGasPhase}\footnote{Jenkins' definition of [C/H]$_{\rm gas}$ does not explicitly include C in CO but it is assumed there is only a negligible amount of CO in the environments observed.} due to the scarcity of data and apparent excess of gas-phase C compared to the amount needed to be locked up in carbonaceous dust. \citet{sofia_2011:DeterminingInterstellarCarbon, parvathi_2012:ProbingRoleCarbon} suggest that these gas-phase C values are too high by a factor of $\sim$2 when comparing C abundances determined from strong and weak \textsc{C\,ii} transition lines. For this reason we decrease all sight line C depletions from \citet{jenkins_2009:UnifiedRepresentationGasPhase} by a factor of 2. We also include observations from \citet{parvathi_2012:ProbingRoleCarbon}\footnote{Parvathi specifically includes CO in their calculations of [C/H]$_{\rm gas}$ but CO takes up $<1\%$ of gas phase C for most sight lines.} of carbon depletion and $N_{\rm H,neutral}$ along 21 sight lines in the Milky Way. Note that since a handful of these sight lines have C abundances greater than the reference \citet{lodders_2003:SolarSystemAbundances} solar abundances (C/H = $288 \pm 26$ ppm) used in Jenkins', we take the maximum abundance from this data set (C/H = $464 \pm 57$ ppm) as the reference abundance. This reduces the resulting depletion values by a factor of ${\sim}40\%$.

\subsubsection{Local Gas Element Depletions}

\begin{figure*}
    \plotsidesize{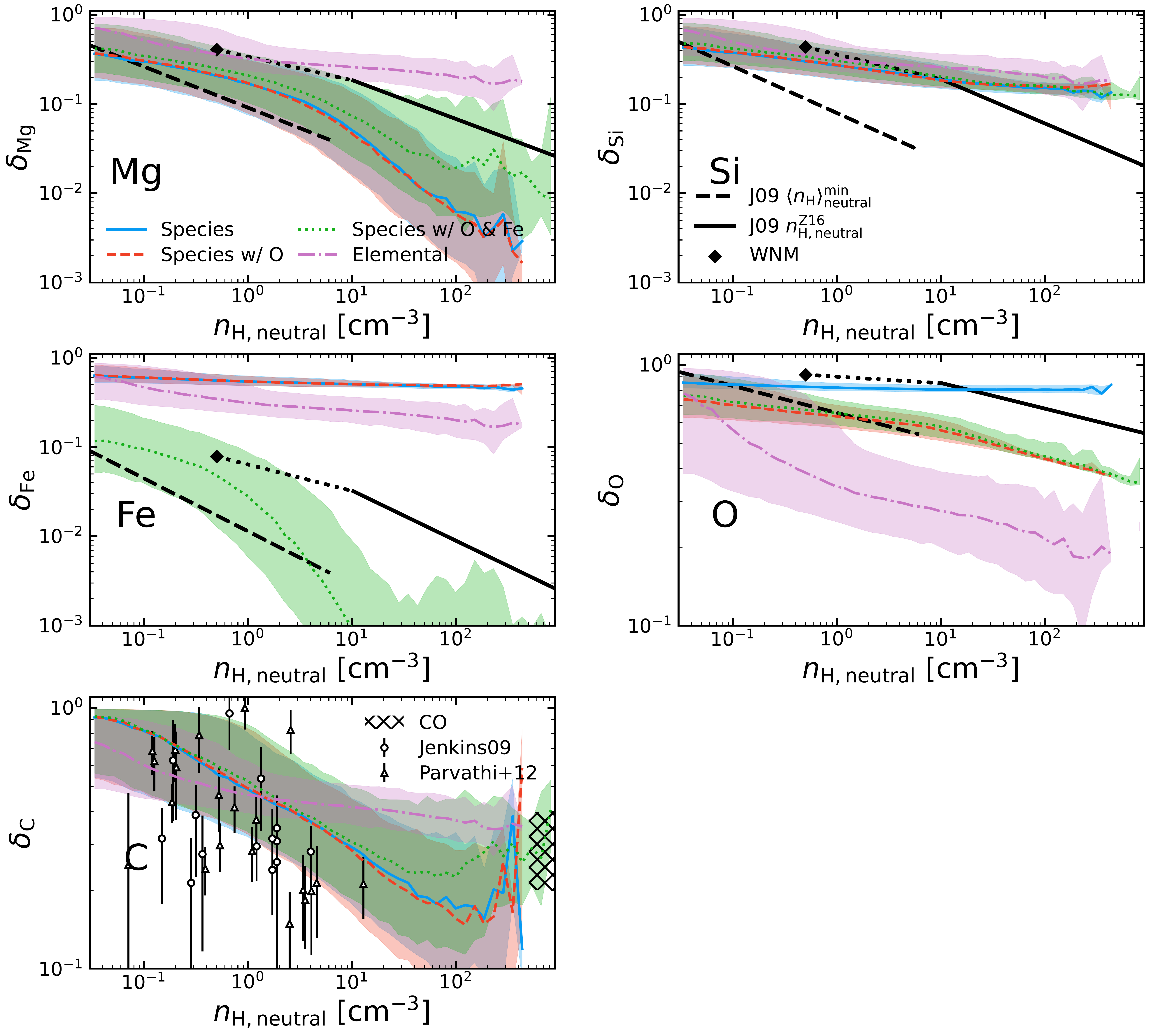}{0.99}
    \vspace{-0.25cm}
    \caption{Predicted C, O, Mg, Si, and Fe depletion versus $n_{\rm H,neutral}$ in an idealized Milky Way-like galaxy for our fiducial ``Species'' model, ``Species'' with O-reservoir dust species, ``Species'' with O-reservoir and Nano-iron dust species, and the fiducial ``Elemental'' implementation. For each, 16-/84-percentile ranges are represented by shaded regions. We compare with observed elemental depletions in the Milky Way from \citet{jenkins_2009:UnifiedRepresentationGasPhase} assuming mean sight line density is the physical density ({\it black-dashed}), this can be treated as a lower limit, and using \citet{zhukovska_2016:ModelingDustEvolution,zhukovska_2018:IronSilicateDust} mean sight line density to physical density fit ({\it black-solid}). For O, Mg, Si, and Fe we include estimates for the WNM depletions ({\it diamond}) along with an interpolation to the Jenkins' relation ({\it black-dotted}). For C we only include the individual sight line depletions from Jenkins ({\it triangles}) decreased by a factor of 2 based on observations from \citet{sofia_2011:DeterminingInterstellarCarbon} and \citet{parvathi_2012:ProbingRoleCarbon}. We also include 21 sight line observations from \citet{parvathi_2012:ProbingRoleCarbon} ({\it circles}) as another lower bound along with a range of expected minimum depletions in dense environments ({\it hatched}) based on observations of 20\%\ to 40\%\ of C in CO. The ``Elemental'' model produces a relatively shallow sloped depletion relation that is near identical for all elements and either under-depletes Mg, Si, and Fe or over-depletes O at high $n_{\rm H,neutral}$. The ``Species'' model produces a relation which transitions from a shallow to steep slope for increasing $n_{\rm H,neutral}$ matching well with observations. Si is the only element which does not show this transition, but this is due to the metal yield prescription in FIRE-2. Updated yields in the next version of FIRE \citep{hopkins_2022:FIRE3UpdatedStellar}) produce much better agreement with observations as shown in Appendix~\ref{Appendix_FIRE23}. For this model, the inclusion of the O-reservoir and Nano-iron dust species are also needed to match the observed strong depletions for both O and Fe respectively. For both models, our sub-resolved CO prescription produces a flattening of the C depletion relation in the densest environments due to gas-phase C becoming locked in CO and halting carbonaceous dust growth.}
    \label{fig:depl_vs_nH} 
\end{figure*}

Individual sight lines probe various gas phases and sight lines with similar $N_{\rm H,neutral}$ but different lengths can probe vastly different gas environments. This makes sight line observations less suited for constraining our dust evolution models since they depend on local gas environments. Since the total element abundances and the amount of each element in dust are tracked for all gas cells in our simulations, we can also directly measure the depletion for each element as a function of physical gas density. This means Eq.~\ref{eq:NH_depletion} becomes 
\begin{equation}
    \left[ \frac{\rm X}{\rm H} \right]_{\rm gas} = \log \left( \frac{n_{\rm X}}{n_{\rm H}} \right)_{\rm gas} - \log \left( \frac{n_{\rm X}}{n_{\rm H}} \right)_{\odot},
\end{equation}
where $n_{\rm X}$ and $n_{\rm H}$ are the local, gas-phase number density of element X and neutral H ($n_{\rm H,neutral}=n_{\rm H}+2n_{\rm \Hmol}$) respectively and $\left( n_{\rm X}/n_{\rm H} \right)_{\odot}$ is the total abundance (gas+dust) of element X in the gas cell. To do this we bin the gas cells in logarithmic neutral gas density and calculate the median values and 16-/84-percentiles for C, O, Mg, Si, and Fe depletions. The resulting relation between element depletion and neutral gas density, $n_{\rm H,neutral}$, for each element can be seen in Fig.~\ref{fig:depl_vs_nH} resulting from the fiducial ``Elemental'' model, and fiducial ``Species'' model along with optional O-reservoir and Nano-iron dust species included.

Comparing to observations, \citet{jenkins_2009:UnifiedRepresentationGasPhase} derived an empirical fit between element depletions and the average neutral gas density along lines of sight $\left< \nH \right>^{\rm min}_{\rm neutral}=N_{\rm H,neutral}/d$, where $d$ is the distance to the background UV source viewed in absorption.\footnote{For clarity, Jenkins does not derive a direct fit for each element depletion but instead for the $F_*$ parameter which represents the total strength of all element depletions along a line of sight ($F_*=0$ is the least depleted and $F_*=1$ is the depletions for $\zeta$ Oph, one of the most depleted sight lines in their sample). This fit also only includes sight lines with $\log [N_{\rm H,neutral}]>19.5$ to avoid contamination from ionized H.} We include these volume- and sight line-averaged relations between element depletions and $\left< \nH \right>^{\rm min}_{\rm neutral}$ in Fig.~\ref{fig:depl_vs_nH}. 

Note for C, due to the paucity of observed data and resulting poor fit in Jenkins we opt to only show the individual sight lines from \citet{jenkins_2009:UnifiedRepresentationGasPhase} and \citet{parvathi_2012:ProbingRoleCarbon} and not the fitted relation\footnote{Note, we do use the fitted relation for C depletion when aggregating element depletions to determine the expected D/Z.}. Allowing for any realistic degree of inhomogeneity in the ISM, $\left< \nH \right>^{\rm min}_{\rm neutral}$ will always be significantly lower than the true physical density of the cold gas in which a majority of the dust and neutral column density along a line of sight resides. Therefore, this should be considered a lower bound for the ``true'' $n_{\rm H,neutral}$ of interest in the simulations. 

To compare to a reasonable estimation of element depletions in dense environments we also compare the estimated relation between [Si/H]$_{\rm gas}$ and $n_{\rm H,neutral}$ from \citet{zhukovska_2016:ModelingDustEvolution}, who used fine-structure excitations of neutral carbon from \citet{jenkins_2011:DistributionThermalPressures} measured for a subset of the sight lines from \citet{jenkins_2009:UnifiedRepresentationGasPhase} to try to infer a better estimate of the true $n_{\rm H,neutral}$ for the same sight lines. This effectively gives the following relationship between $\left< \nH \right>^{\rm min}_{\rm neutral}$ and $n^{\rm Z16}_{\rm H,neutral}$ of $n^{\rm Z16}_{\rm H,neutral} = 147.2 \left(\frac{\left< \nH \right>^{\rm min}_{\rm neutral}}{\rm 1\; cm^{-3}}\right)^{1.05} $ cm$^{-3}$, which is restricted to $n^{\rm Z16}_{\rm H,neutral}=10-10^3$ cm$^{-3}$ since the method used is biased to denser gas. We further modify this by reducing the depletions for Mg, Si, and Fe by 0.2 dex as recommended by \citet{zhukovska_2018:IronSilicateDust} to account for the increase in depletion due to contamination by the warm neutral medium (WNM) along sight lines, since only the high density gas is probed by \textsc{C\,i} which will have lower depletions than the contaminating WNM. We do not account for this for O and C since their depletion slopes are quite shallow compared to the other elements and so the WNM contamination should have a comparatively small effect. 

To gain a general constraint on element depletions in low density environments we included the depletions for O, Mg, Si, and Fe at $F_*=0.12$ from \citet{jenkins_2009:UnifiedRepresentationGasPhase} which they recommend for the WNM based on matching depletion values with those given in \citet{savage_1996:InterstellarAbundancesAbsorptionLine} for the WNM in the Galactic disk. In Fig.~\ref{fig:depl_vs_nH} we place this depletion at the typical density of the WNM, $\nH\sim0.5$ cm$^{-3}$. We exclude this for C due to the reasons mentioned above.

\begin{figure}
    \includegraphics[width=0.99\columnwidth]{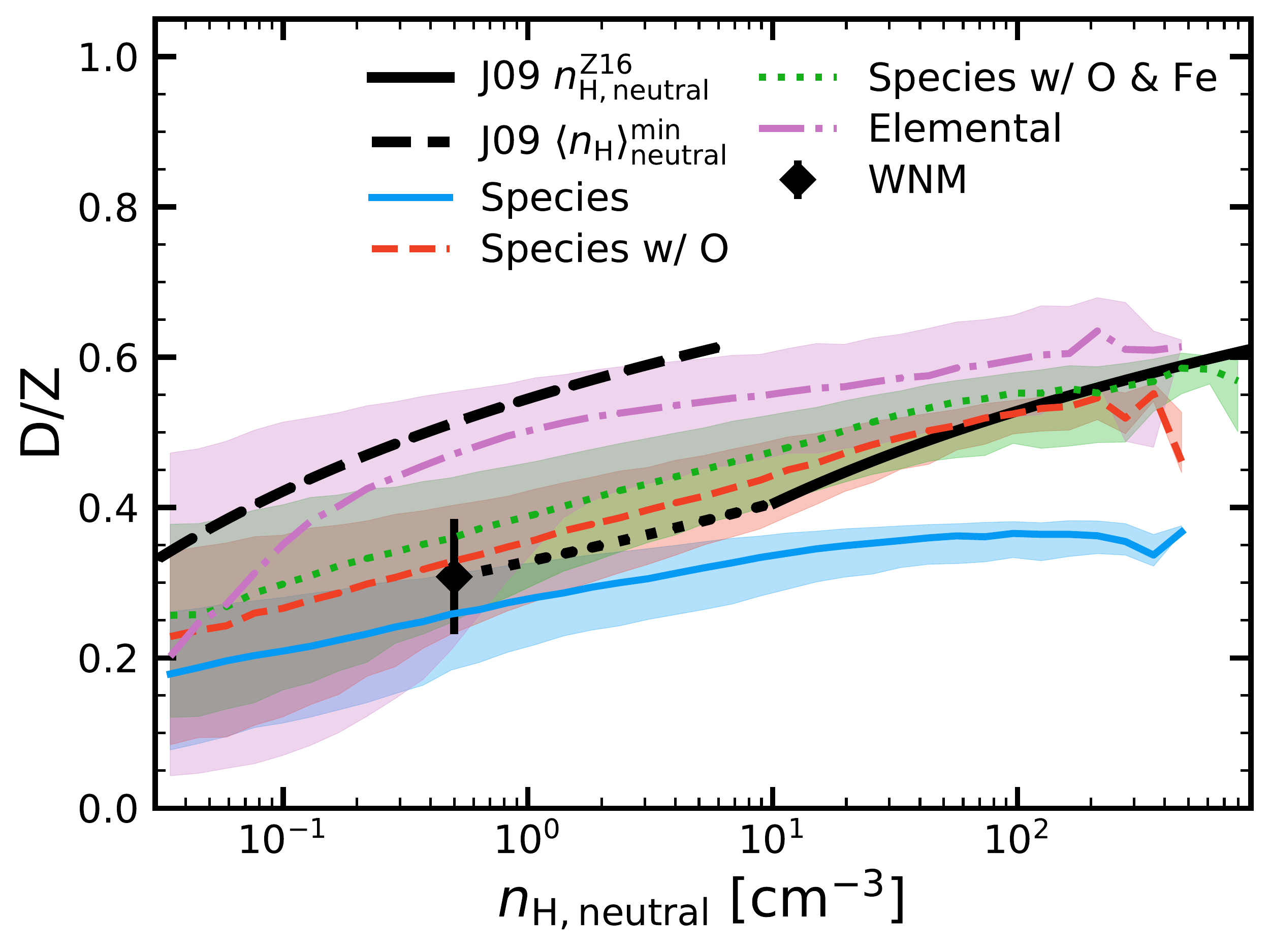}
    \vspace{-0.25cm}
    \caption{Relation between D/Z ratio and $n_{\rm H,neutral}$ in an idealized Milky Way-like galaxy for ``Species'' and ``Elemental'' implementations, with 16-/84-percentiles represented by shaded regions. We compare with observed D/Z values derived from observed elemental depletions (summing depletions from Fig.~\ref{fig:depl_vs_nH})  in the Milky Way from \citet{jenkins_2009:UnifiedRepresentationGasPhase} and assuming mean sight line density is the physical density ({\it black-dashed}), which can be treated as an upper limit, and using \citet{zhukovska_2016:ModelingDustEvolution,zhukovska_2018:IronSilicateDust} mean sight line density to physical density fit ({\it black-solid}). We also include an estimate of the expected D/Z in the WNM ({\it diamond}) with error bars assume  0-80\% of C locked up in dust along with an interpolation to the \citet{jenkins_2009:UnifiedRepresentationGasPhase} fit ({\it black-dotted}). The ``Elemental'' model produces a shallow sloped D/Z relation across the observed range which over-predicts the amount of dust in low-density environments. The ``Species'' model produces a more strongly sloped relation, but the fiducial model does not produce the high D/Z values observed in dense environments. The addition of the O-reservoir dust species largely fixes this issue and steepens the D/Z slope. On the other hand, the inclusion of the Nano-iron dust species only produces a small increase in the overall D/Z relation.}
    \label{fig:DZ_vs_nH} 
\end{figure}

Since the total dust population is an aggregate of all element depletions, we also investigate the resulting distribution of D/Z with neutral gas density for the same models. The median D/Z and 16-/84-percentile are shown in Fig.~\ref{fig:DZ_vs_nH}. Aggregating the observed element depletions, we include an upper bound on the expected D/Z along with an estimate for the WNM\footnote{To account for the uncertainty in the observed WNM D/Z owing to the lack of measured sight line C depletions in this regime we include error bars representing 80\% (assuming 20\% in CO) of C or no C in dust.} 
based on depletions from \cite{jenkins_2009:UnifiedRepresentationGasPhase} along with a reasonable estimate of D/Z in dense environments following the aforementioned relation from \cite{zhukovska_2016:ModelingDustEvolution,zhukovska_2018:IronSilicateDust}.

We will remark here that the large scatter in the simulated depletion values and consequently D/Z arises from the diverse histories of gas with the same density, particularly depending on what stage the gas is in the molecular cloud life cycle. If the gas is being ejected from the molecular cloud phase and on its way to the WNM phase, it will have higher depletions than gas that has resided in the WNM for an extended period, being subject to various destruction processes. Similarly, gas that quickly collapses into the molecular cloud phase or spends a relatively long amount of time in the WNM phase will have lower depletions than gas that takes a long time to collapse or resides in the WNM phase for a short period of time.

\subsubsection{Elemental}

Since this implementation uses the same gas-dust accretion timescales for all elements, any differences between element depletions due to the initial dust population vanish in the long term, producing identical depletion relations with column $N_{\rm H,neutral}$ and physical $n_{\rm H,neutral}$ for each element. This inherently cannot match observed depletions as shown in Fig.~\ref{fig:depl_vs_NH} \&~\ref{fig:depl_vs_nH}, which clearly vary element by element. C is the only element that deviates in this model, due to our accounting of C trapped in CO which produces a $\leq0.15$ increase in the depletion at any given column or physical density. Ultimately this model can only match depletion observations for one element at a time, while heavily over- or under-depleting the rest of the elements. This is especially problematic for O since it makes up a large fraction of the ISM metal mass. The resulting relation also has a relatively shallow slope across the observable range, with depletions increasing by $\sim$0.1 or $\sim$0.2 from the WNM to the dense molecular regime for the column and physical density respectively, compared to the steeper relations observed in the MW for Mg, Si, and Fe.  This shallow relation arises predominantly from the lack of a temperature restriction on dust growth and subordinately on the identical gas-dust accretion treatment for each element and generalized dust chemical composition. Dust residing in diffuse, warm gas for significant amounts of time grows appreciably, replacing dust destroyed by SNe remnants and thermal sputtering and leading to relatively small changes in element depletions for all but the most diffuse gas.

In regards to the total dust population, this produces reasonable D/Z values at high densities but the slope of the relation is again relatively shallow as seen in Fig.~\ref{fig:DZ_vs_nH}. While this does produce a range of D/Z values in the WNM regime which overlap with observations, the typical D/Z is still well above this.

\subsubsection{Species}

Since many details and optional dust physics modules for this implementation are motivated by observed depletions of individual elements, we review each element depletion relation with column $N_{\rm H,neutral}$ and physical $n_{\rm H,neutral}$ as shown in Fig.~\ref{fig:depl_vs_NH} \&~\ref{fig:depl_vs_nH} in detail below. A brief overview of which dust physics modules for this implementation affect individual element depletions and/or overall D/Z is provided in Table \ref{tab:DustOptions}. 

{\bf Magnesium and Silicon:}
Mg and Si are expected to have nearly identical depletion relations and almost entirely reside in silicate dust grains, so we will examine both elements together. Focusing at first on Mg, our fiducial model is able to reasonably match observed depletion trends with respect to both H column density and local gas density, with the predictions transitioning from a shallow to steep slope with increasing density. This change in slope arises from the transition between the diffuse medium where dust is destroyed by SNe and thermal sputtering to dense neutral gas where the dust rapidly grows via gas-dust accretion. On the other hand, our model does not produce a similar change in slope for Si depletion, instead exhibiting a constant shallow slope (shallower than inferred from observations). This result stems not from a failing of our dust model per se, but from the metal yield prescription in FIRE-2. The IMF-integrated SNe yields in FIRE-2 produce more Si than Mg, leading to a galaxy wide overabundance of Si compared to Mg after ${\sim}100$ Myr into our simulations. This, in turn, leads to Mg being the key element for silicate dust growth. The next version of the FIRE model \citep{hopkins_2022:FIRE3UpdatedStellar} uses an updated set of metal yields which predict modestly higher Mg production. This has the effect of making Si the key element, and, as we show in Appendix \ref{Appendix_FIRE23}, producing much better agreement with observations.

The inclusion of either the O-reservoir and/or Nano-iron dust species has only modest effects on the resulting Mg and Si depletion relations. It should again be noted that the Nano-iron module assumes the metallic iron nanoparticle inclusions in silicate dust provide the needed Fe for the silicate chemical structure instead of atomic, gas-phase Fe. Without this assumption, Fe becomes the key element for silicate dust growth and the resulting Mg and Si depletions end up far too low compared to observations.

{\bf Oxygen:}
Our fiducial model demonstrates that, as expected, pure silicate dust alone cannot reproduce the large dispersion in observed O depletions in dense environments, sequestering $\sim $20\% of available O and producing a relatively flat depletion relation with almost no scatter. Adding an O-reservoir dust species partly rectifies this issue, producing a similar change in slope to Mg and Si and a larger scatter, which produce a better match to both H column density and local gas density observations. While the O-reservoir species parameters are implicitly designed to match observed depletions at high densities, the shallow slope in diffuse environments arises purely from gas cycling out of neutral gas and being exposed to SNe dust destruction and thermal sputtering.

{\bf Iron:}
It is already known that purely silicate dust alone cannot reproduce the extreme depletions of gas-phase iron. Adding the Normal-iron species, as used in our fiducial model, does not fix this issue. Such metallic iron dust grows too slowly, even with accounting for Coulomb enhancement in our model. The addition of the Nano-iron dust species produces a depletion relation more in line with observations, with a similar transition from a shallow to steep slope as Mg and Si but with a far steeper slope in dense environments with both high H column density and local gas density. The combination of small grain sizes and Coulomb enhancement gives this metallic iron species an extremely short accretion timescale depleting nearly all gas-phase Fe in dense environments. The Coulomb enhancement term being especially important, as shown in Appendix~\ref{Appendix_Coulomb}. In combination, the reduced destruction efficiency of these grains due to them being modeled as shielded inclusions allows for the still relatively high depletions in the low density regime, compared to other elements.

{\bf Carbon:}
Of the elements we track in dust, carbon is the least constrained by observations, with only a handful of sight line depletion observations over a narrow range of $N_{\rm H}$ and observed fractions of C in CO (${\sim}20\%{-}40\%$) providing a constraint in dense molecular environments\footnote{In diffuse and dense neutral gas regime CO is directly observed via absorption or emission features \citep[e.g][]{sheffer_2008:UltravioletSurveyCO}. These CO observations give a lower limit to the gas-phase C abundance.}. Moreover, our treatment of carbonaceous dust is quite restrictive compared to silicate, with growth via accretion only occurring in CNM and diffuse molecular environments before the gas transitions to the dense molecular phase, where we assume CO takes up any remaining gas-phase C, halting carbon dust growth. Even with these constraints, our predicted C depletions fall well within observational bounds, producing a steep slope between the WNM and CNM phases ($0.1$ cm$^{-3}<n_{\rm H,neutral}<50$ cm$^{-3}$), which in turn yields a steep slope for $\NHn>2\times10^{20}$ cm$^{-2}$, while still leaving enough gas-phase C at high densities to match observed CO abundances. The scatter in C depletions at high density also matches surprisingly well with the range of observed values for C in CO. This dispersion arises from the variable history of individual gas parcels in the simulation, with gas that quickly transitions to the dense molecular phase forming less dust and allowing a larger amount of C to be locked in CO, while gas which slowly transitions forms more dust and leaves less gas-phase C to form CO. This suggests that the large range in observed C in CO fractions (20\%-40\%) can be attributed to the history of the gas in question. We also tested our ``Species'' model without any accounting for CO, allowing carbonaceous dust to grow in dense molecular gas akin to the other (non-C) dust species above, but the resulting C depletions are too low, leaving far too little gas-phase C in dense environments compared to observations of CO as shown in Appendix~\ref{Appendix_Coulomb}. This highlights the need for an accounting of C in CO to accurately model the evolution of carbonaceous dust. Our results also suggest very little carbonaceous dust should exist at low local gas densities ($n_{\rm H,neutral}<0.1$ cm$^{-3}$) compared to silicate dust, although this prediction is sensitive to our assumptions about dust sizes and destruction in SNe. This could have considerable effects on the effective attenuation law in such environments. However, these results may be due to this being an idealized galaxy without a realistic corona/disc-halo interface, or due to the details of cooling/heating and neutral gas physics used in FIRE-2 (Appendix \ref{Appendix_FIRE23}). Investigation with fully cosmological simulations will be needed to explore this further.

{\bf D/Z:} Our fiducial model produces too little dust in dense environments, per Fig.~\ref{fig:DZ_vs_nH}, leading to the low galaxy-integrated D/Z shown in Sec.~\ref{Species Free Parameters}. This failing is a direct result of our adopted dust composition constraints, specifically of O. When we include the O-reservoir dust species this issue is largely resolved, with the D/Z slope steepening for $n_{\rm H,neutral}>1$ cm$^{-3}$, in plausible consistency with observations and increasing the galaxy-integrated value to D/Z $\sim0.34$. Adding the Nano-iron dust species produces an overall shift in D/Z of $\leq$ 0.04 but does not change the shape of the relation, suggesting that tracking a separate metallic iron dust species is not essential when modeling steady-state dust populations. In any case, all versions of this implementation produce a non-negligible median D/Z $\geq$ 0.2 even in the most diffuse gas. This suggests a sizeable fraction of metals are trapped in dust no matter the gas phase, but again this may be a consequence of this being an idealized galaxy or due to details of the gas phase structure.

In summary, the ``Elemental'' implementation's near identical treatment of all refractory elements in dust prevents it from matching observed variable element depletions and its allowance for unrestricted gas-dust accretion produces relatively flat element depletions and D/Z across the observed range of H column and local gas densities. Conversely, the ``Species'' implementation's accounting for dust chemical composition is able to match observed element depletion, but the inclusion of some additional theoretical O-reservoir and Nano-iron dust species are needed to match O and Fe depletions respectively. In addition, the T$<$300K restriction on gas-dust accretion along with Coulomb enhancement produces a steep slope in element depletion and D/Z relations with both H column density and local gas density.

\begin{table}
    \centering
    \renewcommand{\arraystretch}{1.2}
	\begin{tabular}{c c c}
		 \hline\hline
			 & D/Z & Element Depletion \\ [0.5ex] 
			\hline
			O-reservoir & Yes & Yes \\
			Nano-iron & No & Yes \\
			C in CO & No & Yes \\
			Coulomb Enh. & Yes & Yes \\
		\hline
	\end{tabular}
	\caption{Table summarizing whether or not a given piece of our assumed dust physics strongly influences either the D/Z ratio or element-by-element depletion trends.}
	\label{tab:DustOptions}
\end{table}

\subsection{Spatially Resolved D/Z Beyond the MW} \label{Extragalactic}

\begin{figure*}
    \plotsidesize{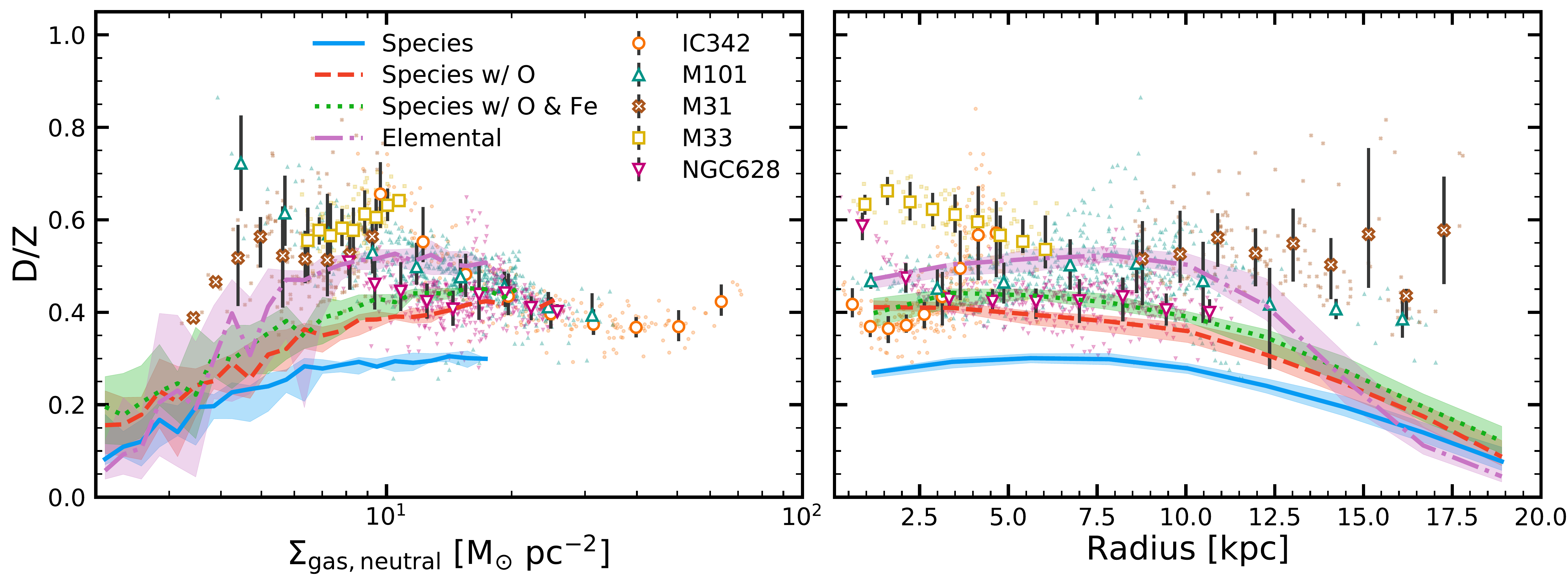}{0.99}
    \vspace{-0.25cm}
    \caption{Relation between median D/Z ratio and neutral gas surface density ({\it left}) and galactocentric radius ({\it right}) in 2 kpc bins at simulation end for the ``Elemental'' and ``Species'' implementations with 16-/84-percentiles represented by the shaded regions. We compare with dust emission based observations of spatially-resolved D/Z for a few local galaxies (IC 342, M101, M31, M33, and NGC 628) from \citet{chiang_2021:ResolvingDusttoMetalsRatio} with 2 kpc resolution and $\alpha_{\rm CO}^{\rm B13}$ conversion factor. We also show each observed galaxy's binned median and 16-/84-percentile ranges for D/Z with respect to each given property. We emphasize that the observed galaxies' physical sizes and metallicities do not closely correspond to our idealized galaxy and so this comparison is only an illustration of these dependencies and should not be used for strong quantitative comparison without proper matching of galactic properties which we will do with fully cosmological simulations in future work.}
    \label{fig:obs_DZ_vs_gas}
\end{figure*}

Looking to extragalactic observations of dust, direct measurements of element depletions are very challenging with current instruments (i.e. key refractory elements, notably carbon, are not observable via absorption outside the MW; \citealt{roman-duval_2019:MetalAbundancesDepletions,roman-duval_2019:METALMetalEvolution,peroux_2020:CosmicBaryonMetal}). An alternative, albeit somewhat model-dependant, method for estimating D/Z is to combine separate, multi-wavelength estimates of dust mass, gas mass, and metallicity. This method relies on matching dust emission spectra to infer a dust mass and so mainly probes denser environments compared to depletion-based observations, and does build in some implicit dependence on assumed dust chemistry and size distributions. While this method has yielded a plethora of galaxy-integrated studies of D/Z \citep[e.g.][]{remy-ruyer_2014:GastodustMassRatios,devis_2019:SystematicMetallicityStudy}, these observations are less suited for constraining our dust evolution models since our models depend on the local gas environments within the galaxy and we only simulate one Milky Way-like galaxy. A more useful constraint for our purposes here is spatially resolved D/Z studies of individual galaxies, but few of these studies exist with only the Magellanic Clouds \citep{jenkins_2017:InterstellarGasphaseElement,roman-duval_2014:DustGasMagellanic,roman-duval_2017:DustAbundanceVariations} and M31 \citep{draine_2014:AndromedaDust} being mapped until recently. Recent work by \citet{chiang_2021:ResolvingDusttoMetalsRatio} investigated the spatially resolved D/Z-environment relations (using the technique above) for five nearby galaxies: IC 342, M31, M33, M101, and NGC 628. We compare our simulations with these observations in Fig.~\ref{fig:obs_DZ_vs_gas} examining the relation between D/Z, neutral gas surface density $(\Sigma_{\rm gas,neutral})$, and galactocentric radius. We specifically show their derived D/Z values using the \citet{bolatto_2013:COtoH2ConversionFactor} $\alpha_{\rm CO}$ prescription ($\alpha_{\rm CO}^{\rm B13}$), which \citet{chiang_2021:ResolvingDusttoMetalsRatio} argued yields the most reasonable D/Z. To match the observational resolution we bin the simulation gas in 2 kpc face-on square pixels and calculate D/Z $=\Sigma_{\rm dust}/\Sigma_{\rm metals}$ for each pixel. We then group these pixels across $\Sigma_{\rm gas,neutral}$ and galactocentric radius and calculate the median D/Z values and 16-/84-percentiles for each. 

Both the fiducial ``Elemental'' implementation and ``Species'' implementation with included O-reservoir and Nano-iron dust species are consistent with observations, falling near the middle and lower end of the observed range respectively for D/Z relative to both $\Sigma_{\rm gas,neutral}$ and galactocentric radius. On the other hand, the fiducial ``Species'' implementation produces too low D/Z in all gas environments, again emphasizing the importance of an additional O depletor beyond purely silicate dust. \citet{chiang_2021:ResolvingDusttoMetalsRatio} also points out that there is an offset between emission based and depletion based observations, with emission producing higher D/Z across all gas metallicities they observe. This offset is either due to them probing different gas phases (\textsc{Hi}- vs $\Hmol$-dominated), or due to any of the many systematic uncertainties in both methods, most notably the assumed dust population/emissivity model \citep{chastenet_2021:BenchmarkingDustEmission}. Further study utilizing fully cosmological simulations with a sample of different galaxies is needed to better compare both our models with these observations due to the variable properties and histories of these galaxies.

\section{Conclusions} \label{Conclusions}

In this work we implemented two separate dust evolution models, labeled ``Elemental'' and ``Species'', into the GIZMO code base and coupled them with FIRE-2 stellar feedback and ISM physics. Both models account for dust creation in stellar outflows, dust growth from gas-phase accretion, dust destruction from SNe shocks, thermal sputtering, and astration, and turbulent dust diffusion in gas. The ``Elemental'' model tracks the dust yields of individual elements incorporated into carbonaceous and generalized silicate dust which are treated near identically in all physical processes and utilizes a `tunable' dust growth routine. The ``Species'' model tracks the yields of specific dust species (silicates, carbon, silicon carbide, and iron), treating each uniquely depending on their chemical composition, along with optional nano-particle metallic iron (Nano-iron) dust species and an unknown oxygen based (O-reservoir) dust species, and incorporates a physically motivated dust growth routine. We also devised and integrated a sub-resolution dense molecular gas scheme with both models to account for different efficiencies of Coulomb enhancement for gas-dust accretion (in the ``Species'' model) and the reduction in carbon dust accretion due to the lock-up of gas-phase C into CO in dense molecular gas (Appendix~\ref{Appendix_MC}).

Using both dust models, we ran idealized non-cosmological simulations of a Milky Way-mass galaxy to test their sensitivity to free parameters and compare to observations of D/Z and elemental depletions. We summarize our findings below:

\begin{enumerate}

    \item Both implementations reaffirm that the steady-state galaxy-integrated D/Z ratio depends on the balance between gas-phase accretion and dust destruction by SNe, with the efficiency of initial stellar dust production having little effect as long as some ``seeds'' exist so that accretion can take over as the dominant dust source (Appendix~\ref{Appendix_Stardust}). The fiducial ``Species'' implementation is able to produce a reasonable, but slightly low, D/Z $\sim$ 0.27 (Fig.~\ref{fig:spec_time_evo}), which increases to D/Z $\sim$ 0.34 or D/Z $\sim$ 0.38 with the inclusion of either the O-reservoir dust species or both the O-reservoir and Nano-iron dust species respectively (Fig.~\ref{fig:DZ_vs_nH}). The fiducial ``Elemental'' model produces a reasonable D/Z $\sim$ 0.47 (Fig.~\ref{fig:elem_time_evo}), but this requires manually ``tuning'' the gas-phase accretion rate for our simulation. While both models can produce, or be `tuned' to produce, reasonable galaxy-integrated D/Z ratios, the predicted relations between element depletions and local gas properties vary dramatically.
    
    \item The ``Elemental'' implementation is inherently unable to reproduce the variation in observed MW element depletions (Fig.~\ref{fig:depl_vs_NH} \&~\ref{fig:depl_vs_nH}), owing to its uniform treatment of accretion for each element in dust. This is especially problematic for O, which makes up a large portion of the metal mass. Furthermore, since there are no restrictions on gas-dust accretion for this implementation, dust that resides in hot gas for long periods can grow faster than it is destroyed by SNR or thermal sputtering, thus producing a relatively flat D/Z-$n_{\rm H,neutral}$ relation in all but the most diffuse gas (Fig.~\ref{fig:DZ_vs_nH}).
    
    \item The fiducial ``Species'' model is only able to match observed Mg, Si, and C depletions in the Milky Way using our default standard model for silicates and carbonaceous grains with fixed chemical compositions. The inclusions of some additional theoretical O-reservoir and Nano-iron dust species are needed to match observed O and Fe depletions respectively for the model variations we study (Fig.~\ref{fig:depl_vs_NH} \&~\ref{fig:depl_vs_nH}). This additional O depletion is also critical to match observed D/Z ratios in the Milky Way, with the resulting D/Z-$n_{\rm H,neutral}$ relation being consistent with observations (Fig.~\ref{fig:DZ_vs_nH}). In this model, a temperature restriction on gas-dust accretion produces low D/Z ratios in diffuse environments, while high gas-phase accretion rates in the cold ISM (in conjunction with Coulomb enhancement; Appendix~\ref{Appendix_Coulomb}) yield large D/Z ratios in dense environments.

    \item Extragalactic observations of spatially-resolved D/Z are, at present, roughly consistent with both models, provided these models also reproduce the MW D/Z (Fig.~\ref{fig:obs_DZ_vs_gas}).
        
    \item An accounting of C locked in CO ($\fCO$; Appendix~\ref{Appendix_MC}) can have important effects on depletion patterns, especially for C (Appendix~\ref{Appendix_Coulomb}).

\end{enumerate}

Our results show that while a simplistic one-phase ``dust by element'' evolution model can produce reasonable galaxy-integrated dust properties, a more complex, chemically motivated two-phase ``dust by species'' evolution model is needed to reproduce observed spatial dust variability, in both amount and composition, within a galaxy. In a companion paper, we will further investigate and compare both models in a fully cosmological context for a wide range of galaxy halo masses to study the relation between dust and various galactic properties and the effects of integrating `live' dust evolution with radiative transfer/feedback and cooling and heating ISM physics on galaxy evolution (as opposed to assuming a constant D/Z). These studies will provide further tests for the current dust population and chemical composition paradigm.

We stress that our dust models are in no way ``complete'' and, beyond the major uncertainties we detail, our models lack important dust physics which could have drastic effects on our results. 
First and foremost we do not track the evolution of the dust grain size distribution, and with it grain coagulation and shattering physics, which can greatly affect the accretion rate (small grains dominate gas-dust accretion) and SNe destruction efficiency (small grains are more easily destroyed compared to large grains). 
We also do not account for polycyclic aromatic hydrocarbons (PAHs), a subspecies of carbonaceous dust grains which are extremely small (<1 nm) and could dominate carbonaceous gas-dust accretion (but these may only be produced via grain shattering and would require tracking the aromatization of dust grains due to the local radiation field; \citealt{rau_2019:ModellingEvolutionPAH}). 
Our dust models also do not incorporate a full, non-equilibrium chemical network even though molecular formation depends on the exact amount and size of dust grains either directly, by forming on grain surfaces (e.g. $\rm H_2$), or indirectly, by depending on molecules which form on dust grains (e.g. CO).
With these in mind, our goal here is to lay a solid foundation for the incorporation and investigation of such physics in future works.

\section*{Acknowledgements}
We thank I-Da Chiang, Lichen Liang, and Alexander J. Richings for insightful suggestions and advice.
CC and DK were supported by NSF grants AST-1715101 and AST2108314 and the Cottrell Scholar Award from the Research Corporation for Science Advancement. 
Numerical calculations were run on the UC San Diego Triton Shared Computing Cluster, and allocations AST20016 \&\ TG-AST140023 supported by the NSF and TACC.
Support for PFH was provided by NSF Research Grants 1911233 \&\ 20009234, NSF CAREER grant 1455342, NASA grants 80NSSC18K0562, HST-AR-15800.001-A. 
KS was supported by National Science Foundation grant No. 1615728.
CAFG was supported by NSF through grants AST-1715216, AST-2108230,  and CAREER award AST-1652522; by NASA through grant 17-ATP17-0067; by STScI through grant HST-AR-16124.001-A; and by the Research Corporation for Science Advancement through a Cottrell Scholar Award. 
The data used in this work were, in part, hosted on facilities supported by the Scientific Computing Core at the Flatiron Institute, a division of the Simons Foundation. 
This work also made use of MATPLOTLIB \citep{hunter_2007:Matplotlib2DGraphics}, NUMPY \citep{harris_2020:ArrayProgrammingNumPy}, SCIPY \citep{virtanen_2020:SciPyFundamentalAlgorithms}, the \yt\ project \citep{turk_2011:YtMulticodeAnalysis}, and NASA’s Astrophysics Data System.

\datastatement{The data supporting the plots within this article are available on reasonable request to the corresponding author. A public version of the \GIZMO\ code is available at \gizmourl.}



\bibliographystyle{mnras}
\bibliography{references}


\appendix

\section{Effects of variations in initial conditions} \label{Appendix_ICs}

\begin{figure*}
	\plotsidesize{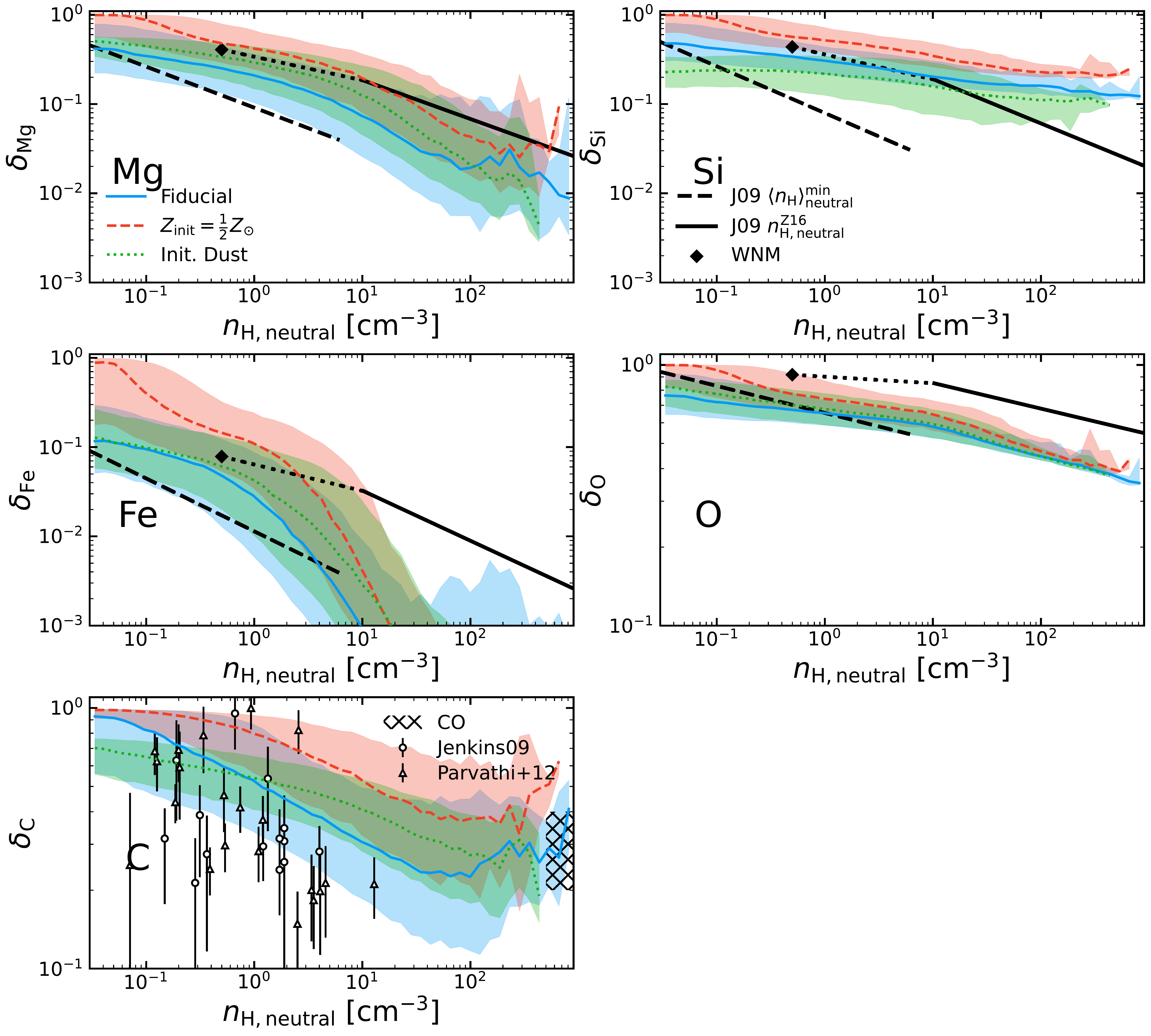}{0.99}
    \vspace{-0.25cm}
    \caption{Same as Fig.~\ref{fig:depl_vs_nH} comparing our ``Species'' implementation (including Nano-iron and O-reservoir dust species) with our fiducial initial conditions ({\it solid}), with initial gas and stellar metallicity of $Z_{\rm init}=0.5 Z_{\sun}$ ({\it dashed}), and with an initial dust population in all gas cells ({\it dotted}). Overall the resulting element depletions are quite similar for the different initial conditions, but with a small systematic offsets due to differences in gas-dust accretion rates which depend on the gas metallicity. The large differences for the $Z_{\rm init}=0.5 Z_{\sun}$ run at low densities is due to the reduced efficiency of metal-line cooling which reduces the cycling of gas into and out of cool, dense environments.}
    \label{fig:ICs_depl_vs_nH}
\end{figure*}

\begin{figure}
	\includegraphics[width=0.99\columnwidth]{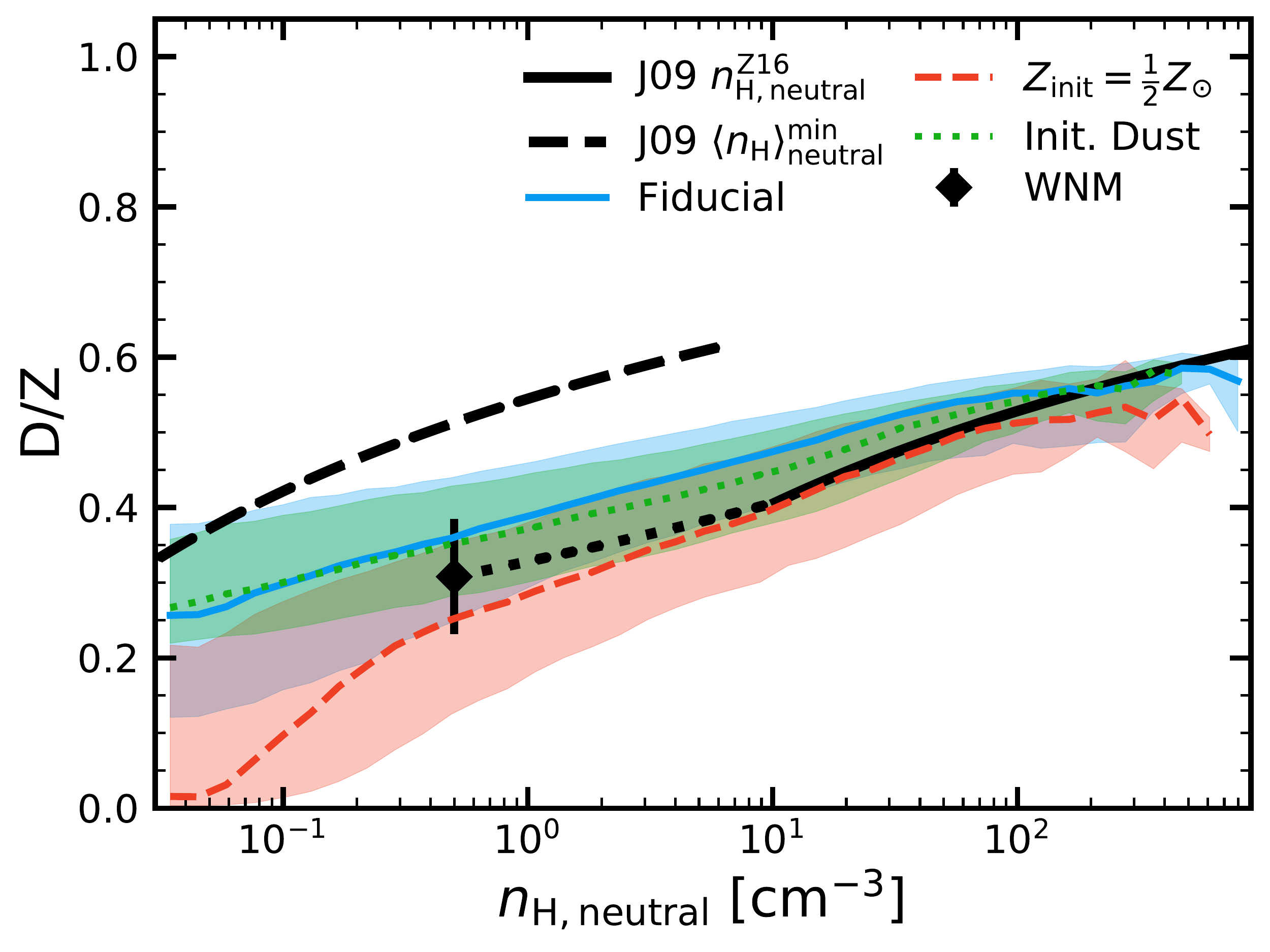}
    \vspace{-0.25cm}
    \caption{Same as Fig.~\ref{fig:DZ_vs_nH} comparing our ``Species'' implementation (including Nano-iron and O-reservoir dust species) with our fiducial initial conditions ({\it solid}), with initial gas and stellar metallicity of $Z_{\rm init}=0.5 Z_{\sun}$ ({\it dashed}), and with an initial dust population in all gas cells ({\it dotted}). The small systematic offsets between the runs for $n_{\rm H,neutral}\geq0.5$ is due to differences in gas-dust accretion rates which depend on the gas metallicity. The sharp decrease in D/Z for the $Z_{\rm init}=0.5 Z_{\sun}$ run at $n_{\rm H,neutral}<0.5$ is due to the reduced efficiency of metal-line cooling which reduces the cycling of gas into and out of cool, dense environments.}
    \label{fig:ICs_DZ_vs_nH}
\end{figure}

In Fig.~\ref{fig:ICs_depl_vs_nH} and Fig.~\ref{fig:ICs_DZ_vs_nH} we show the resulting element depletions and D/Z relation versus $n_{\rm H,neutral}$ at the end of the simulation for the ``Species'' implementation (including Nano-iron and O-reservoir dust species) with our fiducial initial conditions, initial gas cell and stellar metallicity of $Z_{\rm init}=0.5 Z_{\sun}$ with no initial dust population, and initial gas cell and stellar metallicity of $Z_{\rm init}=Z_{\sun}$ with an initial dust population for all gas cells. Specifically the initial dust population is assumed to be entirely from SNe II and is set by $\delta_{\rm Si}=0.5$ with $\delta_{\rm Mg}$ and $\delta_{\rm O}$ set to match our defined silicate dust composition, $\delta_{\rm Fe} = 0.5$, and  $\delta_{\rm C} = 0.25$ such that the silicate-to-carbon ratio ${\sim}2.5$. This results in an initial D/Z${\sim}0.2$. The $Z_{\rm init}=0.5 Z_{\sun}$ simulation was evolved for 1.5 Gyr and resulted in a median gas metallicity of $Z = 0.75 Z_{\sun}$ while the initial dust population simulation was evolved for 0.7 Gyr (the time at which $\sim90\%$ of the galactic dust mass is composed of ``new'' dust produced by gas-phase accretion compared to the initial SNe II dust population) and resulted in a median gas metallicity of $Z = 1.15 Z_{\sun}$. 
The results are roughly consistent across the observed range with similar depletion trends for all elements, but there are small systematic offsets compared to our fiducial run largely due to the overall lower median gas metallicity at simulation end which reduces the gas-dust accretion rate.
The largest difference is the drastically reduced element depletions and D/Z in the most diffuse gas for the $Z_{\rm init}=0.5 Z_{\sun}$ simulation. This is due to the reduced cycling of gas into and out of cool, dense environments. The reduced metallicity reduces the efficiency of metal-line cooling in hot gas which in turn reduces the amount of cool, dense gas that is formed in the galaxy by a factor of ${\sim}0.5$.
Fully cosmological simulations will be needed to further investigate the effects of galactic metallicity on the resulting element depletions and D/Z.

\section{Sub-Resolution treatment of dense molecular gas chemistry} \label{Appendix_MC}

In order to accurately model gas-dust accretion for carbonaceous dust grains and account for Coulomb enhancement terms we must track the mass fraction of the gas that is in the dense molecular phase ($\fdens$), where we assume that (1) nearly all gas-phase metals are neutral (so Coulomb enhancement terms are negligible) and (2) specifically gas-phase carbon is almost completely molecular in the form of CO \citep[e.g.][]{snow_2006:DiffuseAtomicMolecular} and unavailable for carbonaceous dust growth.\footnote{In extremely dense, cold environments ($\nH\geq10^5$ cm$^{-3}$, $T<20\,$K) CO does `freeze-out', forming icy mantles on the surface of dust grains \citep[e.g.][]{boogert_2015:ObservationsIcyUniverse} but we do not track this.} Note in this work we do not use $\fdens$ to account for sub-resolution density/temperature structure in our gas-dust accretion routine.

To calculate $\fdens$ we employ a method similar to~\citet{krumholz_2011:ComparisonMethodsDetermining}, which is used in FIRE-2 to estimate the molecular mass fraction ($f_{\Hmol}$) of gas cells. Specifically, this method assumes the gas cell is an idealized spherical cloud immersed in a isotropic dissociating radiation field, with an assumed shielding length ($r_{\rm shield}$) and metallicity used to estimate the total integrated column density of the gas cell to this radiation. Based on the column density, a depth into the cloud is then determined beyond which the gas is self-shielded and molecular, which in turn determines $f_{\Hmol}$. FIRE-2 uses a Sobolev+cell approximation for $r_{\rm shield}$ which accounts for the average contribution of neighboring gas cells and actual contribution of the main cell to the column depth respectively (see~\citealt{hopkins_2018:FIRE2SimulationsPhysics}). In a similar fashion, for determining $\fdens$ we assume a spherical gas cell with radius equal to the Sobolev+cell shielding length. We then look to observations to determine what column depths are typical for gas to transition between the diffuse/C-rich and dense molecular/CO-rich phase. Observations have found that CO quickly takes over as the dominant form of gas-phase C over a very narrow range of sight line $\Hmol$ column densities \citep{liszt_2007:FormationFractionationExcitation, sheffer_2008:UltravioletSurveyCO}, and so we assume a critical $\Hmol$ column density ($N_{\Hmol}^{\rm crit}$) above which carbon immediately converts to CO and the gas is in dense molecular phase (we also for simplicity assume the same threshold for where the ionized metal fraction becomes negligible, although this may occur at different column densities in reality). With this assumption, along with assuming the $\Hmol$ tracked by $f_{\Hmol}$ in FIRE is evenly distributed within the gas cell, we calculate the depth into the gas cloud needed to reach this critical $\Hmol$ column density as
\begin{equation} \label{eq:dense_mol}
    d = \frac{2 N_{\Hmol}^{\rm crit}}{f_{\Hmol}\nH}
\end{equation}
where $\nH$ is the hydrogen number density of the gas cell tracked in our simulation. We then assume that all gas past this depth within the cell is in the dense molecular phase. Thus we determine the fraction of gas cell in the dense molecular phase as
\begin{equation}
    \fdens = \frac{(r_{\rm shield}-d)^3}{r_{\rm shield}^3}.
\end{equation}
We then track and update $\fdens$ at each time-step in our simulation and with this we also track the fraction of total C (in gas and dust) locked in CO ($\fCO$) for each gas cell based on the current fraction of atomic gas-phase C (not in dust or CO), $f_{\rm C,gas}$, specifically $\fCO^{\rm new} = \fCO^{\rm old} + \frac{(\fdens^{\rm new}-\fdens^{\rm old})}{1-\fdens^{\rm old}}f_{\rm C,gas}$. Note $f_{\rm C,gas}$ can change between time steps due to the injection of gas-phase C from metal producers or turbulent diffusion and/or the accretion of gas-phase C onto dust. Also, since we do not follow the exact physical location of CO in our gas cells, when $\fdens$ decreases between time steps we simply reduce $\fCO$ such that $\fCO^{\rm new} = \frac{\fdens^{\rm new}}{\fdens^{\rm old}} \fCO^{\rm old}$. It should also be noted that since we average the contribution of neighboring gas cells and assume they have similar $\Hmol$ densities, this could overestimate $\fdens$, and thus $\fCO$, in complex configurations such as near the edge of molecular clouds where there could be a sharp gradient in $f_{\Hmol}$ and $\nH$ for neighboring gas cells.

\begin{figure*}
	\plotsidesize{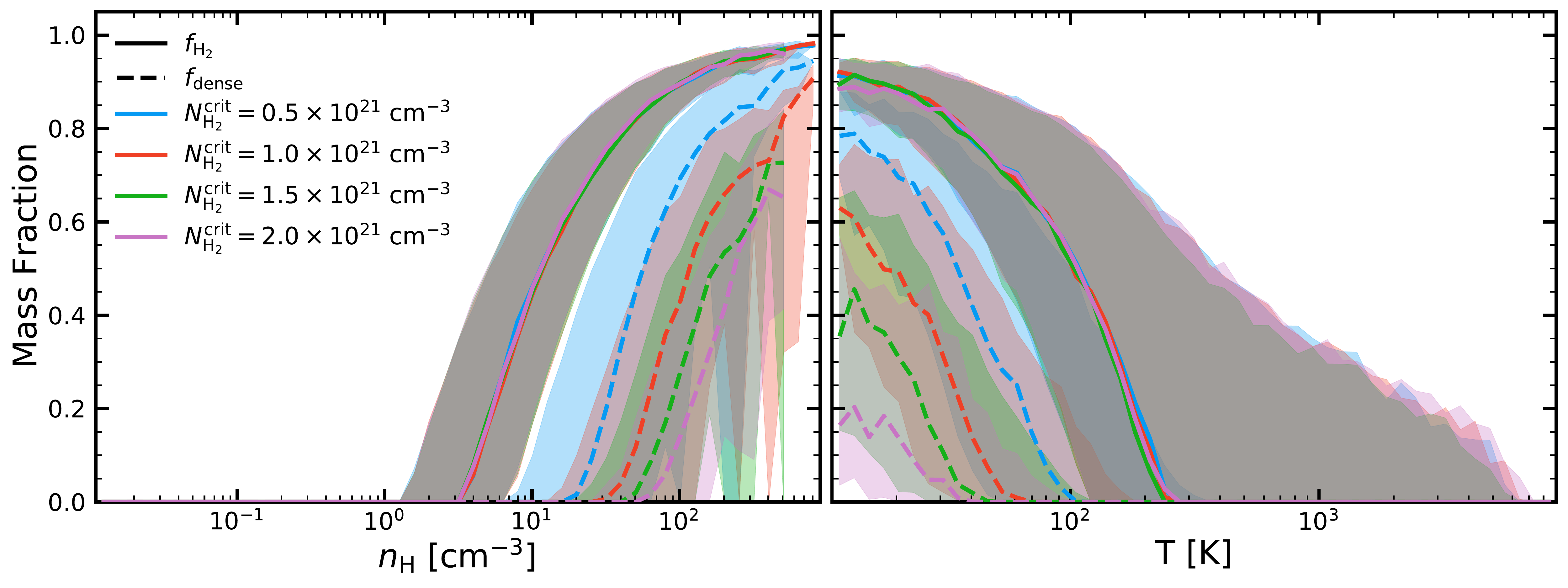}{0.99}
    \vspace{-0.25cm}
    \caption{Resulting median molecular mass fraction ($f_{\Hmol}$) ({\it solid}) predicted in our simulated galaxy's ISM gas and median mass fraction of gas in dense molecular phase ($\fdens$) ({\it dashed}) produced by our model (Appendix~\ref{Appendix_MC}) versus $\nH$ ({\it left}) and T ({\it right}) of all simulation gas cells, with 16-/84-percentile represented by shaded regions. We show the sensitivity of these results to our choice of $N_{\Hmol}^{\rm crit}$ (Eq.~\ref{eq:dense_mol}). Note $f_{\Hmol}$ does not depend on $N_{\Hmol}^{\rm crit}$ and so any differences between runs is purely stochastic.}
    \label{fig:NH2_crit} 
\end{figure*}

We tested the sensitivity of $\fdens$ to the $N_{\Hmol}^{\rm crit}$ parameter, with the resulting relations between $f_{\Hmol}$, $\fdens$, gas number density, and temperature given in Fig. \ref{fig:NH2_crit}. Using the ``Species'' dust model, we decided on a $N_{\Hmol}^{\rm crit} = 1.5 \times 10^{21}$ cm$^2$ which produces an average $\fCO\approx30\%$ in the densest environments at simulation end, which falls in the middle of the observed $\fCO$ range \citep[e.g.][]{irvine_1987:ChemicalAbundancesMolecular, vandishoeck_1993:ChemicalEvolutionProtostellar, vandishoeck_1998:ChemicalEvolutionStarForming,lacy_1994:DetectionAbsorptionH2}. This choice of $N_{\Hmol}^{\rm crit}$ is also in good agreement with observations, falling roughly in the middle of the observed transition between diffuse and dense molecular gas and low to high $N_{\rm CO}$ \citep[][see Fig. 7]{sheffer_2008:UltravioletSurveyCO}.

\section{Effects of variations in stellar dust formation prescriptions} \label{Appendix_Stardust}

\begin{figure}
	\includegraphics[width=0.99\columnwidth]{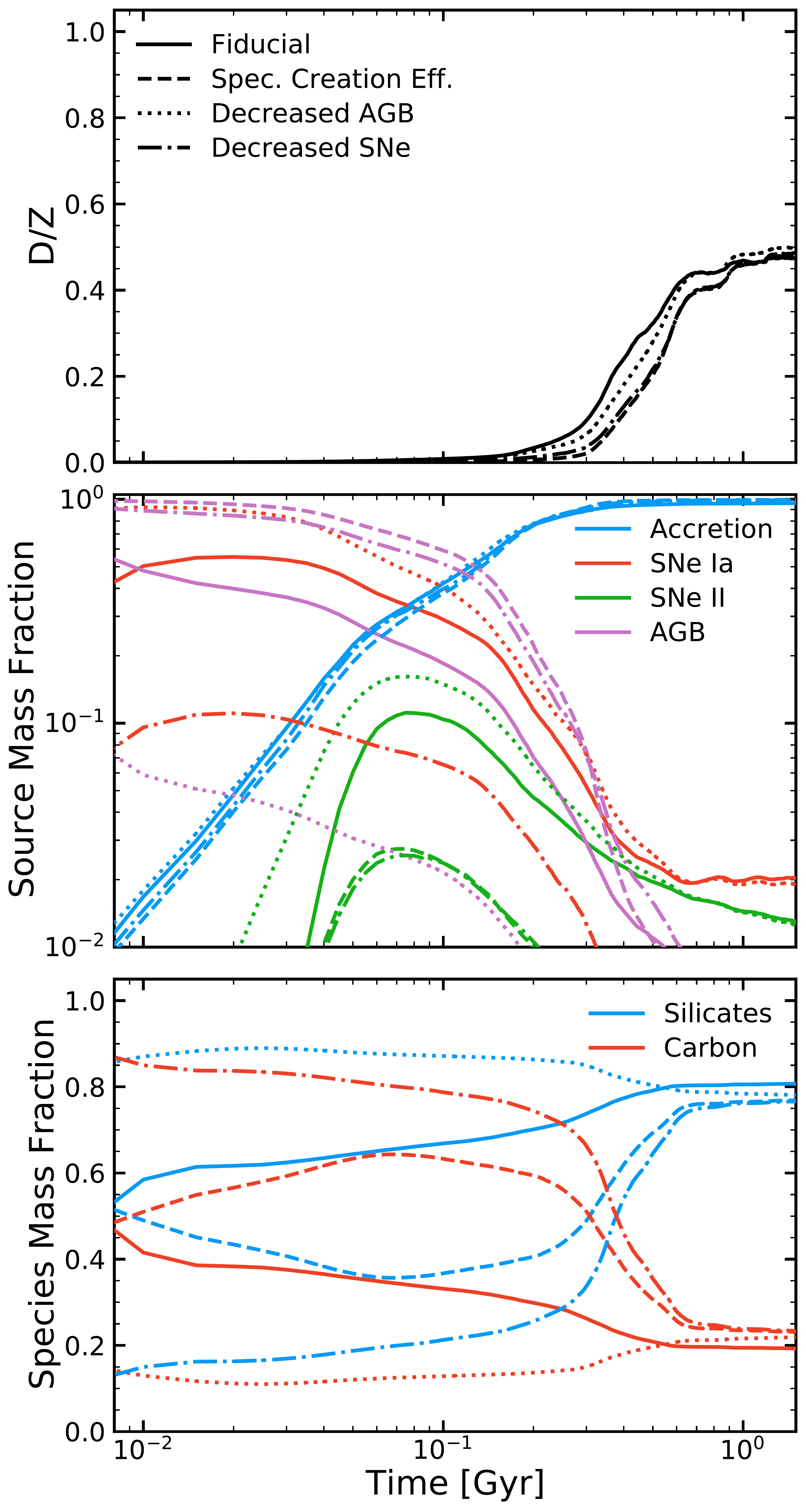}
    \vspace{-0.25cm}
    \caption{Same as Fig.~\ref{fig:elem_time_evo} but varying our assumed creation/formation efficiencies for the ``Elemental'' implementation of dust. We compare the ``Elemental'' implementation with the fiducial model ({\it solid}), using default ``Species'' SNe and AGB dust production rates ({\it dashed}), factor of 10 decrease in AGB dust production ({\it dotted}), and factor of 10 decrease in SNe dust production ({\it dash-dotted}). The effects at simulation end are small.}
    \label{fig:elem_vary_creation}
\end{figure}

\begin{figure}
	\includegraphics[width=0.99\columnwidth]{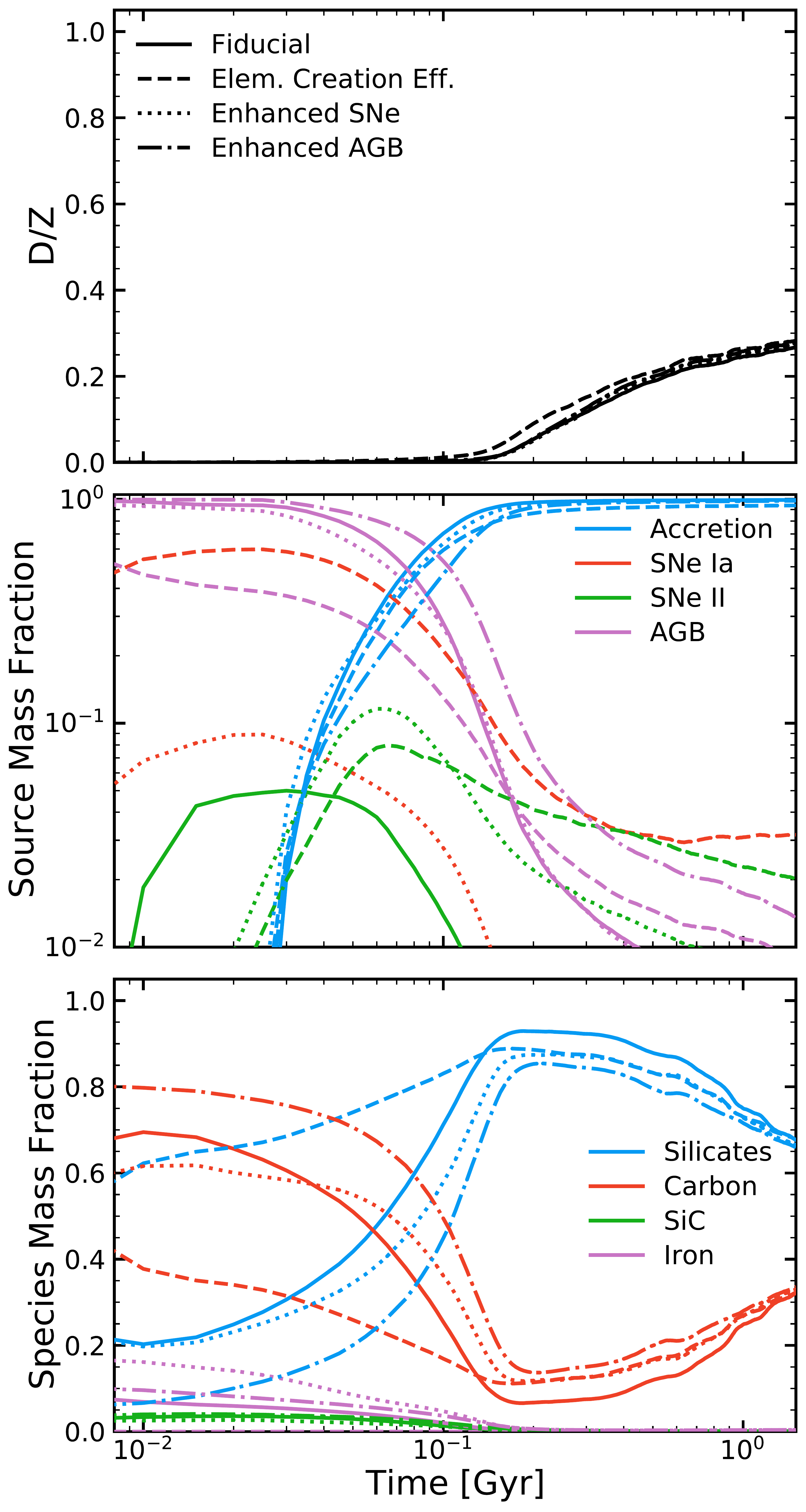}
    \vspace{-0.25cm}
    \caption{Same as Fig.~\ref{fig:spec_time_evo} but varying our assumed dust creation/formation efficiencies for the ``Species'' implementation of dust. We compare the ``Species'' implementation with the fiducial model ({\it solid}), using default ``Elemental'' SNe and AGB dust production rates ({\it dashed}), factor of 10 increase in SNe dust production ({\it dotted}), and factor of 10 increase in AGB dust production ({\it dash-dotted}). The effects are generally small, as the system rapidly reaches steady-state in which gas-dust accretion dominates. Note the differences in SNe II dust source fractions at early times is due to run-to-run variations in SNe II events and our galaxy being initially free of dust.}
    \label{fig:species_vary_creation}
\end{figure}

Fig.~\ref{fig:elem_vary_creation} and~\ref{fig:species_vary_creation} show resulting galaxy-integrated D/Z and dust mass fractions (as Fig.~\ref{fig:elem_time_evo} and~\ref{fig:spec_time_evo}) for models where we arbitrarily vary the rates of dust formation/creation from stellar AGB outflows and SNe. For the ``Elemental'' implementation we tested a model in which we (1) decreased the mass of dust created in SNe by a systematic factor of 10, (2) did the same for dust created by AGB stars, and (3) replaced our default ``Elemental'' dust creation rates with the default creation rates from the ``Species'' model. For the ``Species'' implementation we tested (1) increasing the mass of dust formed by SNe by a systematic factor of 10, (2) did the same for dust produced by AGB stars, and (3) replaced our default ``Species'' dust creation rates with the default creation rates from the ``Elemental'' model. These variations cause the initial dust population to differ drastically, in composition and amount, at early times but these differences quickly subside as gas-dust accretion takes over, becoming the dominant source of dust mass growth and producing near identical galaxy-integrated D/Z and dust composition at simulation end.

\section{Importance of the Coulomb Enhancement and Molecular CO Terms} \label{Appendix_Coulomb}

\begin{figure*}
	\plotsidesize{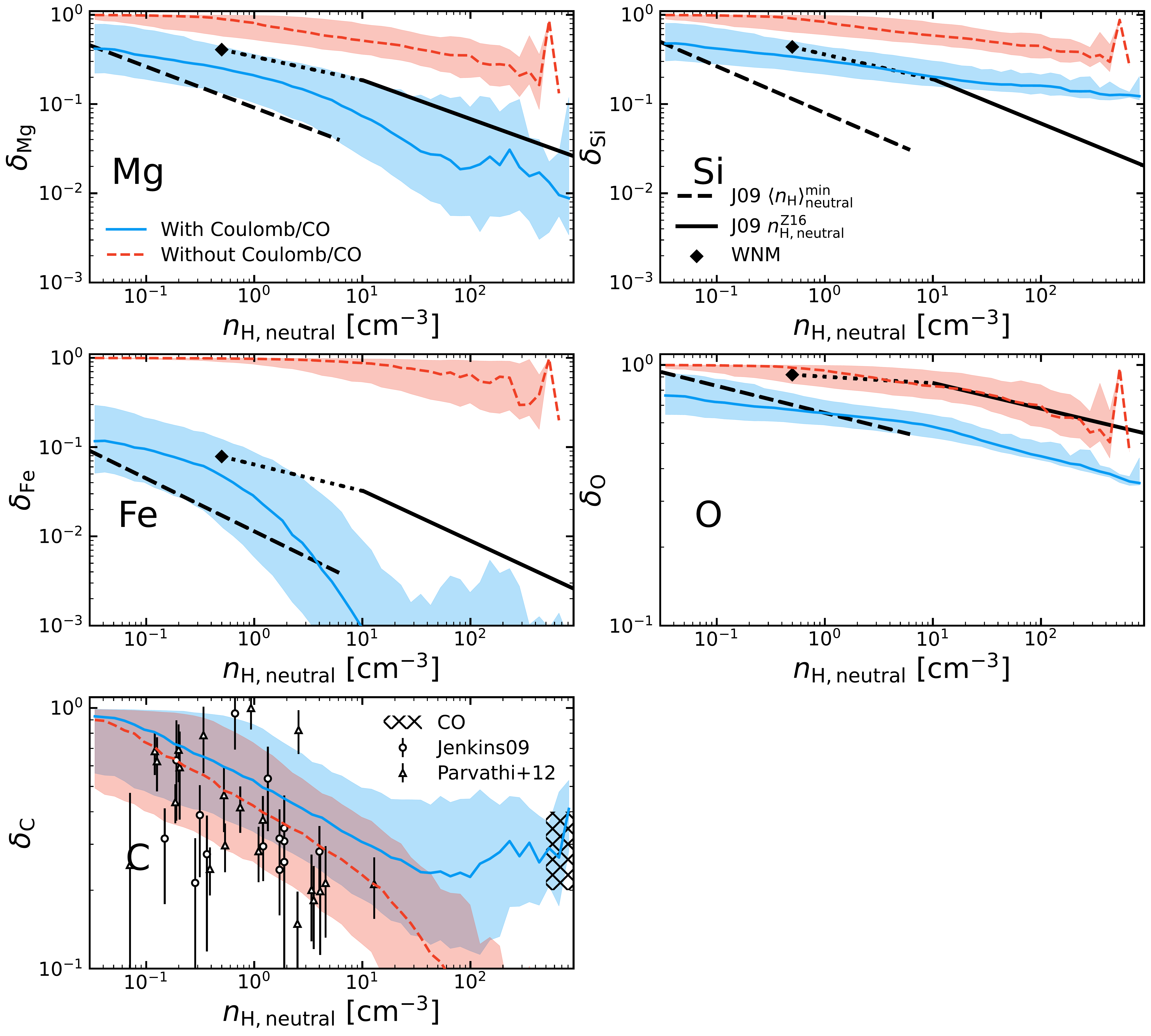}{0.99}
    \vspace{-0.25cm}
    \caption{Same as Fig.~\ref{fig:depl_vs_nH} comparing our default ``Species'' implementation (including Nano-iron and O-reservoir dust species) with ({\it solid}) and without ({\it dashed}) the default terms which attempt to account for Coulomb enhancement of gas-phase accretion rates, and the fraction of C unavailable to dust as it is locked into CO. The former has an appreciable effects on the predicted depletion of Mg, Si, Fe, and O. The latter only influences C depletion at high densities: without it, all C is locked in dust which of course would be inconsistent with observed CO abundances.}
    \label{fig:Coulomb_depl_vs_nH}
\end{figure*}

\begin{figure}
	\includegraphics[width=0.99\columnwidth]{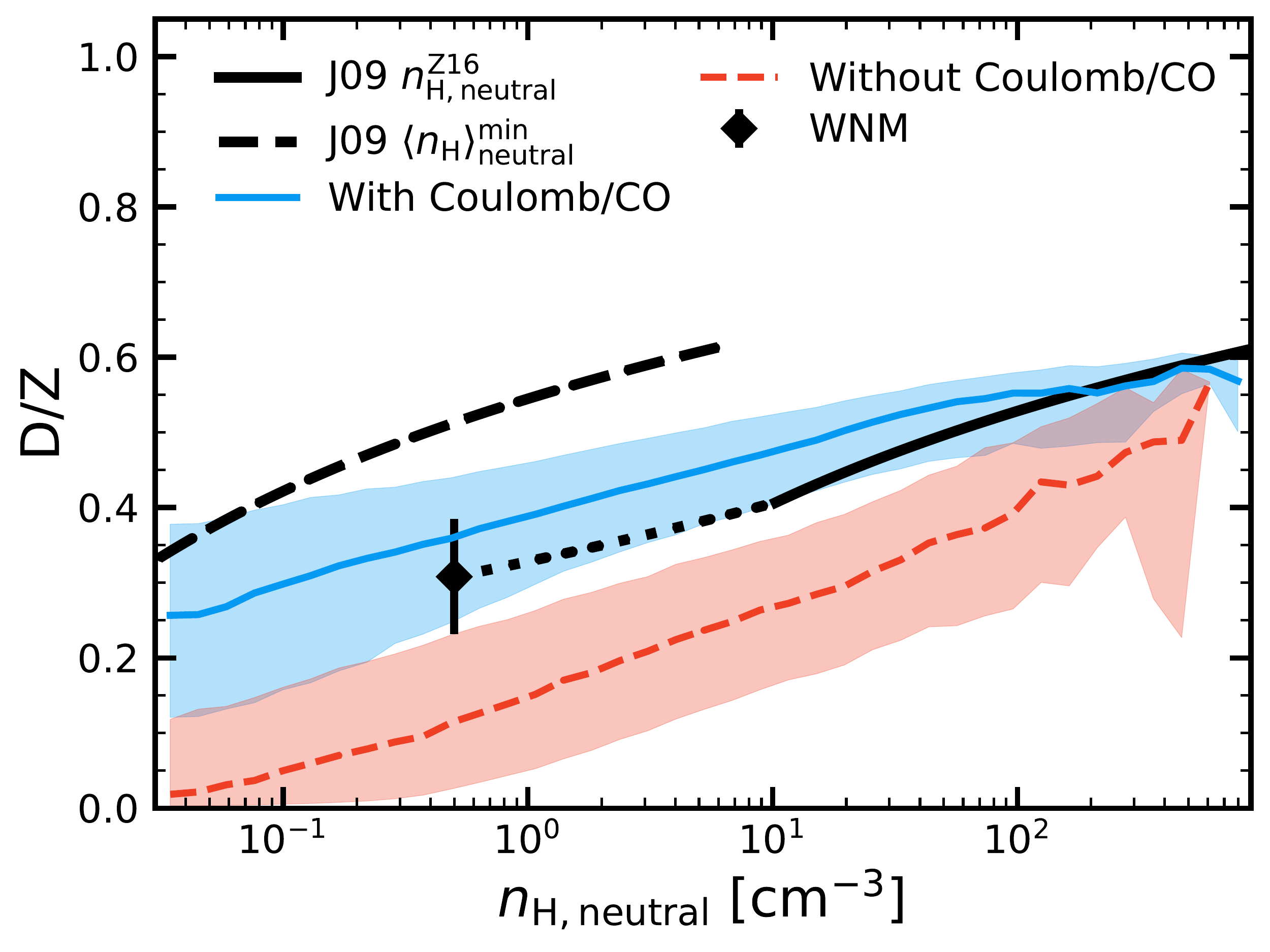}
    \vspace{-0.25cm}
    \caption{Same as Fig.~\ref{fig:DZ_vs_nH} comparing our default ``Species'' implementation (including Nano-iron and O-reservoir dust species) with ({\it solid}) and without ({\it dashed}) the default terms which attempt to account for Coulomb enhancement of gas-phase accretion rates, and the fraction of C unavailable to dust as it is locked into CO. The former systematically lowers D/Z at all densities. The latter increases the D/Z in dense gas but is the result of near all C locked in dust.}
    \label{fig:Coulomb_DZ_vs_nH}
\end{figure}

In Fig.~\ref{fig:Coulomb_depl_vs_nH} and Fig.~\ref{fig:Coulomb_DZ_vs_nH} we show the resulting element depletions and D/Z relation versus $n_{\rm H,neutral}$ at the end of the simulation for the ``Species'' implementation (including Nano-iron and O-reservoir dust species) with and without including our default Coulomb enhancement term and removing the fraction of C locked in CO. Examining the element depletions for Mg, Si, and Fe it is apparent that at least for our default assumptions, without accounting for the Coulomb enhancement term, accretion onto silicates and metallic iron dust is too slow, sequestering too little metal mass into dust. In the case of C, accretion rates for carbonaceous dust are hardly changed by Coulomb enhancement (see Table~\ref{tab:AccretionSummary}), but explicitly accounting for the fraction of C in CO has noticeable effects. When C in CO is not accounted for, the expected C depletion in dense environments is too high, consuming nearly all C (so, by construction, not enough residual C would be available for the observed CO). These changes to the element depletions result in a systematically lower D/Z relation for all but the densest gas, but even here this is a result of too much C being locked up in dust.

\section{Effects of Different Metal Yields - A Comparison between FIRE-2 and FIRE-3} \label{Appendix_FIRE23}

\begin{figure}
	\includegraphics[width=0.99\columnwidth]{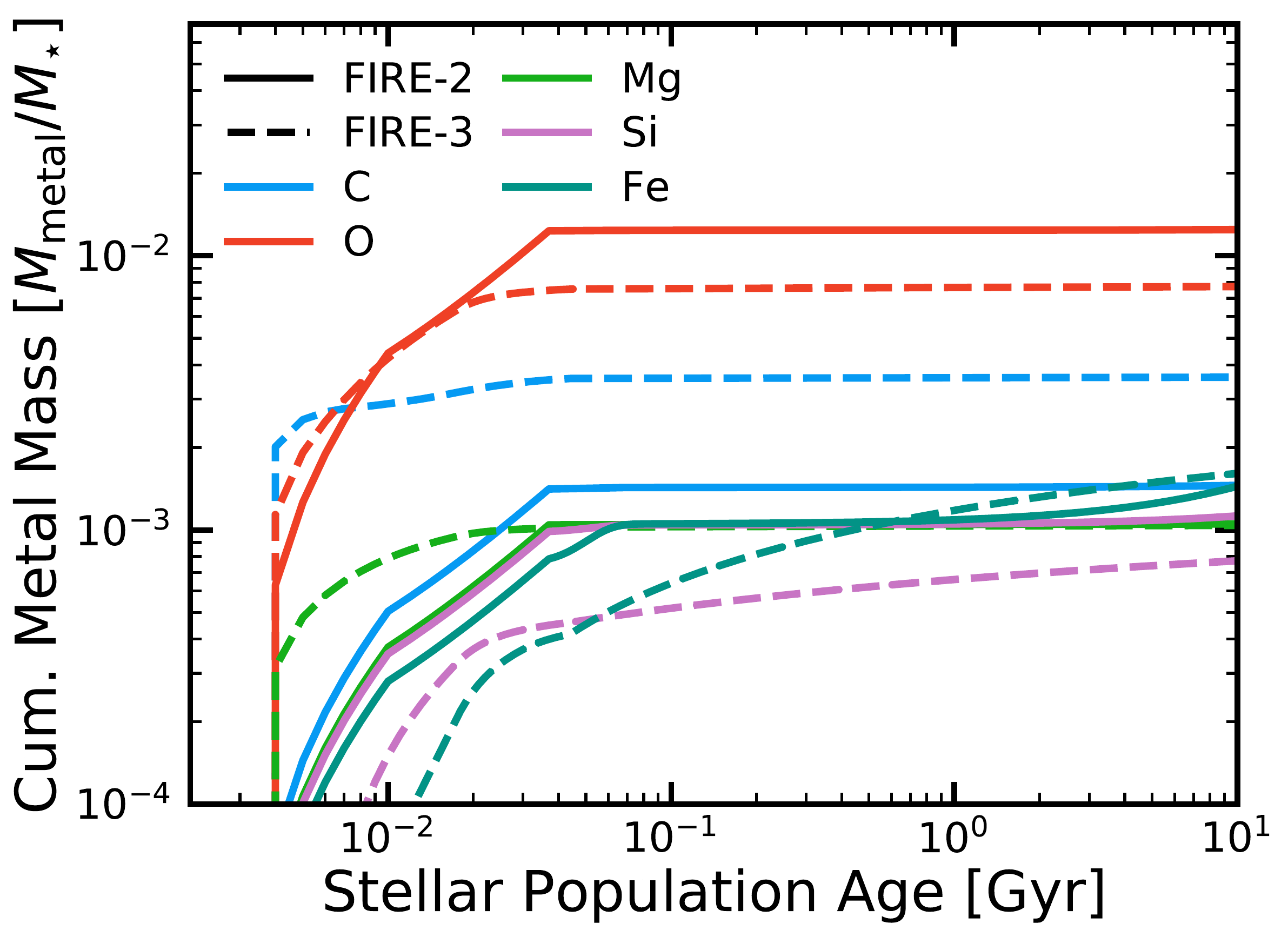}
    \vspace{-0.25cm}
    \caption{Cumulative SNe metal yields per stellar mass for main refractory elements in dust over a stellar population's life for the assumed yield models in FIRE-2 (from \citealt{nomoto_2006:NucleosynthesisYieldsCorecollapse}) and FIRE-3 (from the synthesis of \citealt{nomoto_2013:NucleosynthesisStarsChemical,pignatari_2016:NuGridStellarData,sukhbold_2016:CorecollapseSupernovae120, limongi_2018:PresupernovaEvolutionExplosive,prantzos_2018:ChemicalEvolutionRotating}). Note these yields are not metallicity dependant.}
    \label{fig:FIRE_metal_yields} 
\end{figure}

\begin{figure*}
    \plotsidesize{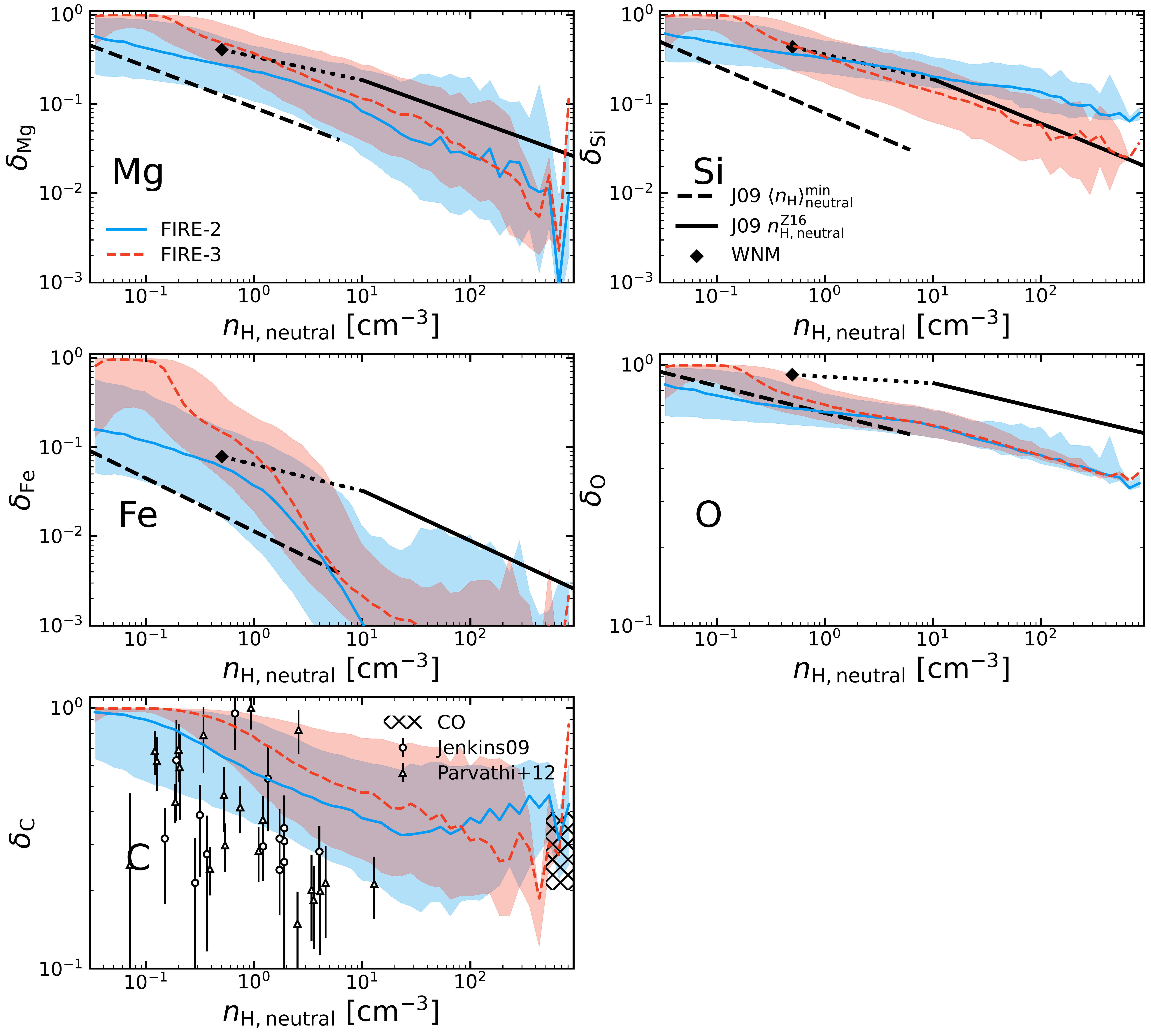}{0.99}
    \vspace{-0.25cm}
    \caption{Same as Fig.~\ref{fig:depl_vs_nH} comparing the ``Species'' implementation with O-reservoir and Nano-iron dust species integrated with FIRE-2 or FIRE-3 stellar feedback and ISM physics. The updated nucleosynthetic yields of FIRE-3 produce a better match to expected depletion trends for Si while not appreciably changing the depletion trends of the other elements over the range of observations.}
    \label{fig:FIRE23_depl_vs_nH}
\end{figure*}

\begin{figure}
	\includegraphics[width=0.99\columnwidth]{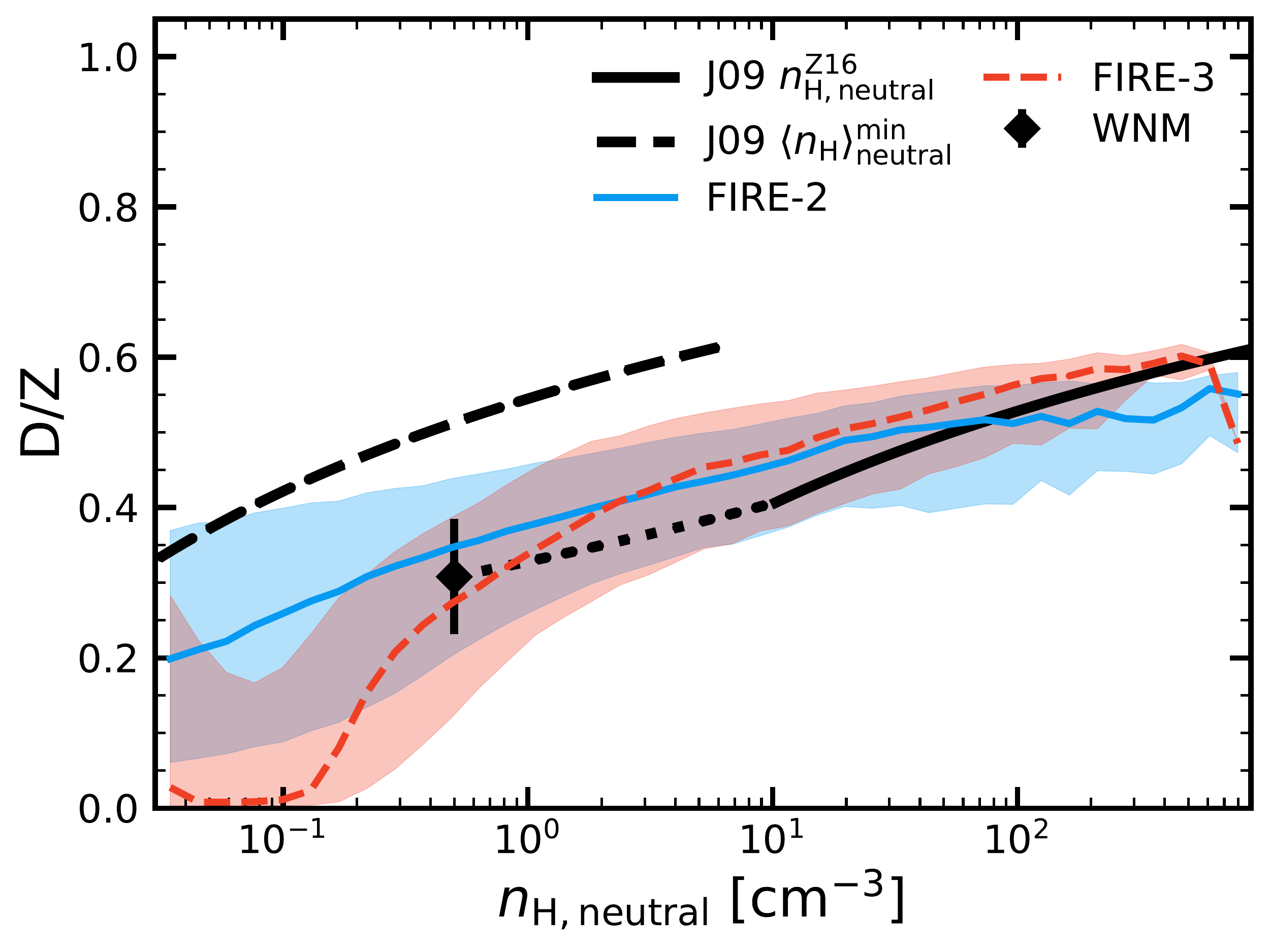}
    \vspace{-0.25cm}
    \caption{Same as Fig.~\ref{fig:DZ_vs_nH} comparing the ``Species'' implementation including Nano-iron and O-reservoir dust species integrated with FIRE-2 and FIRE-3 stellar feedback and ISM physics. The resulting D/Z trends do not diverge appreciably over the range of observations.}
    \label{fig:FIRE23_DZ_vs_nH}
\end{figure}

All our simulations in the main text used the FIRE-2 version of the FIRE code, following \citet{hopkins_2018:FIRE2SimulationsPhysics} with minor modifications as described in Section~\ref{GFM}. The next version of FIRE, FIRE-3 \citep{hopkins_2022:FIRE3UpdatedStellar}, makes a variety of improvements to the stellar inputs and numerical methods, focusing in particular on updating the stellar evolution tracks used for stellar feedback and nucleosynthesis with newer, more detailed models, as well as improving the detailed thermochemistry of cold atomic and molecular ISM gas, and adopting the newer \citet{asplund_2009:ChemicalCompositionSun} proto-solar reference abundances with $Z\sim0.014$.
We have made a preliminary comparison, running simulations with our dust models (specifically the ``Species'' model including both O-reservoir and Nano-iron species) coupled to the FIRE-3 instead of FIRE-2 models. While there are a variety of small differences, we find that the most important is related to the updated nucleosynthetic yields. The FIRE-3 yields include updated AGB mass loss rates which are reduced compared to FIRE-2 in better agreement with recent observational constraints \citep[e.g.][]{kriek_2010:SpectralEnergyDistribution,melbourne_2012:ContributionTPAGBRHeB,zibetti_2013:NearinfraredSpectroscopyPoststarburst,smith_2014:MassLossIts,hofner_2018:MassLossStars}, making the dust creation somewhat more dominated by SNe (though as with our default model, this has weak overall effect). The primary difference comes from the FIRE-3 core-collapse SNe yields, based on a synthesis of the updated yield models in \citet{nomoto_2013:NucleosynthesisStarsChemical,pignatari_2016:NuGridStellarData,sukhbold_2016:CorecollapseSupernovae120, limongi_2018:PresupernovaEvolutionExplosive,prantzos_2018:ChemicalEvolutionRotating}. These are compared to FIRE-2 in Fig.~\ref{fig:FIRE_metal_yields}. While C and O yields differ from FIRE-2 at an appreciable level, these actually have little effect on the steady-state dust population (influencing only the early-time production, for the reasons in Section~\ref{Free Parameters}). The most subtle but interesting change is that Mg is produced more promptly while Si is slightly reduced: this slightly increases the ratio of Mg to Si in FIRE-3, making Si instead of Mg the key element for silicate dust growth. In Fig.~\ref{fig:FIRE23_depl_vs_nH} \&~\ref{fig:FIRE23_DZ_vs_nH}, we see this produces a significantly steeper depletion-density relation for Si  while not appreciably changing the other element depletions and D/Z trends across the range of observations. It is however important to note that this is somewhat degenerate with our assumed dust chemical compositions: we could also make Si the key element in FIRE-2 by decreasing the assumed Mg-to-Si ratio ``$A_{\rm Mg}/A_{\rm Si}$'' for the mean silicate composition (see Section~\ref{Dust_by_Species}).

\bsp	
\label{lastpage}
\end{document}